\documentclass{article}
\usepackage[utf8]{inputenc}

\newcommand{\Gardenfors}{G\"ardenfors}

\newcommand{\St}{\mathsf{St}} %Todo: update this prob. 
\usepackage{amsmath}
\usepackage{amsthm}
\usepackage{amssymb}
\usepackage{stmaryrd}  % for ovee

\newcommand{\kto}{\to} 
\theoremstyle{definition}
\newtheorem{Def}{Definition}
\newtheorem{definition}[Def]{Definition}

\newtheorem{remark}{Remark}
\newtheorem{example}{Example}

\usepackage{tikz-cd}
\usepackage{stackengine}

\newcommand{\cat}[1]{\ensuremath{\mathbf{#1}}}
\newcommand{\catC}{\cat{C}}

\newcommand{\id}[1]{\ensuremath{\mathrm{id}_{#1}}}

\newcommand{\Quant}{\cat{Quant}}

\newcommand{\hilbH}{\mathcal{H}} %Hilbert space
\newcommand{\hilbK}{\mathcal{K}} %Hilbert space

\newcommand{\discard}[1]{\ensuremath{\tinygroundnew_{#1}}}

\usepackage{tikz,xypic}
\usetikzlibrary{decorations.pathreplacing,decorations.markings,arrows.meta,backgrounds}
\pgfdeclarelayer{edgelayer}
\pgfdeclarelayer{nodelayer}
\pgfsetlayers{background,edgelayer,nodelayer,main}
\tikzstyle{whitedot}=[circle, draw=black, fill=white, inner sep=.4ex]
\tikzstyle{none}=[inner sep=0mm]

\usepackage{tikz,xypic}

\usetikzlibrary{decorations.pathreplacing,decorations.markings,arrows.meta,backgrounds}
\pgfdeclarelayer{edgelayer}
\pgfdeclarelayer{nodelayer}
\pgfsetlayers{background,edgelayer,nodelayer,main}

\tikzstyle{cdot}=[circle, draw=black, fill=black!25, inner sep=.4ex] %Chris dot style 
\tikzstyle{bigdot}=[dot, inner sep=0pt]
\tikzstyle{whitedot}=[circle, draw=black, fill=white, inner sep=.4ex]
\tikzstyle{greydot}=[circle, draw=black, fill=black!25, inner sep=.4ex] %Added by Sean
\tikzstyle{blackdot}=[circle, draw=black, fill=black, inner sep=.4ex]
\tikzset{arrow/.style={decoration={
    markings,
    mark=at position #1 with \arrow{>[length=2pt, width=3pt]}},
    postaction=decorate},
    reverse arrow/.style={decoration={
    markings,
    mark=at position #1 with {{\arrow{<[length=2pt, width=3pt]}}}},
    postaction=decorate},
    thick/.style = {line width=0.15em}
}

\newcommand{\tinycomult}[1][cdot]{
\smash{\raisebox{-2pt}{\hspace{-5pt}\ensuremath{\begin{pic}[scale=0.4,yscale=1]
    \node (0) at (0,0) {};
    \node[#1, inner sep=1.5pt] (1) at (0,0.55) {};
    \node (2) at (-0.5,1) {};
    \node (3) at (0.5,1) {};
    \draw (0.center) to (1.center);
    \draw (1.center) to [out=left, in=down, out looseness=1.5] (2.center);
    \draw (1.center) to [out=right, in=down, out looseness=1.5] (3.center);
    \node[#1, inner sep=1.5pt] (1) at (0,0.55) {};
\end{pic}
}\hspace{-3pt}}}}

\newenvironment{pic}[1][] {\begin{aligned}\begin{tikzpicture}[scale=2.0, font=\tiny,#1]}{\end{tikzpicture}\end{aligned}} %Hard coded picture 

\newif\ifvflip\pgfkeys{/tikz/vflip/.is if=vflip}
\newif\ifhflip\pgfkeys{/tikz/hflip/.is if=hflip}
\newif\ifhvflip\pgfkeys{/tikz/hvflip/.is if=hvflip}

\newenvironment{picc}[1][]
{\begin{aligned}\begin{tikzpicture}[font=\tiny,#1]}
{\end{tikzpicture}\end{aligned}}

\newlength\minimummorphismwidth
\newlength\stateheight
\newlength\minimumstatewidth
\newlength\connectheight
\tikzset{colour/.initial=white}

\tikzstyle{pure}=[line width=.7pt]

\makeatletter

\pgfdeclareshape{groundd}
{
    \savedanchor\centerpoint
    {
        \pgf@x=0pt
        \pgf@y=0pt
    }
    \anchor{center}{\centerpoint}
    \anchorborder{\centerpoint}

    \anchor{north}
    {
        \pgf@x=0pt
        \pgf@y=0.16\stateheight
    }
    \anchor{south}
    {
        \pgf@x=0pt
        \pgf@y=0pt
    }
    \saveddimen\overallwidth
    {
        \pgfkeysgetvalue{/pgf/minimum width}{\minwidth}
        \pgf@x=\minimumstatewidth
        \ifdim\pgf@x<\minwidth
            \pgf@x=\minwidth
        \fi
    }
    \backgroundpath
    {
        \begin{pgfonlayer}{main} %Sean: this used to be 'foreground', swtich back if trouble
        \pgfsetstrokecolor{black}
        \pgfsetlinewidth{1.25pt}
        \ifhflip
            \pgftransformyscale{-1}
        \fi
        \pgftransformscale{0.5}
        \pgfpathmoveto{\pgfpoint{-0.5*\overallwidth}{0pt}}
        \pgfpathlineto{\pgfpoint{0.5*\overallwidth}{0pt}}
        \pgfpathmoveto{\pgfpoint{-0.33*\overallwidth}{0.33*\stateheight}}
        \pgfpathlineto{\pgfpoint{0.33*\overallwidth}{0.33*\stateheight}}
        \pgfpathmoveto{\pgfpoint{-0.16*\overallwidth}{0.66*\stateheight}}
        \pgfpathlineto{\pgfpoint{0.16*\overallwidth}{0.66*\stateheight}}
        \pgfpathmoveto{\pgfpoint{-0.02*\overallwidth}{\stateheight}}
        \pgfpathlineto{\pgfpoint{0.02*\overallwidth}{\stateheight}}
        \pgfusepath{stroke}
        \end{pgfonlayer}
    }
}

\usepackage{tikz,xypic}
\usetikzlibrary{decorations.pathreplacing,decorations.markings,arrows.meta,backgrounds,shapes}
\usetikzlibrary{circuits.ee.IEC}
\pgfdeclarelayer{edgelayer}
\pgfdeclarelayer{nodelayer}
\pgfsetlayers{background,edgelayer,nodelayer,main}
\tikzstyle{none}=[inner sep=0mm]
\tikzstyle{every loop}=[]
\tikzstyle{mark coordinate}=[inner sep=0pt,outer sep=0pt,minimum size=3pt,fill=black,circle]

\tikzset{arrow/.style={decoration={
    markings,
    mark=at position #1 with \arrow{>[length=2pt, width=3pt]}},
    postaction=decorate},
    reverse arrow/.style={decoration={
    markings,
    mark=at position #1 with {{\arrow{<[length=2pt, width=3pt]}}}},
    postaction=decorate}
}

\tikzstyle{upground}=[circuit ee IEC,thick,ground,rotate=90,scale=1.5]
\tikzstyle{upgroundwhite}=[circuit ee IEC,thick,ground,rotate=90,scale=1.5, fill=white]
\tikzstyle{downground}=[circuit ee IEC,thick,ground,rotate=-90,scale=1.5]
\tikzstyle{downgroundnorm}=[circuit ee IEC,thick,ground,rotate=-90,scale=1.5, fill=white]
\newcommand{\mapminh}{5mm} % Aleks and Bob have 6mm

\newcommand{\maplw}{0.7pt} % Alesk and Bob previously just left as default
\tikzstyle{box}=[map]
\tikzstyle{medium box}=[medium map]
\tikzstyle{dot}=[inner sep=0mm,minimum width=2mm,minimum height=2mm,draw,shape=circle]  
\tikzstyle{black dot}=[dot,fill=black]
\tikzstyle{white dot}=[dot,fill=white,,text depth=-0.2mm]
\tikzstyle{grey dot}=[dot,fill=black!25] %Added by Sean

\tikzstyle{corner1}=[box,fill=white, font=\footnotesize] %
\tikzstyle{corner2}=[dot,fill=white, font=\footnotesize] %
\tikzstyle{corner3}=[dot,fill=black!25, font=\footnotesize] %
\tikzstyle{corner4}=[dot,fill=black, font=\footnotesize] %

\tikzstyle{scalar}=[circle,draw,inner sep=2pt, line width=\maplw] %,font=\small] %Consider changing to a circle!

\usetikzlibrary{shapes.misc, positioning}

\tikzset{stateshape/.style={append after command={
   \pgfextra
        \draw[sharp corners, fill=white, line width = \maplw]% 
    (\tikzlastnode.west)% 
    [rounded corners=0pt] |- (\tikzlastnode.north)% 
    [rounded corners=0pt] -| (\tikzlastnode.east)% 
    [rounded corners=5pt] |- (\tikzlastnode.south)% 
    [rounded corners=5pt] -| (\tikzlastnode.west);
   \endpgfextra}}}

\tikzset{effectshape/.style={append after command={
   \pgfextra
        \draw[sharp corners, fill=white, line width = \maplw]% 
    (\tikzlastnode.west)% 
    [rounded corners=0pt] |- (\tikzlastnode.south)% 
    [rounded corners=0pt] -| (\tikzlastnode.east)% 
    [rounded corners=5pt] |- (\tikzlastnode.north)% 
    [rounded corners=5pt] -| (\tikzlastnode.west);
   \endpgfextra}}}

 \tikzstyle{map}=[draw,shape=rectangle, inner sep=2pt,minimum height=\mapminh, minimum width=5mm,fill=white]

\tikzstyle{point}=[fill=white,draw,shape=isosceles triangle,shape border rotate=-90,isosceles triangle stretches=true,inner sep=0.2pt,minimum width=0.5cm,minimum height=0.8mm,yshift=-0.0mm]
\tikzstyle{copoint}=[fill=white,draw,shape=isosceles triangle,shape border rotate=90,isosceles triangle stretches=true,inner sep=0.2pt,minimum width=0.5cm,minimum height=0.8mm,yshift=-0.0mm]
\tikzstyle{wide point}=[point, minimum width=12mm]
\tikzstyle{wide copoint}=[copoint, minimum width=12mm]

\tikzstyle{decomp}=[fill=white,draw,shape=isosceles triangle,shape border rotate=-90,isosceles triangle stretches=true,inner sep=0pt,minimum width=0.75cm,minimum height=4mm,yshift=-0.0mm]

\tikzstyle{decompwide}=[fill=white,draw,shape=isosceles triangle,shape border rotate=-90,isosceles triangle stretches=true,inner sep=0pt,minimum width=1.4cm,minimum height=5mm,yshift=-0.0mm]

\tikzstyle{decompflip}=[fill=white,draw,shape=isosceles triangle,shape border rotate=90,isosceles triangle stretches=true,inner sep=0pt,minimum width=0.75cm,minimum height=4mm,yshift=-0.0mm]

\tikzstyle{decompwideflip}=[fill=white,draw,shape=isosceles triangle,shape border rotate=90,isosceles triangle stretches=true,inner sep=0pt,minimum width=1.4cm,minimum height=5mm,yshift=-0.0mm]

\tikzstyle{medium map} = [map, minimum width = 12mm] 
\tikzstyle{semilarge map} = [map, minimum width = 15mm] 
\tikzstyle{large map} = [map, minimum width = 18mm] 

\tikzstyle{kpoint} =[point]
\tikzstyle{kpointadj} =[copoint]
\tikzstyle{kpointconj}=[dagpointconj] %Need daggers for conjugates
\makeatletter
\newcommand{\boxshape}[3]{%
\pgfdeclareshape{#1}{
\inheritsavedanchors[from=rectangle] % this is nearly a rectangle
\inheritanchorborder[from=rectangle]
\inheritanchor[from=rectangle]{center}
\inheritanchor[from=rectangle]{north}
\inheritanchor[from=rectangle]{south}
\inheritanchor[from=rectangle]{west}
\inheritanchor[from=rectangle]{east}
\backgroundpath{% this is new
\southwest \pgf@xa=\pgf@x \pgf@ya=\pgf@y
\northeast \pgf@xb=\pgf@x \pgf@yb=\pgf@y

\@tempdima=#2
\@tempdimb=#3

\pgfpathmoveto{\pgfpoint{\pgf@xa - 5pt + \@tempdima}{\pgf@ya}}
\pgfpathlineto{\pgfpoint{\pgf@xa - 5pt - \@tempdima}{\pgf@yb}}
\pgfpathlineto{\pgfpoint{\pgf@xb + 5pt + \@tempdimb}{\pgf@yb}}
\pgfpathlineto{\pgfpoint{\pgf@xb + 5pt - \@tempdimb}{\pgf@ya}}
\pgfpathlineto{\pgfpoint{\pgf@xa - 5pt + \@tempdima}{\pgf@ya}}
\pgfpathclose
}
}}

\boxshape{NEbox}{0pt}{3pt} %My setting of 3pt
\boxshape{SEbox}{0pt}{-3pt}
\boxshape{NWbox}{3pt}{0pt}
\boxshape{SWbox}{-3pt}{0pt}
\boxshape{rec-box}{0pt}{0pt}
\makeatother

\tikzstyle{cloud}=[shape=cloud,draw,minimum width=1.5cm,minimum height=1.5cm]

\tikzstyle{dagmap}=[draw,shape=NEbox,inner sep=2pt,minimum height=\mapminh,fill=white, line width = \maplw] %
\tikzstyle{dashedmap}=[draw,dashed,shape=NEbox,inner sep=2pt,minimum height=\mapminh,fill=white, line width = \maplw]
\tikzstyle{mapdag}=[draw,shape=SEbox,inner sep=2pt,minimum height=\mapminh,fill=white, line width = \maplw]
\tikzstyle{mapadj}=[draw,shape=SEbox,inner sep=2pt,minimum height=\mapminh,fill=white, line width = \maplw]
\tikzstyle{maptrans}=[draw,shape=SWbox,inner sep=2pt,minimum height=\mapminh,fill=white, line width = \maplw]
\tikzstyle{mapconj}=[draw,shape=NWbox,inner sep=2pt,minimum height=\mapminh,fill=white, line width = \maplw]

\tikzstyle{medium dagmap}=[draw,shape=NEbox,inner sep=2pt,minimum height=\mapminh,fill=white,minimum width=7mm, line width = \maplw]
\tikzstyle{semilarge dagmap}=[draw,shape=NEbox,inner sep=2pt,minimum height=\mapminh,fill=white,minimum width=9.5mm, line width = \maplw]
\tikzstyle{large dagmap}=[draw,shape=NEbox,inner sep=2pt,minimum height=\mapminh,fill=white,minimum width=12mm, line width = \maplw]
\makeatletter

\pgfdeclareshape{cornerpoint}{
\inheritsavedanchors[from=rectangle] % this is nearly a rectangle
\inheritanchorborder[from=rectangle]
\inheritanchor[from=rectangle]{center}
\inheritanchor[from=rectangle]{north}
\inheritanchor[from=rectangle]{south}
\inheritanchor[from=rectangle]{west}
\inheritanchor[from=rectangle]{east}
\backgroundpath{% this is new
\southwest \pgf@xa=\pgf@x \pgf@ya=\pgf@y
\northeast \pgf@xb=\pgf@x \pgf@yb=\pgf@y

\pgfmathsetmacro{\pgf@shorten@left}{\pgfkeysvalueof{/tikz/shorten left}}
\pgfmathsetmacro{\pgf@shorten@right}{\pgfkeysvalueof{/tikz/shorten right}}

\pgfpathmoveto{\pgfpoint{0.5 * (\pgf@xa + \pgf@xb)}{\pgf@ya - 5pt}}
\pgfpathlineto{\pgfpoint{\pgf@xa - 8pt + \pgf@shorten@left}{\pgf@yb - 1.5 * \pgf@shorten@left}}
\pgfpathlineto{\pgfpoint{\pgf@xa - 8pt + \pgf@shorten@left}{\pgf@yb}}
\pgfpathlineto{\pgfpoint{\pgf@xb + 8pt - \pgf@shorten@right}{\pgf@yb}}
\pgfpathlineto{\pgfpoint{\pgf@xb + 8pt - \pgf@shorten@right}{\pgf@yb - 1.5 * \pgf@shorten@right}}
\pgfpathclose
}
}

\pgfdeclareshape{cornercopoint}{
\inheritsavedanchors[from=rectangle] % this is nearly a rectangle
\inheritanchorborder[from=rectangle]
\inheritanchor[from=rectangle]{center}
\inheritanchor[from=rectangle]{north}
\inheritanchor[from=rectangle]{south}
\inheritanchor[from=rectangle]{west}
\inheritanchor[from=rectangle]{east}
\backgroundpath{% this is new
\southwest \pgf@xa=\pgf@x \pgf@ya=\pgf@y
\northeast \pgf@xb=\pgf@x \pgf@yb=\pgf@y

\pgfmathsetmacro{\pgf@shorten@left}{\pgfkeysvalueof{/tikz/shorten left}}
\pgfmathsetmacro{\pgf@shorten@right}{\pgfkeysvalueof{/tikz/shorten right}}

\pgfpathmoveto{\pgfpoint{0.5 * (\pgf@xa + \pgf@xb)}{\pgf@yb + 5pt}}
\pgfpathlineto{\pgfpoint{\pgf@xa - 8pt + \pgf@shorten@left}{\pgf@ya + 1.5 * \pgf@shorten@left}}
\pgfpathlineto{\pgfpoint{\pgf@xa - 8pt + \pgf@shorten@left}{\pgf@ya}}
\pgfpathlineto{\pgfpoint{\pgf@xb + 8pt - \pgf@shorten@right}{\pgf@ya}}
\pgfpathlineto{\pgfpoint{\pgf@xb + 8pt - \pgf@shorten@right}{\pgf@ya + 1.5 * \pgf@shorten@right}}
\pgfpathclose
}
}

\makeatother

\pgfkeyssetvalue{/tikz/shorten left}{0pt}
\pgfkeyssetvalue{/tikz/shorten right}{0pt}

\tikzstyle{dagpoint common}=[draw,fill=white,inner sep=1pt, line width = \maplw, minimum height = 4mm, yshift=1.2pt] %SEAN: I ADDED THIS yshift to make it line up with maps. 1pt seems to work.
\tikzstyle{dagpoint sc}=[shape=cornerpoint,dagpoint common]
\tikzstyle{dagpoint adjoint sc}=[shape=cornercopoint,dagpoint common]
\tikzstyle{dagpoint}=[shape=cornerpoint,shorten left=4pt,dagpoint common]
\tikzstyle{dagpointadj}=[shape=cornercopoint,shorten left=5pt,dagpoint common]
\tikzstyle{dagpointconj}=[shape=cornerpoint,shorten right=5pt,dagpoint common]
\tikzstyle{dagpointtrans}=[shape=cornercopoint,shorten right=5pt,dagpoint common]
\tikzstyle{dagpointsymm}=[shape=cornerpoint,shorten left=5pt,shorten right=5pt,dagpoint common]

\tikzstyle{widedagpoint}=[dagpoint, minimum width=1 cm, inner sep=2pt]%, text depth=-0.7 mm]
\tikzstyle{widedagpointadj}=[dagpointadj, minimum width=1 cm, inner sep=2pt]%, text depth=0.7 mm]

\tikzstyle{every picture}=[baseline=-0.25em,scale=0.5]
\tikzstyle{label}=[font=\footnotesize,text height=1ex, text depth=0.15ex]

\usetikzlibrary[shapes]

\tikzset{
sidetriangle/.style = {regular polygon, regular polygon sides = 3, aspect = 1, shape border rotate = 90, draw, inner sep = 0, minimum width = 1.2cm}
}

\tikzset{
isoc/.style = {shape=isosceles triangle, shape border rotate = 180, isosceles triangle stretches = true, minimum width = 1.2cm, minimum height= 1.5cm, inner sep = 0.3}}

\tikzset{
coarse/.style = {shape = circle, fill = white, draw, inner sep = 0, minimum width =0.125cm}
}
\tikzset{
coarsesymbol/.style = {shape = circle, fill = white, inner sep = -0.7, minimum width = 0.125cm}
}
\tikzstyle{sidetriangle2}=[sidetriangle, minimum width = 2cm, fill=white]
\tikzstyle{sideisocsmall}]=[style=isoc, minimum width = 1cm, minimum height = 0.8cm, draw, fill=white, font=\Large]
\tikzstyle{sideisoc}]=[style=isoc, minimum width = 2cm, draw, fill=white, font=\Large]
\tikzstyle{sideisocmid}]=[style=isoc, minimum width = 2.5cm, draw, fill=white, font=\Large]
\tikzstyle{sideisocmedium}]=[style=isoc, minimum width = 3cm, draw, fill=white, font=\Large]

\newcommand{\tinygroundnew}{
\smash{
{\hspace{-3pt}
\ensuremath{
\begin{picc}[scale=1.0] 
    \node[upground, xscale=0.8, yscale=0.7] (1) at (0,0.16) {};
    \draw (0,0.03) to (0,-0.25);
\end{picc}
}\hspace{-1pt}}}}

\tikzstyle{label}=[font=\footnotesize,text height=1ex, text depth=0.15ex]

\tikzstyle{box}=[map]
\tikzstyle{medium box}=[medium map]
\tikzstyle{dot}=[inner sep=0mm,minimum width=2mm,minimum height=2mm,draw,shape=circle]  
\tikzstyle{black dot}=[dot,fill=black]
\tikzstyle{white dot}=[dot,fill=white,,text depth=-0.2mm]
\tikzstyle{grey dot}=[dot,fill=black!25] %Added by Sean

\tikzstyle{corner1}=[box,fill=white, font=\footnotesize] %
\tikzstyle{corner2}=[dot,fill=white, font=\footnotesize] %
\tikzstyle{corner3}=[dot,fill=black!25, font=\footnotesize] %
\tikzstyle{corner4}=[dot,fill=black, font=\footnotesize] %

\tikzstyle{scalar}=[circle,draw,inner sep=2pt, line width=\maplw] %,font=\small] %Consider changing to a circle!

\tikzstyle{sharpstate}=[fill=white,draw,shape=isosceles triangle,shape border rotate=-90,isosceles triangle stretches=true,inner sep=0.2pt,minimum width=0.5cm,minimum height=0.8mm,yshift=-0.0mm]
\tikzstyle{sharpeffect}=[fill=white,draw,shape=isosceles triangle,shape border rotate=90,isosceles triangle stretches=true,inner sep=0.2pt,minimum width=0.5cm,minimum height=0.8mm,yshift=-0.0mm]
\tikzstyle{wide sharpstate}=[point, minimum width=12mm]
\tikzstyle{wide sharpeffect}=[copoint, minimum width=12mm]

\tikzstyle{point}=[sharpstate]

\usepackage{tikzit}
\usepackage{graphicx}
\usepackage{apacite}
\usepackage{booktabs}

\newcommand{\cc}{C} %Fuzzy Concept notation in formal section 
\newcommand{\crc}{C} %Crisp Concept notation in formal section 

\newcommand{\Class}{\cat{Class}}
\newcommand{\Prob}{\cat{Prob}}
\newcommand{\ConSp}{\cat{ConSp}}
\newcommand{\Tr}{\mathsf{Tr}}

\newcommand{\RXYZ}{R}
\newcommand{\RZYX}{R'}

\newcommand{\RX}[1]{R^X_{\theta_{#1,X}}}
\newcommand{\RY}[1]{R^Y_{\theta_{#1,Y}}}
\newcommand{\RZ}[1]{R^Z_{\theta_{#1,Z}}}

\newcommand{\RXC}[1]{R^X_{\thetac_{#1,X}}}
\newcommand{\RYC}[1]{R^Y_{\thetac_{#1,Y}}}
\newcommand{\RZC}[1]{R^Z_{\thetac_{#1,Z}}}

\newcommand{\Hdom}[1]{\hilbH_{\text{#1}}}

\newcommand{\thetac}{\phi^C} % Params for a concept
\newcommand{\Ha}{\hilbH_1} % Params for a concept
\newcommand{\Hb}{\hilbH_2} % Params for a concept
\newcommand{\Hc}{\hilbH_3} % Params for a concept
\newcommand{\Hd}{\hilbH_4} % Params for a concept

\usepackage{braket}

\title{Formalising and Learning a Quantum Model of Concepts}
\author{Sean Tull, Razin A. Shaikh, Sara Sabrina Zemlji\v{c} and Stephen Clark\\
Quantinuum\\
17 Beaumont Street, Oxford, UK\\
\texttt{\normalsize \{sean.tull,razin.shaikh,sara.zemljic,steve.clark\}@quantinuum.com}}

\begin{document}

\maketitle 

\begin{abstract}
In this report we present a new modelling framework for concepts based on quantum theory, and  demonstrate how the conceptual representations can be learned automatically from data. A contribution of the work is a thorough category-theoretic formalisation of our framework. We claim that the use of category theory, and in particular the use of string diagrams to describe quantum processes, helps elucidate some of the most important features of our quantum approach to concept modelling. Our approach builds upon \Gardenfors' classical framework of \emph{conceptual spaces}, in which cognition is modelled geometrically through the use of convex spaces, which in turn factorise in terms of simpler spaces called \emph{domains}. We show how concepts from the domains of \textsc{shape}, \textsc{colour}, \textsc{size} and \textsc{position} can be learned from images of simple shapes, where individual images are represented as quantum states and concepts as quantum effects. Concepts are learned by a hybrid classical-quantum network trained to perform concept classification, where the classical image processing is carried out by a convolutional neural network and the quantum representations are produced by a parameterised quantum circuit. We also use discarding to produce mixed effects, which can then be used to learn concepts which only apply to a subset of the domains, and show how entanglement (together with discarding) can be used to capture interesting correlations across domains. Finally, we consider the question of whether our quantum models of concepts can be considered conceptual spaces in the \Gardenfors\ sense.
\end{abstract}

\section{Introduction}

The study of concepts has a long history in a number of related fields, including philosophy, linguistics, psychology and cognitive science \cite{murphy_concepts,conceptual_mind}. More recently, researchers have begun to consider how mathematical tools from quantum theory can be used to model cognitive phenomena, including conceptual structure. The general use of quantum formalism in psychology and cognitive science has led to an emerging area called quantum cognition \cite{aerts2009,pothos2013bbs}. The idea is that some of the features of quantum theory, such as entanglement, can be used to account for psychological data which can be hard to model classically. Examples include ordering effects in how subjects answer questions \cite{Trueblood2011} and concept combination \cite{aerts_gabora2005,tomas2015}.\footnote{Note that no claims are typically being made in such work (and this report) regarding the existence of quantum processes in the brain, only that some cognitive processes can be effectively modelled at an abstract level using the quantum formalism.}

Another recent development in the study of concepts has been the application of machine learning to the problem of how artificial agents can automatically learn concepts from raw perceptual data \shortcite{beta-vae,SCAN,shaikh2022}. The motivation for endowing an agent with conceptual representations, and learning those representations automatically from the agent's environment, is that this will enable it to reason and act more effectively in that environment, similar to how humans use concepts \shortcite{lake_thinking_machines}. One hope is that the explicit use of concepts will ameliorate some of the negative consequences of the ``black-box" nature of neural architectures currently being used in AI.

In this report we present a new modelling framework for concepts based on the mathematical formalism used in quantum theory, and demonstrate how the conceptual representations can be learned automatically from data. A contribution of the work is a thorough category-theoretic formalisation of our framework, following \citeA{bolt2019interacting} and \citeA{tull2021categorical}. Formalisation of conceptual models is not new \cite{ganter:2016}, but we claim that the use of category theory \cite{applied_cat_theory}, and in particular the use of string diagrams to describe quantum processes \cite{coecke_kissinger_2017}, helps elucidate some of the most important features of our quantum approach to concept modelling. This aspect of our work also fits with the recent push to introduce category theory into machine learning and AI more broadly. The motivation is to make deep learning less ad-hoc and less driven by heuristics, by viewing deep learning models through the compositional lens of category theory \shortcite{cat_ml}.

\citeA[p.1]{murphy_concepts} describes concepts as ``the glue that holds our mental world together". But how should concepts be modelled and represented mathematically? There are many modelling frameworks in the literature, including the \emph{classical theory} \cite{margolis_stanford}, the \emph{prototype theory} \cite{rosch1973natural}, and the \emph{theory theory} \cite{gopnik}.
Here we build upon \Gardenfors' framework of \emph{conceptual spaces} \cite{gardenfors2004conceptual,gardenfors2014geometry}, in which cognition is modelled geometrically through the use of convex spaces, which in turn factorise in terms of simpler spaces called \emph{domains}. We show how concepts from the domains of \textsc{shape}, \textsc{colour}, \textsc{size} and \textsc{position} can be learned from images of simple shapes, where individual images are represented as quantum states and concepts as quantum effects. The factoring of the conceptual space is represented naturally in our models through the use of the tensor product as the monoidal product. We also show how discarding---which produces mixed effects---can be used when the concept to be learned only applies to a subset of the domains, and how entanglement (together with discarding) can be used to capture interesting correlations across domains.

We choose to implement our modelling framework using a hybrid classical-quantum network trained to perform concept classification, where the classical image processing is carried out by a convolutional neural network \shortcite[Ch.9]{deep_learning} and the quantum representations are produced by a parameterised quantum circuit \shortcite{Benedetti2019}. Even though the framework has been described at an abstract level independent of any particular implementation, the use-case we have in mind is one in which the models are (eventually) run on a quantum computer, exploiting the potential advantages such computers may bring. Here the implementation is a classical simulation of a quantum computation.\footnote{Note that we are not making any claims of ``quantum advantage" \cite{preskill} for the particular set of models that we implement in this report. However, we do anticipate the possibility of quantum models of concepts satisfying our framework which require quantum hardware for their efficient training and deployment, especially as we scale to more realistic datasets and larger quantum circuits.}

This report is a natural successor to our earlier work \shortcite{shaikh2022} in which we described how a Variational Autoencoder (VAE) framework for learning concepts \cite{beta-vae,kingma14} could be adapted so that concepts such as \emph{red} have an explicit representation as a Gaussian on a factored domain. We also provided a short section on how that model could be formalised using tools from category theory and string diagrams. This work differs in that, whilst it still uses Spriteworld shapes as data \shortcite{spriteworld19}, it departs significantly from the probabilistic VAE learning paradigm, representing concepts as quantum effects rather than Gaussians. This report also contains a substantially more detailed theoretical treatment of the models.

What are some of the main reasons for applying the formalism of quantum theory to the modelling of concepts? First, it provides an alternative, and interesting, mathematical structure to the convex structure of conceptual spaces (see Section~\ref{sec:Is-Quantum-Conceptual}). Second, this structure comes with features which are well-suited to modelling concepts, such as entanglement for capturing correlations, and partial orders for capturing conceptual hierarchies.\footnote{Section~\ref{sec:entangled_concepts} describes entanglement; the use of partial orders in experiments we leave for future work.} Third, the use of the tensor product for combining domains leads to machine learning models with different characteristics to those typically employed in concept learning (i.e. neural networks which use direct sum as the monoidal product plus non-linearities to capture interactions between features) \cite{quantum_ml_features,PhysRevLett.122.040504}. The advantages this may bring, especially with the advent of larger, fault-tolerant quantum computers in the future, is still being worked out by the quantum machine learning community, but the possibilities are intriguing at worst and transformational at best.

This work is of a highly inter-disciplinary nature, and hence could appeal to a variety of audiences. These include researchers who are interested in: the formalisation of concepts, and the use of category theory in AI and cognitive science; the question of how the quantum formalism can be used to model concepts; alternatives to conceptual spaces for modelling concepts; the question of how explicit models of concepts can be learned automatically from data; and finally the application of quantum machine learning to problems in AI and cognitive science. We note that our goal is to investigate how the use of the quantum formalism can potentially help with the modelling and learning of concepts, rather than the modelling of data from a psychology lab, with the ultimate goal of building more effective artificial agents. Hence this work is perhaps better placed in the emerging field of quantum AI, rather than quantum cognition, although there are obvious connections to the latter.

%There are two difficulties with inter-disciplinary work of this nature. The first is selecting an intended audience, and the second is providing enough motivation for a new mathematical approach to concept modelling, as well as the application of quantum theory to this problem. Some reasons for a new mathematical approach include the possibility of offering a new perspective on some long-standing research questions, and providing an alternative mathematical structure to the convex spaces of conceptual spaces theory. In  terms of the use of quantum theory, this provides natural solutions to the question of how to model some features of concepts, such as entanglement for capturing correlations, and it opens up the possibility of preparing concept state representations on quantum hardware which cannot be efficiently simulated classically. 

%In terms of the intended audience, this is broad, and includes researchers who are interested in: 
%\begin{itemize}
%\item the formalisation of concepts, and the use of category theory in AI; 
%\item how quantum theory can be used to model concepts; 
%\item alternatives to conceptual spaces for modelling concepts; 
%\item how explicit models of concepts can be learned automatically from data;
%\item the application of quantum machine learning to problems in AI and cognitive science more broadly.
%\end{itemize}
%\scc{anything i missed here?}

The rest of the report is structured as follows. Section~\ref{sec:conspaces} provides an introduction to the formalisation of conceptual spaces. This will appeal to researchers interested in conceptual space theory. Section~\ref{sec:cats} then introduces the categorical framework on which the rest of the formalisation depends, including some example categories. This section, and the following three, will be of particular interest to researchers applying category theory to problems in AI and cognitive science, as well as those interested in the formalisation of conceptual spaces. Section~\ref{sec:cat-con-spaces} defines a conceptual model in terms of the categorical features introduced so far, including some examples of classical concept models. Section~\ref{sec:quantum-models} provides the category of quantum processes which is then used to define our quantum model of concepts, and Section~\ref{sec:entangled_concepts} shows how a special feature of quantum theory, namely entanglement, can be used to capture correlated concepts. Section~\ref{sec:expts} is the experimental part of the report, showing how  quantum concepts can be learned in practice, using a hybrid network which is trained to perform concept classification. This section will be of particular interest to researchers using machine learning to learn conceptual representations from data. Section~\ref{sec:Is-Quantum-Conceptual} 
brings us back to \Gardenfors{} by
addressing the question of whether the quantum models we have defined are convex in the conceptual spaces sense, which again will be of interest to researchers working in conceptual spaces theory. And finally, Section~\ref{sec:conclusion} offers some conclusions and describes some possible future work.

%Some reasons for using quantum theory, and string diagrams, to model concepts include: 
%\begin{itemize}
%\item it offers an alternative mathematical formalism which may offer a new perspective on some long-standing research questions; 
%\item it provides an alternative mathematical structure to the convex spaces of conceptual spaces theory; 
%\item it provides natural solutions to the question of how to model some features of concepts, such as entanglement for capturing correlations; 
%\item it opens up the possibility of preparing concept state representations on quantum hardware which cannot be efficiently simulated classically. 
%\end{itemize}

\section{Conceptual Spaces Formalised} \label{sec:conspaces}

\Gardenfors' framework of \emph{conceptual spaces} \cite{gardenfors2004conceptual,gardenfors2014geometry} models conceptual reasoning in both human and artificial cognition. In this approach, cognition is described geometrically through the use of convex spaces, which in turn factorise in terms of more elementary spaces called \emph{domains}. Examples include the domains of \textsc{colour}, \textsc{taste} and \textsc{sound}, as well as more abstract domains such as \textsc{power} or \textsc{agency}. A central claim of \Gardenfors' framework is that concepts should be represented as convex regions of a space. More generally, graded or `fuzzy' concepts can be modelled as suitable positive functions defined over the space.

The conceptual space framework has been formalized in a variety of ways \shortcite{aisbett2001general,rickard2007reformulation,lewis2016hierarchical,bechberger2017thorough}. Here we draw on the treatment of fuzzy concepts from \citeA{tull2021categorical} which builds on the category-theoretic approach of \shortciteA{bolt2019interacting}.

\begin{definition} \label{def:convex-space}
A \emph{convex space} is a set $Z$ coming with operations which allow us to take convex combinations of elements in the following sense. That is, for all $z_1,\dots,z_n \in Z$ and $p_1, \dots, p_n \in [0,1]$ with $\sum^n_{i=1} p_i = 1$, there is an element of $Z$ denoted
\[
\sum^n_{i=1} p_i \cdot z_i 
\]
These convex combinations satisfy the rules one might expect; for example the order in the sums is irrelevant, elements weighted by $0$ do not affect the sum, and combinations of combinations are calculated by multiplying weights. For a precise formulation we refer to \citeA{bolt2019interacting}. Additionally we here require that $Z$ forms a \emph{measurable space}, meaning it comes with a $\sigma$-algebra $\Sigma_Z$ of measurable subsets $M \subseteq Z$.  
\end{definition}

% \begin{definition} \label{def:convex-space}
% A \emph{convex space} is a set $Z$ in which we may take convex combinations of elements. 
% That is, for all $z_1, z_2 \in Z$ and $p \in [0,1]$ there is an element of $Z$ denoted: 
% \[
% p \cdot z_1 + (1-p) \cdot z_2 
% \]
% These combinations satisfy the rules one might expect
% % , in particular allowing for finite convex combinations $\sum^n_{i=1} p_i \cdot z_i$ where $\sum p_i = 1$
% ; see \citeA{bolt2019interacting} for a more precise formulation. Additionally we here require that $Z$ forms a \emph{measurable space}, meaning it comes with a $\sigma$-algebra $\Sigma_Z$ of measurable subsets $M \subseteq Z$.  
% \end{definition} 

\begin{definition} \label{def:consp}
A \emph{conceptual space} is a convex space $Z$ which is given as a subset of a product of convex spaces:
\[
Z \subseteq Z_1 \times \dots \times Z_n
\]
Here the product is equipped with element-wise convex operations. Note that in particular any single factor $Z_i$ can be considered a conceptual space itself. 

We call an element $z = (z_1,\dots, z_n) \in Z$ an \emph{instance} of the conceptual space, following the terminology of \citeA{clark_concepts2021}. In particular each $z_i$ is an instance of the factor $Z_i$.
\end{definition}

There are two kinds of factor $Z_i$ which appear in the conceptual spaces framework. At a high level, a conceptual space is often written as a product of domains, such as the domains of \textsc{colour} or \textsc{sound}. Secondly, each domain is often described as a product, or subset of a product, of \emph{dimensions}, which provide co-ordinates describing the domain, but cannot be cognitively separated. For example, the \textsc{sound} domain is described in terms of the inseparable dimensions of \textsc{pitch} and \textsc{volume}. Here we simply use the neutral term `factor' to treat either dimensions or domains.

% The overall space forms (a subset of) a product of domains, and the domains decompose further in terms of dimensions. 
Next let us consider concepts themselves.

\begin{definition} \label{def:crisp-concept}
A \emph{crisp concept} in a conceptual space $Z$ is a measurable  subset $\crc \subseteq Z$ which is \emph{convex}, meaning it is closed under convex combinations. When $z \in \crc$ we say $z$ is an \emph{instance of \crc}.\footnote{This is consistent with the fact that if $z$ is an instance of the concept $C \subseteq Z$ it is also an instance of the convex space $C$ viewed as a conceptual space itself.} 
\end{definition} 

Convexity means that any point lying ``in-between" two instances of a concept will again form an instance of the concept. \citeA{gardenfors2004conceptual} justifies convexity based on cognitive experiments, including evidence from the division of colour space, as well as from the relative ease of learning convex regions.

% Let us now meet some examples of conceptual spaces and concepts.

% Todo: Add more examples of conceptual spaces.

\begin{example} \label{ex:RnCS}
Any convex subset $Z \subseteq \mathbb{R}^d$ forms a conceptual space, by taking $\Sigma_Z$ to be the Lebesgue measurable subsets. Thus any product $Z = Z_1 \times \dots \times Z_n$ of convex subsets $Z_i \subseteq \mathbb{R}^{d_i}$ forms a conceptual space also.
\end{example} 

\begin{example} A simple example of a domain is the \textsc{taste} simplex depicted below (taken from \shortciteA{bolt2019interacting}). This forms a convex subset of $\mathbb{R}^3$, generated by the extremal points \emph{sweet, bitter, salt} and \emph{sour}. Highlighted in red is a convex region describing a crisp concept for \emph{sweet}.
\[
\includegraphics[scale=0.2]{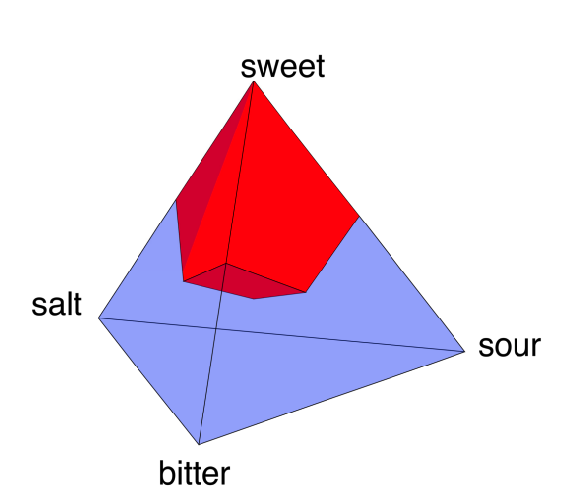}
\]
\end{example} 

By definition, for a crisp concept, each point $z$ either is or is not an instance of the concept.  However, it is natural to more generally consider concepts $\cc$ which are instead graded or `fuzzy', so that membership can take any value $\cc(z)$ between $0$ and $1$. The value $\cc(z)$ represents the degree to which $z$ is an instance of the concept, with $0$ meaning that the concept is not at all satisfied, and $1$ meaning the concept is perfectly satisfied. The membership function should respect the existing convex structure in an appropriate sense, such as in the following definition proposed by one of the authors in \citeA{tull2021categorical}.

\begin{definition} \label{def:fuzzy-concept}
A \emph{fuzzy concept} of $Z$ is a measurable function $\cc \colon Z \to [0,1]$ which is \emph{log-concave}, meaning that 
\begin{equation} \label{eq:lc}
\cc(p \cdot z + (1-p) \cdot z') \geq \cc(z)^p\cc(z')^{1-p}
\end{equation}
for all $z,z' \in Z$ and $p \in [0,1]$. A \emph{prototypical instance} of a fuzzy concept $\cc$ is a point $z$ with $\cc(z) = \max_{w \in Z} \cc(w)$.
\end{definition} 

The prototypical instances of a concept $\cc$ always form a crisp concept of $Z$, and conversely any crisp concept $P \subseteq Z$ forms a special case of a fuzzy concept via its indicator function $\cc = 1_P$. 

The log-concavity condition \eqref{eq:lc} is justified fully in \citeA[Theorem 8]{tull2021categorical}. Briefly, it provides the most general definition of fuzzy concepts which satisfies the natural criterion of `quasi-concavity' identified by \Gardenfors{}, which ensures that the sets of instances fitting a concept to any given extent always form a convex set. It  also includes the special cases of crisp concepts and Gaussians \cite{beta-vae,shaikh2022}, and allows one to form product concepts, as follows.

% \snew{ Log-concavity can be justified as the most general way to form a monoidal category of fuzzy concepts containing some basic examples (such as Gaussians) and which satisfy a natural criterion known as \emph{quasi-concavity} which ensures that the sets of instances fitting a concept to any given extent always form a convex set; see \cite{tull2021categorical} for the details.}

% is compositionally well-behaved, and \footnote{More precisely, the largest class which are \emph{quasi-concave}, closed under element-wise products, and on $\mathbb{R}$ contain all functions $x \mapsto \lambda^{-x}$ for $\lambda \geq 1$.}. In this work we focus solely on the latter kind. 

\begin{example}
Any product of fuzzy concepts
\begin{equation} \label{eq:product-concept}
\cc(z) = \prod^n_{i=1} \cc_i(z_i) 
\end{equation}
over factors $Z_1, \dots, Z_n$ forms a fuzzy concept on any convex subset $Z \subseteq Z_1 \times \dots \times Z_n$.
\end{example} 

\begin{example}
We may define a fuzzy concept on $Z=\mathbb{R}^n$ from any multivariate Gaussian 
with mean $\mu$ and covariance matrix $\Sigma$:
\begin{align} \label{eq:Gaussian} 
\cc(z;\mu,\Sigma) &= e^{-\frac{1}{2}(z - \mu)^{\mathsf{T}}\Sigma^{-1}(z -\mu)} \\ 
& = e^{\sum^n_{i=1}-\frac{1}{2 \sigma_i^2}(z_i - \mu_i)^2}
\end{align}  
In the second line we restrict to the case where $\Sigma$ is diagonal, with $i$-th diagonal entry $\sigma_i^2$. In this case $\cc$ is given as a product of one-dimensional Gaussians $\cc_i(z_i;\mu_i,\sigma_i^2)$ as in \eqref{eq:product-concept}.
\end{example}

\begin{example} \label{ex:fuzzy-concepts}
We give some simple examples of fuzzy concepts from \citeA{tull2021categorical}. On the left we depict a fuzzy concept on $\mathbb{R}^2$. From a set of exemplars (white crosses) we form the convex closure, yielding the crisp concept $P$ given by the inner triangle. We then define a fuzzy concept $C(x) = e^{-\frac{1}{2\sigma^2}d(x,P)^2}$ where $d_H(x,P) = \inf_{p \in P} d(x,p)$, where each point in $P$ is prototypical. Thanks to quasi-concavity the set of points $x$ with $C(x) \geq t$ is convex for any $t$, as shown for $t=1.00,0.20,0.05$. 
\[
\includegraphics[scale=0.5]{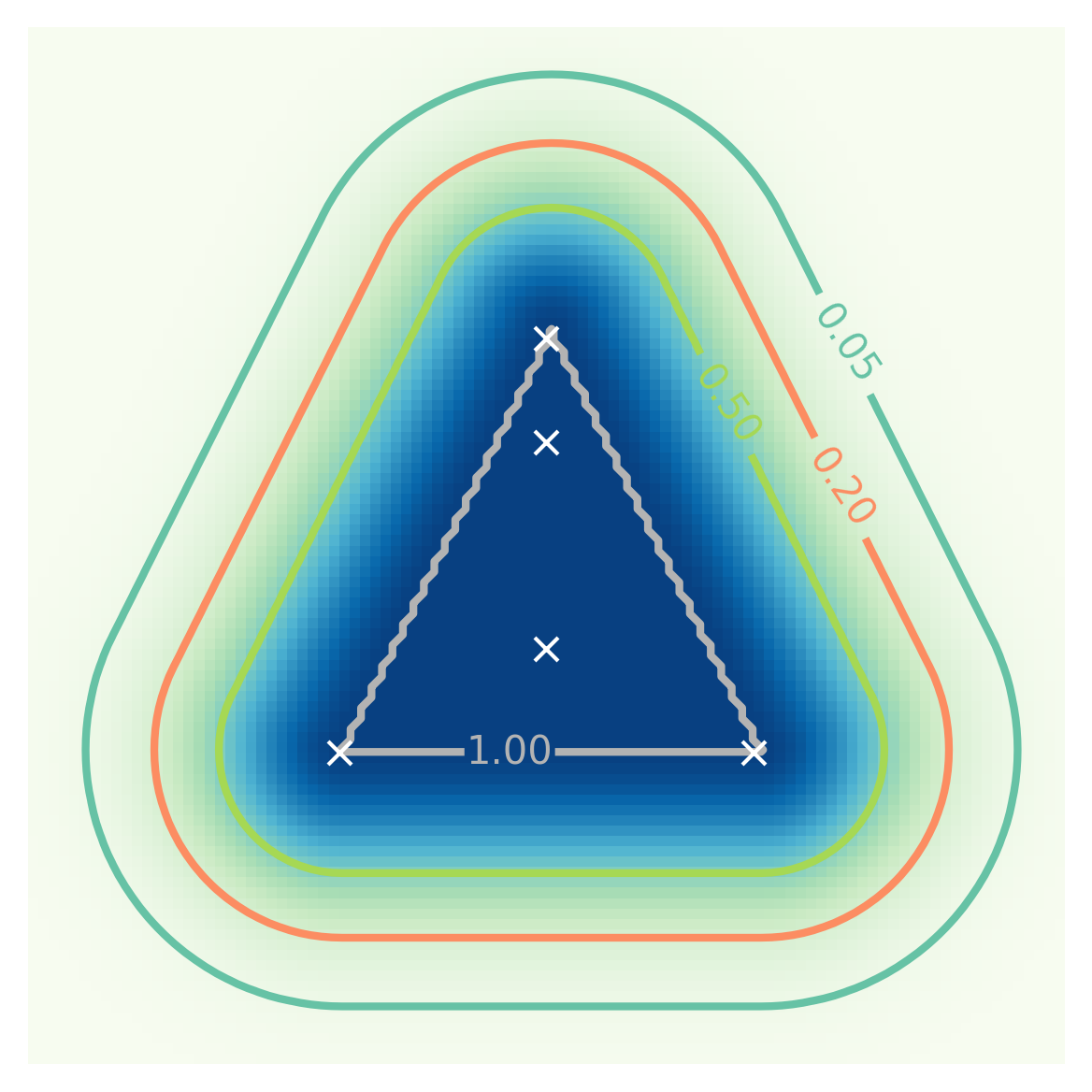}
\qquad 
\includegraphics[scale=0.5]{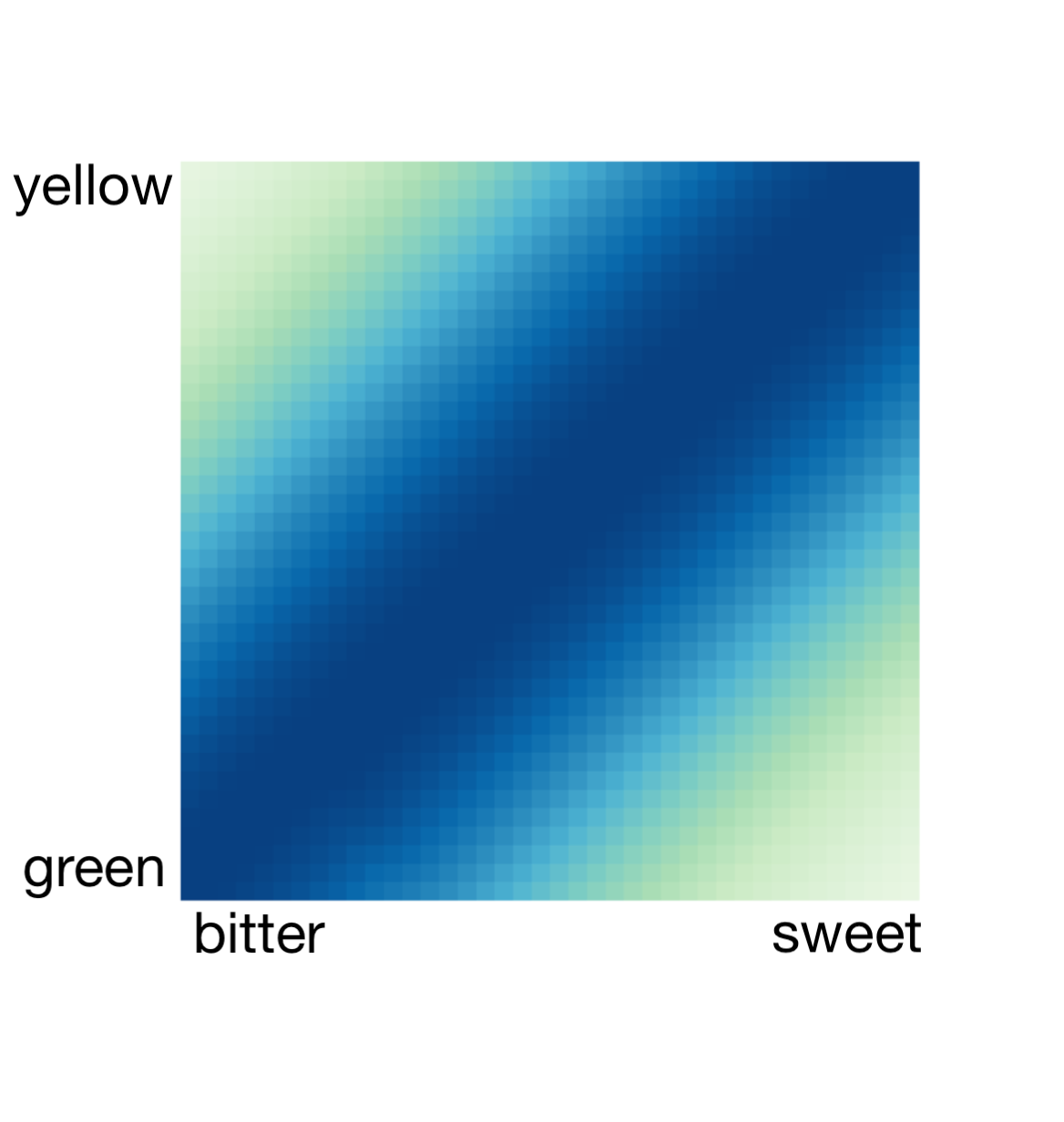}
\]
A similar example exhibiting correlations between domains is shown on the right, adapted from \citeA{bolt2019interacting}. We consider a simple conceptual space for foods, given as a product of domains $F=\textsc{colour} \times \textsc{taste}$. A fuzzy concept `Banana' with prototypical instances $(\text{yellow}, \text{sweet}), (\text{green}, \text{bitter}) \in F$ is plotted over the subset $[\text{green},\text{yellow}] \times [\text{bitter},\text{sweet}] \subseteq F$.
\end{example}

% Todo: Consider writing this way.
% \begin{example}
% We may define a fuzzy concept on $Z=\mathbb{R}^n$ from any multivariate Gaussian 
% with mean $\mu$ and covariance matrix $\Sigma$:
% \begin{align} \label{eq:Gaussian} 
% \cc(z;\mu,\Sigma) &= e^{-\frac{1}{2}(z - \mu)^{\mathsf{T}}\Sigma^{-1}(z -\mu)}
% % \\ 
% % & = e^{\sum^n_{i=1}-\frac{1}{2 \sigma_i^2}(z_i - \mu_i)^2}
% \end{align}  
% In the case where $\Sigma$ is diagonal, with $i$-th diagonal entry $\sigma_i^2$, $\cc$ is given as a product of one-dimensional Gaussians $\cc_i(z_i;\mu_i,\sigma_i^2)$ as in \eqref{eq:product-concept}: 
% \[
% \cc(z;\mu,\Sigma) &= \prod^n_{i=1}e^{-\frac{1}{2 \sigma_i^2}(z_i - \mu_i)^2}
% \]
% \end{example}

%Todo: Fill in examples
%\begin{example}
%\scc{Fuzzy Banana concept example ...}
%\end{example} 

Our aim will now be to lift these basic notions from conceptual space theory beyond the setting of convex spaces, and use them to produce our quantum model of concepts. To do so we will reformulate them in the common language of \emph{category theory}, allowing us to pass notions from the classical to the quantum setting in a principled manner. 

\section{Categorical Setup} \label{sec:cats}

Throughout we work in a \emph{symmetric monoidal category} $(\catC, \otimes ,I)$. Recall that a monoidal category consists of a class of objects $A, B, C\dots$ and morphisms $f \colon A \to B$ between them. We can compose morphisms in  sequence $f, g \mapsto g \circ f$ and both objects and morphisms in parallel via $A, B \mapsto A \otimes B$ and $f,g \mapsto f \otimes g$. 

Monoidal categories are best understood using their \emph{graphical calculus} \cite{selinger2010survey}, in which objects are depicted as labelled wires and morphisms as boxes, with diagrams here read from bottom to top. Identity morphisms and composition are depicted as follows.  
\[
\begin{tikzpicture}
	\begin{pgfonlayer}{nodelayer}
		\node [style=none] (0) at (-3, -1) {};
		\node [style=none] (1) at (-3, 1) {};
		\node [style=map] (2) at (-3, 0) {$\id{A}$};
		\node [style=none] (3) at (-1.5, 0) {$=$};
		\node [style=none] (4) at (-0.25, -1) {};
		\node [style=none] (5) at (-0.25, 1) {};
		\node [style=label] (7) at (-3, -1.5) {$A$};
		\node [style=label] (8) at (-0.25, -1.5) {$A$};
		\node [style=label] (9) at (-0.25, 1.25) {$A$};
		\node [style=label] (10) at (-3, 1.25) {$A$};
		\node [style=none] (11) at (3.5, -1) {};
		\node [style=none] (12) at (3.5, 1) {};
		\node [style=map] (13) at (3.5, 0) {$g \circ f$};
		\node [style=none] (14) at (5.25, 0) {$=$};
		\node [style=none] (15) at (6.75, -2) {};
		\node [style=none] (16) at (6.75, 2) {};
		\node [style=label] (17) at (3.5, -1.5) {$A$};
		\node [style=label] (18) at (6.75, -2.5) {$A$};
		\node [style=label] (19) at (6.75, 2.25) {$C$};
		\node [style=label] (20) at (3.5, 1.25) {$C$};
		\node [style=map] (21) at (6.75, -1) {$f$};
		\node [style=map] (22) at (6.75, 1) {$g$};
		\node [style=label] (23) at (7.25, 0) {$B$};
		\node [style=none] (24) at (11.75, -1) {};
		\node [style=none] (25) at (11.75, 1) {};
		\node [style=map] (26) at (11.75, 0) {$f \otimes g$};
		\node [style=label] (27) at (11.75, -1.5) {$A \otimes B$};
		\node [style=label] (28) at (11.75, 1.5) {$C \otimes D$};
		\node [style=none] (29) at (13.5, 0) {$=$};
		\node [style=none] (30) at (15, -1) {};
		\node [style=none] (31) at (15, 1) {};
		\node [style=map] (32) at (15, 0) {$f$};
		\node [style=label] (33) at (15, -1.5) {$A$};
		\node [style=label] (34) at (15, 1.25) {$C$};
		\node [style=none] (35) at (16.5, -1) {};
		\node [style=none] (36) at (16.5, 1) {};
		\node [style=map] (37) at (16.5, 0) {$g$};
		\node [style=label] (38) at (16.5, -1.5) {$B$};
		\node [style=label] (39) at (16.5, 1.25) {$D$};
	\end{pgfonlayer}
	\begin{pgfonlayer}{edgelayer}
		\draw (1.center) to (0.center);
		\draw (5.center) to (4.center);
		\draw (12.center) to (11.center);
		\draw (16.center) to (15.center);
		\draw (25.center) to (24.center);
		\draw (31.center) to (30.center);
		\draw (36.center) to (35.center);
	\end{pgfonlayer}
\end{tikzpicture}
\]
The (identity on the) monoidal unit $I$ is depicted as the empty diagram. Thus morphisms of the form $\omega \colon I \to A$, which are called \emph{states} of $A$, are drawn with no input, while morphisms $e \colon A \to I$, called \emph{effects} of $A$, have no output, as shown below.
\[
\begin{tikzpicture}
	\begin{pgfonlayer}{nodelayer}
		\node [style=map] (0) at (0, -0.25) {$\omega$};
		\node [style=none] (1) at (0, 0.75) {};
		\node [style=label] (2) at (0, 1.25) {$A$};
	\end{pgfonlayer}
	\begin{pgfonlayer}{edgelayer}
		\draw (1.center) to (0);
	\end{pgfonlayer}
\end{tikzpicture}
 \qquad \qquad \qquad  \begin{tikzpicture}
	\begin{pgfonlayer}{nodelayer}
		\node [style=map] (0) at (0, 0.75) {$e$};
		\node [style=none] (1) at (0, -0.25) {};
		\node [style=label] (2) at (0, -0.75) {$A$};
	\end{pgfonlayer}
	\begin{pgfonlayer}{edgelayer}
		\draw (1.center) to (0);
	\end{pgfonlayer}
\end{tikzpicture}
 \qquad 
\]
Morphisms of the form $s \colon I \to I$, with `no inputs or outputs', are called \emph{scalars}. In our categories of interest here scalars will correspond to positive real numbers $s \in \mathbb{R}^+$, and often in particular to probabilities $p \in [0,1]$. In such a category, composing any effect with a state yields a positive real scalar.
\[
\begin{tikzpicture}
	\begin{pgfonlayer}{nodelayer}
		\node [style=map] (0) at (0, -0.75) {$\omega$};
		\node [style=map] (1) at (0, 1.25) {$e$};
		\node [style=label] (2) at (-0.5, 0.25) {$A$};
	\end{pgfonlayer}
	\begin{pgfonlayer}{edgelayer}
		\draw (1) to (0);
	\end{pgfonlayer}
\end{tikzpicture}
 \in \mathbb{R}^+
\]

We will consider categories  with additional further structure. Firstly, each object will come with a distinguished effect called \emph{discarding}: 
\[
\begin{tikzpicture}
	\begin{pgfonlayer}{nodelayer}
		\node [style=none] (6) at (4.5, 0) {};
		\node [style=none] (7) at (4.5, -1) {};
		\node [style=upground] (8) at (4.5, 0.25) {};
	\end{pgfonlayer}
	\begin{pgfonlayer}{edgelayer}
		\draw (7.center) to (6.center);
	\end{pgfonlayer}
\end{tikzpicture}

\]
which we interpret as `throwing the system away', with $\discard{I} = 1$. 

We call a morphism $f$ a \emph{channel} when it preserves discarding, in that the following holds. 
\[
\begin{tikzpicture}
	\begin{pgfonlayer}{nodelayer}
		\node [style=map] (0) at (0, 0) {$f$};
		\node [style=upground] (1) at (0, 1.25) {};
		\node [style=none] (2) at (0, -1) {};
		\node [style=none] (3) at (1.5, 0) {$=$};
		\node [style=upground] (5) at (3, 0.25) {};
		\node [style=none] (6) at (3, -1) {};
	\end{pgfonlayer}
	\begin{pgfonlayer}{edgelayer}
		\draw (1) to (0);
		\draw (2.center) to (0);
		\draw (6.center) to (5);
	\end{pgfonlayer}
\end{tikzpicture}
\]
In particular, we call a state $\omega$ \emph{normalised} when it satisfies:
\[
\begin{tikzpicture}
	\begin{pgfonlayer}{nodelayer}
		\node [style=map] (0) at (0, -0.5) {$\omega$};
		\node [style=upground] (1) at (0, 0.75) {};
		\node [style=none] (2) at (1.5, 0) {$=$};
		\node [style=none] (3) at (2.75, 0) {$1$};
	\end{pgfonlayer}
	\begin{pgfonlayer}{edgelayer}
		\draw (1) to (0);
	\end{pgfonlayer}
\end{tikzpicture}

\]
Then one may verify that composing any channel $f$ with a normalised state $\omega$ yields a normalised state $f \circ \omega$.

Our categories will also come with the extra structure of an \emph{ordering} relation which makes the collection of morphisms of any given type a partially ordered set, so that we may have: 
\[
\begin{tikzpicture}
	\begin{pgfonlayer}{nodelayer}
		\node [style=map] (0) at (0, 0) {$f$};
		\node [style=map] (1) at (2.5, 0) {$g$};
		\node [style=none] (2) at (0, -1) {};
		\node [style=none] (3) at (0, 1) {};
		\node [style=none] (4) at (2.5, -1) {};
		\node [style=none] (5) at (2.5, 1) {};
		\node [style=none] (6) at (1.25, 0) {$\leq$};
		\node [style=label] (7) at (0, 1.5) {$Y$};
		\node [style=label] (8) at (2.5, 1.5) {$Y$};
		\node [style=label] (9) at (2.5, -1.5) {$X$};
		\node [style=label] (10) at (0, -1.5) {$X$};
	\end{pgfonlayer}
	\begin{pgfonlayer}{edgelayer}
		\draw (4.center) to (5.center);
		\draw (2.center) to (3.center);
	\end{pgfonlayer}
\end{tikzpicture}
\]
with the ordering $s \leq t$ on scalars the usual order in $\mathbb{R}^+$. Moreover the order is respected by composition, so that if $f \leq g$ then $h \circ f \leq h \circ g$, $f \circ h \leq g \circ h$ and $f \otimes h \leq g \otimes h$ for all morphisms $h$.

Finally our categories will come with structure allowing us to `include an object into a larger one', such as when we include a region $C$ into a larger space $Z$ via $ C \hookrightarrow Z$. For this we assume the presence of a distinguished subcategory of channels called \emph{embeddings}. An embedding $e$ of $A$ into $B$ is depicted as a triangle:
\[
\begin{tikzpicture}
	\begin{pgfonlayer}{nodelayer}
		\node [style=decomp] (0) at (0.5, 0) {$e$};
		\node [style=none] (1) at (0.5, -1) {};
		\node [style=none] (2) at (0.5, 1) {};
		\node [style=label] (3) at (0.5, -1.5) {$A$};
		\node [style=label] (4) at (0.5, 1.5) {$B$};
	\end{pgfonlayer}
	\begin{pgfonlayer}{edgelayer}
		\draw (2.center) to (1.center);
	\end{pgfonlayer}
\end{tikzpicture}

\]
Each embedding is associated with a unique morphism $e^\dagger \colon B \to A$ called its \emph{projection} and depicted by flipping the diagram for $e$ upside down. Together the embedding and projection satisfy: 
\[
\begin{tikzpicture}
	\begin{pgfonlayer}{nodelayer}
		\node [style=decomp] (0) at (0, -0.75) {$e$};
		\node [style=decompflip] (1) at (0, 0.75) {$e$};
		\node [style=none] (2) at (0, -1.75) {};
		\node [style=none] (3) at (0, 0) {};
		\node [style=label] (4) at (0, -2.25) {$A$};
		\node [style=label] (6) at (0, 2.25) {$A$};
		\node [style=label] (7) at (-1.25, 0) {$B$};
		\node [style=none] (8) at (0, 1.75) {};
		\node [style=none] (9) at (0, 0) {};
		\node [style=none] (10) at (1.75, 0) {$=$};
		\node [style=none] (11) at (3.5, -1.75) {};
		\node [style=none] (12) at (3.5, 1.75) {};
		\node [style=label] (13) at (3.5, -2.25) {$A$};
		\node [style=label] (14) at (3.5, 2.25) {$A$};
	\end{pgfonlayer}
	\begin{pgfonlayer}{edgelayer}
		\draw (3.center) to (2.center);
		\draw (8.center) to (9.center);
		\draw (12.center) to (11.center);
	\end{pgfonlayer}
\end{tikzpicture}
\]

A special case is when $e$ is an \emph{isomorphism} $A \simeq B$ and we additionally have $e \circ e^\dagger = \id{}$. 
% \[
% \tikzfig{embed-iso}
% \]
Another special case is an embedding from $A=I$, which by definition consists of a normalised state $\psi$ with an effect $\psi^\dagger$  satisfying:
\[
\begin{tikzpicture}
	\begin{pgfonlayer}{nodelayer}
		\node [style=decomp] (0) at (0, -1) {$\psi$};
		\node [style=decompflip] (1) at (0, 1) {$\psi$};
		\node [style=none] (2) at (0, -1.25) {};
		\node [style=none] (3) at (0, 0) {};
		\node [style=none] (8) at (0, 1) {};
		\node [style=none] (9) at (0, 0) {};
		\node [style=none] (10) at (2, 0) {$=$};
		\node [style=none] (11) at (3.75, 0) {$1$};
	\end{pgfonlayer}
	\begin{pgfonlayer}{edgelayer}
		\draw (3.center) to (2.center);
		\draw (8.center) to (9.center);
	\end{pgfonlayer}
\end{tikzpicture}
\] 
We will call such a state or effect a \emph{point}.\footnote{Later we will define instances as special cases of points. Instances and points differ in quantum models, because of entanglement, but coincide classically.}

Let us now meet our two main classical examples of categories: plain classical probability theory (in two forms), and convex spaces as used in conceptual space theory. 

\subsection{Examples: Classical}

A simple discrete classical setting is the following. 

\begin{example} 
In the category $\Class$ the objects are finite sets and the morphisms $M \colon X \to Y$ are `$X \times Y$-matrices' $(M(y,x))_{x \in X, y \in Y}$ with values in $\mathbb{R}^+$. The composition is given by matrix multiplication:
\[
(N \circ M)(z,x) := \sum_{y \in Y} N(z,y) M(y,x)
\]
The identity morphism on $X$ is given by $\id{}(y,x) = \delta_{x,y}$. 

The tensor is given on objects by $X \otimes Y = X \times Y$ and on morphisms by the Kronecker product of matrices. The monoidal unit is a singleton set $I=\{\star\}$, and the scalars are $\mathbb{R}^+$. Thus we can equate states $\omega$ and effects $e$ of $X$ each with functions $X \to \mathbb{R}^+$ via $\omega(x) := \omega(x,\star)$ and $e(x) := e(\star,x)$. In particular, as promised earlier, scalars correspond to positive reals $s \in \mathbb{R}^+$. The discarding effect $\discard{X}$ is given by $\discard{}(x) = 1$ for all $x \in X$. 

We may describe discrete probability theory via the channels in this category. 
Firstly, note that by definition a state $\omega$ of $X$ is normalised iff it describes a probability distribution over $X$:
\[
\sum_{x \in X} \omega(x) = 1
\] 
More generally, a morphism $M \colon X \to Y$ is a channel iff it sends each $x \in X$ to a normalised distribution over $Y$; that is:
\[
\sum_{y \in Y} M(y,x) = 1 
\]
for each $x \in X$, or equivalently the matrix is \emph{stochastic}. Thus $M$ may be seen as a `probabilistic process' from $X$ to $Y$, sending each $x$ to a probability distribution $M(x)$ over $Y$. The partial order $\leq$ on morphisms is the element-wise ordering from $\mathbb{R}^+$. 

The points of an object $X$ are precisely the point distributions $\delta_x$ for $x \in X$, with $\delta_x(y) = 1$ iff $y = x$ and $0$ otherwise. 

% Sean: Just left this as these examples are about defining the categorical notions (e.g. states, effects, points..) not the conceptual modelling ones (instances, concepts...) % \scc{not the same as instances? - clarify again?} 

Finally, an embedding $X \hookrightarrow Y$ is given by an inclusion of a subset $X \subseteq Y$ via the map $x \mapsto \delta_x$. Its projection $Y \to X$ is given by $y \mapsto \delta_y$ when $y \in X$ and $y \mapsto 0$ otherwise.
\end{example} 

We can upgrade this example beyond the discrete setting to describe general  (measure-theoretic) probability theory as follows. 

% \scc{why the scare quotes on continuous?} 
% Sean: continuity requires a topology to define so is actually more specific than working measure-theoretically. So topology + continuity is a subset of measure theory. But normally we think of topological examples (e.g. R^n) in this setting.

\begin{example} 
In the category $\Prob$ the objects are measurable spaces $(X, \Sigma_X)$, consisting of a set $X$ along with a $\sigma$-algebra of its measurable subsets $\Sigma_X$. A morphism $f \colon X \to Y$ is a \emph{Markov (sub)kernel}, a function sending each $x \in X$ to a subprobability measure $f(x)$ over $Y$. Furthermore this assignment is required to be `measurable' in an appropriate sense; see e.g. \cite{panangaden1998probabilistic,cho2019disintegration}. 

Composition of $f \colon X \to Y$ and  $g \colon Y \kto Z$ is given by integration:
 \[
(g \circ f)(x,A) := \int_{y \in Y} g(y,A) df(x)(y)
\] 
for each $x \in X, A \in \Sigma_Z$. The identity on $X$ sends each $x$ to the point measure $\delta_x$ with $\delta_x(A) =1$ iff $x \in A$. 
% \[
% \delta_x(M) = \begin{cases} 1 & x \in M \\ 0 & x \notin M \end{cases}
% \]
We set $X \otimes Y = X \times Y$, with $I$ being the singleton set, and define $f \otimes g$ to send each pair $(x,y)$ to the \emph{product measure} $f(x) \otimes g(y)$ of the measures $f(x)$ and $g(y)$. 

States of $X$ may be identified with sub-probability measures $\omega$ over $X$, while effects $e$ correspond to measurable functions: 
% \[
% \omega \colon \Sigma_X \to [0,1]
% \]
% while effects correspond to measurable functions: 
\[
e \colon X \to [0,1]
\] 
and scalars $I \kto I$ to probabilities $p \in [0,1]$. In particular, composing a state with an effect yields the expectation value: 
\begin{equation} \label{eq:exp-value-cts}
\begin{tikzpicture}
	\begin{pgfonlayer}{nodelayer}
		\node [style=map] (0) at (0, -0.75) {$\omega$};
		\node [style=map] (1) at (0, 1.25) {$e$};
		\node [style=label] (2) at (-0.5, 0.25) {$A$};
	\end{pgfonlayer}
	\begin{pgfonlayer}{edgelayer}
		\draw (1) to (0);
	\end{pgfonlayer}
\end{tikzpicture}
 = \int_{x \in X} e(x) d\omega(x) \in \mathbb{R}^+
\end{equation}
As before $\discard{X}$ is given by $x \mapsto 1$ for all $x \in X$. Normalised states correspond to probability measures $\omega$ over $X$, and a morphism  $f \colon X \to Y$ is a channel iff it sends each $x \in X$ to an actual normalised probability measure. We order morphisms $X \to Y$ via 
\[
f \leq g \iff f(x,A) \leq g(x,A)  \ \ \ \forall x \in X, A \in \Sigma_Y
% for all $x \in X, A \in \Sigma_Y$}. 
\]
An embedding $X \hookrightarrow Y$ is an inclusion of a subset $X \subseteq Y$ via $x \mapsto \delta_x$ for $x \in X$, with the projection $Y \to X$ given by $y \mapsto \delta_y$ when $y \in X$ and $y \mapsto 0$ otherwise.

% In this category every morphism is a sub-channel in that $\discard{} \circ f \leq \discard{}$. Restricting the objects of this category to finite discrete spaces yields the category of such sub-channels in $\Class$. 
\end{example}

\subsection{Example: Convex Spaces}

We can restrict our previous example to give a categorical setting for the study of conceptual spaces. 

\begin{example} 
The category $\ConSp$ \cite{tull2021categorical} is defined just like $\Prob$ except that the objects $X, Y$ are now convex spaces and the morphisms $f \colon X \to Y$ are those (sub)kernels which are \emph{log-concave}, meaning that they satisfy
\begin{equation} \label{eq:log-conc-han}
f(p x + (1-p)y, p A + (1-p) B) \geq f(x,A)^p f(y,B)^{1-p}
\end{equation}
for all $p \in [0,1], x, y \in X$ and $A, B \in \Sigma_Y$.

Here $X \otimes Y = X \times Y$ is the product of convex spaces, with element-wise convex operations. 
In particular, normalised states of $Z$ are given by log-concave probability measures $\omega$ over $Z$, while effects correspond to log-concave measurable functions $\cc \colon Z \to [0,1]$, i.e. fuzzy concepts. The log-concavity condition for channels \eqref{eq:log-conc-han} comes from the fact that it provides the most general way to form a monoidal category containing all such fuzzy concepts as effects; again we refer to \citeA{tull2021categorical} for a full derivation.
 \end{example} 

\section{Categorical Conceptual Models} \label{sec:cat-con-spaces}

Let us now see how each of our earlier features from conceptual space theory can be described at the categorical level. We work in a general category $\catC$ with the structure outlined in Section \ref{sec:cats}.

Firstly, our formulation in terms of monoidal categories already makes it immediate for us to describe the factors $Z_i$ appearing in a conceptual space, and their composites. A single factor $Z_i$ can be described simply as an object in $\catC$, while the analogue of a conceptual space is the following. 

\begin{definition} \label{Def:conceptual-model}
A \emph{conceptual  model}\footnote{Henceforth we use the generic term `model' rather than `space' since a conceptual model can be defined in a category without any spatial character.} is given by an object $Z$ along with an indexed collection of objects 
$Z_1,\dots, Z_n$, called the \emph{factors}, and an embedding
\[
\begin{tikzpicture}
	\begin{pgfonlayer}{nodelayer}
		\node [style=decompwide] (0) at (0, 0) {};
		\node [style=none] (1) at (0, -1.25) {};
		\node [style=none] (2) at (-1, 0.25) {};
		\node [style=none] (3) at (1, 0.25) {};
		\node [style=none] (4) at (-1, 1.25) {};
		\node [style=none] (5) at (1, 1.25) {};
		\node [style=none] (6) at (0, 0.75) {$\dots$};
		\node [style=label] (7) at (-1, 1.75) {$Z_1$};
		\node [style=label] (8) at (1, 1.75) {$Z_n$};
		\node [style=label] (9) at (0, -1.75) {$Z$};
	\end{pgfonlayer}
	\begin{pgfonlayer}{edgelayer}
		\draw (5.center) to (3.center);
		\draw (4.center) to (2.center);
		\draw (1.center) to (0);
	\end{pgfonlayer}
\end{tikzpicture}
\]
of $Z$ into $Z_1 \otimes \dots \otimes Z_n$.
\end{definition}

For simplicity we usually refer to a model by the object $Z$, with the factors and embedding left implicit. Often the embedding is in fact an isomorphism $Z \simeq Z_1 \otimes \dots \otimes Z_n$ exhibiting $Z$ as a product of the factors. In this case we omit the isomorphism in diagrams and simply draw 
\[
\begin{tikzpicture}
	\begin{pgfonlayer}{nodelayer}
		\node [style=none] (0) at (0, 0.75) {};
		\node [style=none] (1) at (0, -0.75) {};
		\node [style=none] (2) at (1.5, 0) {$=$};
		\node [style=none] (3) at (3, 0.75) {};
		\node [style=none] (4) at (3, -0.75) {};
		\node [style=none] (5) at (5, 0.75) {};
		\node [style=none] (6) at (5, -0.75) {};
		\node [style=none] (7) at (4, 0) {$\dots$};
		\node [style=label] (8) at (0, 1.25) {$Z$};
		\node [style=label] (9) at (0, -1.25) {$Z$};
		\node [style=label] (10) at (3, -1.25) {$Z_1$};
		\node [style=label] (11) at (3, 1.25) {$Z_1$};
		\node [style=label] (12) at (5, 1.25) {$Z_n$};
		\node [style=label] (13) at (5, -1.25) {$Z_n$};
	\end{pgfonlayer}
	\begin{pgfonlayer}{edgelayer}
		\draw (1.center) to (0.center);
		\draw (4.center) to (3.center);
		\draw (6.center) to (5.center);
	\end{pgfonlayer}
\end{tikzpicture}
\]
Next we consider the basic notions of concepts and instances of a model.

\begin{definition} 
A \emph{concept} $C$ of a model $Z$ is an effect on $Z$:
\[
\begin{tikzpicture}
	\begin{pgfonlayer}{nodelayer}
		\node [style=map] (0) at (0, 0.75) {$C$};
		\node [style=none] (1) at (0, -0.25) {};
		\node [style=label] (2) at (0, -0.75) {$Z$};
	\end{pgfonlayer}
	\begin{pgfonlayer}{edgelayer}
		\draw (1.center) to (0);
	\end{pgfonlayer}
\end{tikzpicture}
 
\]
The partial order on morphisms means that the concepts on $Z$ are automatically partially ordered, and we interpret
\begin{equation} \label{eq:concept-order}
\begin{tikzpicture}
	\begin{pgfonlayer}{nodelayer}
		\node [style=map] (0) at (-0.25, 0.75) {$C$};
		\node [style=none] (1) at (-0.25, -0.25) {};
		\node [style=label] (2) at (-0.25, -0.75) {$Z$};
		\node [style=map] (3) at (2.25, 0.75) {$D$};
		\node [style=none] (4) at (2.25, -0.25) {};
		\node [style=label] (5) at (2.25, -0.75) {$Z$};
		\node [style=none] (6) at (1, 0.25) {$\leq$};
	\end{pgfonlayer}
	\begin{pgfonlayer}{edgelayer}
		\draw (1.center) to (0);
		\draw (4.center) to (3);
	\end{pgfonlayer}
\end{tikzpicture}
\end{equation}
as stating that the concept $D$ is `more general' than $C$. 
\end{definition}

\begin{definition}
An \emph{instance} of a model is a point of $Z$ which forms a product of points over the embedding, that is:
\begin{equation} \label{eq:instance-factors}
\begin{tikzpicture}
	\begin{pgfonlayer}{nodelayer}
		\node [style=decompwide] (0) at (1.5, 0) {};
		\node [style=none] (1) at (1.5, -1.25) {};
		\node [style=none] (2) at (0.5, 0.25) {};
		\node [style=none] (3) at (2.5, 0.25) {};
		\node [style=none] (4) at (0.5, 1.25) {};
		\node [style=none] (5) at (2.5, 1.25) {};
		\node [style=none] (6) at (1.5, 0.75) {$\dots$};
		\node [style=label] (7) at (0.5, 1.75) {$Z_1$};
		\node [style=label] (8) at (2.5, 1.75) {$Z_n$};
		\node [style=label] (9) at (4.75, 0) {$=$};
		\node [style=point] (10) at (1.5, -1.5) {$z$};
		\node [style=none] (12) at (7, -0.25) {};
		\node [style=none] (13) at (9, -0.25) {};
		\node [style=none] (14) at (7, 0.75) {};
		\node [style=none] (15) at (9, 0.75) {};
		\node [style=none] (16) at (8, 0.25) {$\dots$};
		\node [style=label] (17) at (7, 1.25) {$Z_1$};
		\node [style=label] (18) at (9, 1.25) {$Z_n$};
		\node [style=point] (19) at (7, -0.5) {$z_1$};
		\node [style=point] (20) at (9, -0.5) {$z_n$};
		\node [style=label] (21) at (0.75, -0.75) {$Z$};
	\end{pgfonlayer}
	\begin{pgfonlayer}{edgelayer}
		\draw (5.center) to (3.center);
		\draw (4.center) to (2.center);
		\draw (1.center) to (0);
		\draw (15.center) to (13.center);
		\draw (14.center) to (12.center);
	\end{pgfonlayer}
\end{tikzpicture}
\end{equation}
for points $z_i$ of $Z_i$. 
%We denote the set of instances of a model $Z$ by $\Inst(Z)$. 
% \scc{of a model?}
\end{definition} 

\begin{remark}
The factorisation property \eqref{eq:instance-factors} generalises the fact that in a conceptual space every instance $z=(z_1,\dots,z_n)$ factors as a product of one instance $z_i$ per factor $Z_i$. In fact in our classical models every point automatically satisfies this, but in quantum theory we will see that there are points which are not instances.
\end{remark} 
% Todo: Could discuss other states more here. 
Thanks to the structure of the scalars in $\catC$, composing a concept $C$ with any input state will yield a positive real scalar. In particular applying $C$ to an instance $z$ yields 
\[
\begin{tikzpicture}
	\begin{pgfonlayer}{nodelayer}
		\node [style=point] (0) at (0, -0.75) {$z$};
		\node [style=map] (1) at (0, 1.25) {$C$};
		\node [style=label] (2) at (-0.5, 0.25) {$Z$};
	\end{pgfonlayer}
	\begin{pgfonlayer}{edgelayer}
		\draw (1) to (0);
	\end{pgfonlayer}
\end{tikzpicture}
 \in \mathbb{R}^+
\]
which we interpret as specifying how well the instance fits the concept. As before, we say that an instance $z$ is \emph{prototypical} for a concept $C$ when $C \circ w  \leq C \circ z$ for all instances $w$. Note that whenever $C \leq D$ we have $C(z) \leq D(z)$ for all instances $z$. 

\begin{remark}
 In each of our example categories, general normalised states $\rho$  typically have an interpretation as states of uncertainty over instances of $Z$, e.g. finite probability distributions in $\Class$. The scalar $C \circ \rho \in \mathbb{R}^+$ again can be interpreted as a measure of how well the concept $C$ fits this uncertain input $\rho$. 
 \end{remark}

% A more general state $\rho$ of $Z$ is typically interpreted in each of our example categories as a kind of probabilistic mixture over points such  

% \scc{define Inst(z) here or replace with Instances}

% \subsection{Crisp Concepts}

% The above notions generalise our previous notion of fuzzy concepts. As such it remains for us to identify which concepts can be considered crisp. 

It remains for us to identify which concepts can be considered crisp. 

\begin{definition} 
A concept $C$ on $Z$ is \emph{crisp} when it is of the form
\[
\begin{tikzpicture}
	\begin{pgfonlayer}{nodelayer}
		\node [style=decompflip] (0) at (0, 0) {};
		\node [style=none] (1) at (0, -1) {};
		\node [style=upground] (2) at (0, 1.25) {};
		\node [style=label] (3) at (-0.75, 0.75) {$K$};
		\node [style=label] (4) at (0, -1.5) {$Z$};
		\node [style=map] (5) at (-4, 0.5) {$C$};
		\node [style=none] (6) at (-4, -0.75) {};
		\node [style=none] (7) at (-2, 0) {$=$};
		\node [style=label] (8) at (-4, -1.25) {$Z$};
	\end{pgfonlayer}
	\begin{pgfonlayer}{edgelayer}
		\draw (2) to (1.center);
		\draw (6.center) to (5);
	\end{pgfonlayer}
\end{tikzpicture}
\]
for some projection morphism $Z \to K$. 
\end{definition} 

Intuitively we can identify the crisp concept with $K$ (more precisely, the embedding and projection from $K$). Indeed, for any instance $z$ of $Z$ given by embedding a point of $K$, we will have $C(z) = 1$: 
\[
\begin{tikzpicture}
	\begin{pgfonlayer}{nodelayer}
		\node [style=point] (0) at (1.5, -1) {$z$};
		\node [style=map] (1) at (1.5, 1) {$C$};
		\node [style=label] (2) at (1, 0) {$Z$};
		\node [style=point] (3) at (-6.5, -0.5) {$z$};
		\node [style=none] (4) at (-6.5, 0.75) {};
		\node [style=label] (5) at (-6.5, 1.25) {$Z$};
		\node [style=none] (6) at (-5.25, 0) {$=$};
		\node [style=point] (7) at (-3.75, -1.25) {$k$};
		\node [style=none] (8) at (-3.75, 0) {};
		\node [style=label] (9) at (-3, -0.5) {$K$};
		\node [style=decomp] (10) at (-3.75, 0) {};
		\node [style=none] (11) at (-3.75, 1) {};
		\node [style=label] (12) at (-3.75, 1.75) {$Z$};
		\node [style=none] (13) at (-1, 0) {$\implies$};
		\node [style=none] (14) at (2.75, 0) {$=$};
		\node [style=point] (15) at (4.5, -2) {$k$};
		\node [style=none] (16) at (4.5, -0.75) {};
		\node [style=decomp] (18) at (4.5, -0.75) {};
		\node [style=upground] (19) at (4.5, 2) {};
		\node [style=decompflip] (20) at (4.5, 0.75) {};
		\node [style=none] (21) at (6.25, 0) {$=$};
		\node [style=point] (22) at (7.5, -0.75) {$k$};
		\node [style=none] (23) at (7.5, 0.75) {};
		\node [style=none] (24) at (9.5, 0) {$1$};
		\node [style=upground] (26) at (7.5, 0.75) {};
		\node [style=none] (27) at (8.5, 0) {$=$};
	\end{pgfonlayer}
	\begin{pgfonlayer}{edgelayer}
		\draw (1) to (0);
		\draw (4.center) to (3);
		\draw (8.center) to (7);
		\draw (11.center) to (10);
		\draw (16.center) to (15);
		\draw (19) to (18);
		\draw (23.center) to (22);
	\end{pgfonlayer}
\end{tikzpicture}
\]
Conversely, in our example categories, a point of $z$ is of the above form for some $k$ iff $C(z)=1$. Thus we can identify $K$ with the prototypical instances of $C$. 

As a special case of crisp concepts, we can see each point (and in particular each instance) $z$ of $Z$ as defining a crisp concept via its projection $z^\dagger$. We call such a concept $z^\dagger$ a \emph{pure concept} of $Z$. We think of $z^\dagger$ as the `maximally sharp' concept which picks out the single point $z$, satisfying 
\[
\begin{tikzpicture}
	\begin{pgfonlayer}{nodelayer}
		\node [style=point] (0) at (0.25, -0.5) {$z$};
		\node [style=none] (1) at (0.25, 0.5) {};
		\node [style=copoint] (3) at (0.25, 0.75) {$z$};
		\node [style=none] (4) at (1.75, 0) {$=$};
		\node [style=none] (5) at (3, 0) {$1$};
	\end{pgfonlayer}
	\begin{pgfonlayer}{edgelayer}
		\draw (1.center) to (0);
	\end{pgfonlayer}
\end{tikzpicture}

\]

\subsection{Examples}

Let us unpack these categorical definitions in each of our motivating `classical' categories. 

\begin{example}
A conceptual model in $\Class$ is precisely a finite set $Z$ given as a subset $Z \subseteq Z_1 \times \dots \times Z_n$. A concept is an arbitrary function $C \colon Z \to \mathbb{R}^+$, ordered pointwise. An instance is an element $z \in Z$, with \eqref{eq:instance-factors} holding automatically since always $z = (z_1, \dots, z_n)$ for some $z_i \in Z_i$. Applying a concept $C$ to an instance $z$ evaluates the function as $C(z) \in \mathbb{R}^+$. Crisp concepts correspond to arbitrary subsets $K \subseteq Z$, or more precisely their indicator functions $1_K \colon Z \to \{0, 1\}$. Pure concepts are indicator functions $1_z$ of instances $z \in Z$. 
\end{example}

\begin{example}
In $\Prob$, a model is a measurable space given as a measurable subset $Z \subseteq Z_1 \times \dots \times Z_n$ of measurable spaces $Z_i$. Concepts are arbitrary measurable functions $C \colon Z \to [0,1]$, instances are (the point measures defined by) points $z \in Z$, crisp concepts $1_K$ correspond to arbitrary measurable subsets $K \subseteq Z$, and pure concepts correspond to points $z \in Z$. 
\end{example} 

In the category $\ConSp$ our definitions yield precisely the conceptual space framework as formalised in Section \ref{sec:conspaces}. 

\begin{example}
A model in $\ConSp$ is precisely a conceptual space, i.e. a convex space viewed as a convex subset $Z \subseteq Z_1 \times \dots \times Z_n$ of convex spaces $Z_i$. Instances are again points $z \in Z$. Crisp concepts are precisely those of Definition \ref{def:crisp-concept}, namely the indicator functions $1_K$ of convex measurable subsets $K \subseteq Z$, with pure concepts being the indicator functions $1_z$ of points $z \in Z$. More general concepts are fuzzy concepts (log-concave measurable maps) $C \colon Z \to [0,1]$ in the sense of Definition \ref{def:fuzzy-concept}.
\end{example} 

\section{Quantum Models} \label{sec:quantum-models}

We can now define our quantum model of concepts inspired by the conceptual space framework. To do so we will simply unpack our definitions from Section \ref{sec:cat-con-spaces} in the category of quantum processes, which we first introduce.

\begin{example} 
In the category $\Quant{}$ the objects are finite dimensional Hilbert spaces $\hilbH, \hilbK \dots$ and the morphisms $f \colon \hilbH \to \hilbK$ are \emph{completely positive} (CP) maps 
\[
f \colon L(\hilbH) \to L(\hilbK)
\]
where $L(\hilbH)$ denotes the space of linear operators on $\hilbH$.  Recall that such a map is linear and is \emph{positive} in that whenever $a \in L(\hilbH)$ is a positive operator we have that $f(a)$ is positive also. Being completely positive means that $f \otimes \id{\hilbH'}$ is positive for any $\hilbH'$. We order morphisms via $f \leq g$ whenever $g -f$ is a positive map. 

The tensor product $\otimes$ is the usual tensor of Hilbert spaces and linear maps, with $I = \mathbb{C}$. In particular, effects $e \colon \hilbH \to I$ are positive linear 
functionals 
\[
e \colon L(\hilbH) \to \mathbb{C}\]
which we may identify with positive operators $a \in L(\hilbH)$ via $e(b) = \Tr(a \circ b)$. Scalars are $r \in \mathbb{R}^+$ as expected.  

Discarding is given by $\discard{}(a) = \Tr(a)$, corresponding to the identity operator $\id{\hilbH}$. A morphism $f$ is a channel iff it is a completely positive trace-preserving (CPTP) map, with $\Tr(f(a)) = \Tr(a)$ for all $a \in L(\hilbH)$. 

States of $\hilbH$ can be identified with positive operators $\rho \in L(\hilbH)$ via $\rho = \rho(1)$. A state $\rho$ is normalised precisely when it defines a density matrix, with $\Tr(\rho) = 1$ -- the usual notion of a quantum state.

A special class of morphisms are the \emph{pure} CP maps $\hat f \colon L(\hilbH) \to L(\hilbK)$ induced by linear maps $f \colon \hilbH \to \hilbK$, given by 
\[
\hat f(a) = f \circ a \circ  f^\dagger
\] 
Morphisms, including states and effects, which are not pure are called \emph{mixed}. Any isomorphism on $\hilbH$ which is also a channel is in fact pure, and induced by a matrix $U \in L(\hilbH)$ which is \emph{unitary}, meaning that is invertible with its Hermitian adjoint as the inverse: $U^\dagger=U^{-1}$.

Embedding morphisms are the pure maps induced by inclusions $i \colon \hilbK \hookrightarrow \hilbH$ of subspaces into $\hilbH$. The corresponding projection is the pure map induced by the linear map $i^\dagger \colon \hilbH \to \hilbK$  with $i^\dagger(\psi) = \psi$ for $\psi \in \hilbK$ and $i^\dagger(\psi) = 0$ for $\psi \in \hilbK^\bot$, the orthogonal subspace to $\hilbK$. 

In particular, a point of $\hilbH$ is formally an inclusion of Hilbert spaces $\mathbb{C} \hookrightarrow \hilbH$, which may be identified with a pure quantum state
% \[
$\ket{\psi} \bra{\psi}$
% \]
for some unit vector $\psi \in \hilbH$.\footnote{Here we use the standard `bra-ket' notation whereby vectors and linear functionals on $\hilbH$ are written in the form $\ket{\psi}$, $\bra{\phi}$ respectively. Then for a unit vector $\psi \in \hilbH$, $\ket{\psi}\bra{\psi}$ is the density operator of the corresponding pure state on $\hilbH$.}

% unit vector $\psi \in \hilbH$ via $\psi = \psi(1)$. Thus points correspond to pure quantum states $\psi \in \hilbH$ of the Hilbert space $\hilbH$.
% We adopt the standard bra-ket notation in which the pure state (point) is denoted $\ket{\psi}$ while its corresponding functional $\psi^\dagger$ is denoted $\bra{\psi}$.
\end{example}

% We can now specify our quantum conceptual models. 
Making use of our categorical formulation of conceptual spaces now gives us our notion of a quantum  model.

\begin{definition} 
A \emph{quantum conceptual model} is a conceptual model in $\Quant{}$. 
\end{definition} 

Thus a quantum conceptual model is given by a Hilbert space $\hilbH$ viewed as a subspace of a tensor product 
\[
\hilbH \subseteq \hilbH_1 \otimes \dots \otimes \hilbH_n
\]
of Hilbert spaces $\hilbH_1, \dots, \hilbH_n$, called the factors, with the inclusion drawn as
\[
\begin{tikzpicture}
	\begin{pgfonlayer}{nodelayer}
		\node [style=decompwide] (0) at (0, 0) {};
		\node [style=none] (1) at (0, -1.25) {};
		\node [style=none] (2) at (-1, 0.25) {};
		\node [style=none] (3) at (1, 0.25) {};
		\node [style=none] (4) at (-1, 1.25) {};
		\node [style=none] (5) at (1, 1.25) {};
		\node [style=none] (6) at (0, 0.75) {$\dots$};
		\node [style=label] (7) at (-1, 1.75) {$\hilbH_1$};
		\node [style=label] (8) at (1, 1.75) {$\hilbH_n$};
		\node [style=label] (9) at (0, -1.75) {$\hilbH$};
	\end{pgfonlayer}
	\begin{pgfonlayer}{edgelayer}
		\draw (5.center) to (3.center);
		\draw (4.center) to (2.center);
		\draw (1.center) to (0);
	\end{pgfonlayer}
\end{tikzpicture}
\]
A quantum concept is then precisely a quantum effect, i.e. a positive operator $C \in L(\hilbH)$. The concepts are ordered via $C \leq D$ whenever $D -C$ is positive or equivalently
\[
C \leq D \iff D = C + E \text{ for some concept $E$ }
\]
Here $+$ is the addition of operators, interpreted as a `coarse-graining' of operators (and distinct from superposition, which is described by addition of vectors in $\hilbH)$. An instance of a quantum model is precisely a pure state $\ket{\psi}\bra{\psi}$, given by a unit vector $\psi \in \hilbH$, which furthermore factorises as 
\[
\psi = \psi_1 \otimes \dots \otimes \psi_n
\]
for unit vectors $\psi_i \in \hilbH_i$. By definition, an instance thus has a well-defined pure state value $\psi_i$ on each factor $\hilbH_i$. 

Note that we only allow (certain) pure states as instances, while mixed states $\rho$ can be interpreted as states of uncertainty (i.e. probabilistic mixtures) over pure states such as instances. In contrast we allow concepts to be mixed or pure. The application of a quantum concept given by a positive operator $C$ to an instance $\psi$ is given by 
\[
\begin{tikzpicture}
	\begin{pgfonlayer}{nodelayer}
		\node [style=map] (0) at (0, 1) {$C$};
		\node [style=point] (1) at (0, -1) {$\psi$};
		\node [style=label] (2) at (-0.75, 0) {$\hilbH$};
	\end{pgfonlayer}
	\begin{pgfonlayer}{edgelayer}
		\draw (0) to (1);
	\end{pgfonlayer}
\end{tikzpicture}
 \ \  = \ \  \bra{\psi} C \ket{\psi} \in \mathbb{R}^+
\]
More generally applying $C$ to any (possibly mixed) state $\rho$ yields the scalar $\Tr(C \rho) \in \mathbb{R}^+$, interpreted as how well $C$ fits the ensemble $\rho$.

Crisp concepts correspond to subspaces $\hilbK \subseteq \hilbH$. More precisely, any such subspace defines a crisp concept via the projection operator $P = 1_\hilbK$ with $P(\psi) = \psi$ for $\psi$ in $\hilbK$ and $P(\psi) = 0$ for $\psi$ in $\hilbK^\bot$.
% \[
% \tikzfig{Cpsi} \ \  = \ \  \bra{\psi} C \ket{\psi} \in \mathbb{R}^+
% \]

Pure quantum concepts are precisely those crisp quantum concepts which are themselves pure as effects. For these, $\hilbK$ is given by a one-dimensional subspace $\langle \psi \rangle$ spanned by some unit vector $\psi \in \hilbH$. Thus a pure quantum concept  is of the form $\ket{\psi}\bra{\psi}$ for such a unit vector, given on instances $\phi$ by 
\[
\begin{tikzpicture}
	\begin{pgfonlayer}{nodelayer}
		\node [style=decompflip] (0) at (0, 1) {$\psi$};
		\node [style=none] (1) at (0, -1) {};
		\node [style=label] (2) at (-0.75, 0) {$\hilbH$};
		\node [style=point] (3) at (0, -1) {$\phi$};
	\end{pgfonlayer}
	\begin{pgfonlayer}{edgelayer}
		\draw (1.center) to (0);
	\end{pgfonlayer}
\end{tikzpicture}
 = \ \  |\braket{\psi | \phi}|^2 \in [0,1]
\]
and with the pure state $\ket{\psi}\bra{\psi}$ as its unique prototype. Conversely every unit vector $\psi$ gives a pure concept $\ket{\psi}\bra{\psi}$ in this way.

\begin{remark}
Observe that in each of our examples, crisp concepts correspond (via embeddings) to appropriately `structured' subsets: in $\Class$ all of the subsets, in $\Prob$ the measurable subsets, in $\ConSp$ the convex subsets and in $\Quant{}$ the linear subspaces. 
\end{remark}

\begin{remark}
While we have defined quantum concepts as arbitrary positive operators, of special interest are those effects $e$ which are \emph{sub-normal}, meaning  $e \leq \discard{}$, or equivalently that as operators $0 \leq e \leq \id{\hilbH}$. These are the concepts which when applied to any normalised state $\rho$ yield a probability $e \circ \rho \in [0,1]$. In particular, as concepts these will always send any instance to a value in $[0,1]$.

One may argue that, like fuzzy concepts in conceptual spaces, any concept should have this property, and so require all concepts to be sub-normal. This can be readily achieved in our framework by replacing the category $\catC$ with the sub-category $\catC_{\mathrm{sc}}$ of \emph{sub-channels} $f$, satisfying $\discard{} \circ f \leq \discard{}$. In the case of $\Quant{}$ this means working in the subcategory of \emph{trace non-increasing} CP maps. 

A subtlety however is that, in $\Quant{}$, since instances are a subset of pure states one may have quantum effects which are not sub-normal but still map all instances to $[0,1]$. An example is the unnormalised Bell effect \eqref{eq:Bell-effect} discussed later. For this reason we do not restrict to sub-normal concepts only in the framework. 

Operationally, however, sub-normal effects $e$ are precisely those which can arise as one of the outcomes of a general quantum measurement, known as a \emph{positive operator valued measurement (POVM)}. More concretely, $e$ can be implemented by first preparing a qubit $Q=\mathbb{C}^2$ in a pure state $\ket{0}$, applying a (non-unique) unitary $U$ to both systems and then \emph{post-selecting} $Q$ on the outcome $\bra{0}$, while discarding $\hilbH$:
\begin{equation} \label{eq:concept-from-unitary}
\begin{tikzpicture}
	\begin{pgfonlayer}{nodelayer}
		\node [style=medium map] (4) at (2.25, 0) {$U$};
		\node [style=upground] (5) at (1.5, 1.75) {};
		\node [style=none] (6) at (1.5, -1.75) {};
		\node [style=none] (7) at (3, -1.75) {};
		\node [style=none] (8) at (3, 2) {};
		\node [style=label] (9) at (1.5, -2.25) {$\hilbH$};
		\node [style=label] (10) at (3.75, -1.25) {$Q$};
		\node [style=decomp] (11) at (3, -2) {$0$};
		\node [style=decompflip] (12) at (3, 2) {$0$};
		\node [style=label] (13) at (0.75, 1) {$\hilbH$};
		\node [style=label] (14) at (3.75, 1.25) {$Q$};
	\end{pgfonlayer}
	\begin{pgfonlayer}{edgelayer}
		\draw (6.center) to (5);
		\draw (7.center) to (8.center);
	\end{pgfonlayer}
\end{tikzpicture}
\end{equation}
Note that every crisp quantum concept is sub-normal in this way. An arbitrary quantum concept takes the form $e = r \cdot e'$ where  $e'$ is sub-normal and $r \in \mathbb{R}^+$ is a positive scaling factor. 
\end{remark}

% Todo: Some discussion of general merits of the quantum models.

\section{Entangled Concepts}
\label{sec:entangled_concepts}

Our quantum models come with a feature which distinguishes them from classical models: the existence of pure \emph{entangled} concepts. 

% \snew{
% For these we make use of some extra structure in our example categories. We assume our category $\catC$ has \emph{addition} in that each collection of morphisms of any given type comes with a (commutative, associative) addition operation 
% \[
% \begin{tikzpicture}[tikzfig]
% 	\begin{pgfonlayer}{nodelayer}
% 		\node [style=map] (0) at (-2, 0) {$f + g$};
% 		\node [style=none] (1) at (-2, 1) {};
% 		\node [style=none] (2) at (-2, -1) {};
% 		\node [style=none] (3) at (0, 0) {$=$};
% 		\node [style=map] (4) at (2, 0) {$f$};
% 		\node [style=none] (5) at (2, 1) {};
% 		\node [style=none] (6) at (2, -1) {};
% 		\node [style=map] (7) at (4, 0) {$g$};
% 		\node [style=none] (8) at (4, 1) {};
% 		\node [style=none] (9) at (4, -1) {};
% 		\node [style=none] (10) at (3, 0) {$+$};
% 	\end{pgfonlayer}
% 	\begin{pgfonlayer}{edgelayer}
% 		\draw (2.center) to (1.center);
% 		\draw (6.center) to (5.center);
% 		\draw (9.center) to (8.center);
% 	\end{pgfonlayer}
% \end{tikzpicture}
% \]
% respected by composition, in that we always have $f \circ (g + h) = f \circ g + f \circ h$, $(f + g) \circ h = f \circ h + g \circ h$ and $f \otimes (g + h) = f \otimes g + f \otimes h$ for all morphisms $f, g, h$. 
% }

% \snew{The following definition applies standard notions from quantum information to our framework \cite{entanglemeasures}}.

% Todo: Somewhere in the paper mention the 'separable' terminology from quantum information. 
% Todo: We require sums now in the framework! 
\begin{definition}
We say that a concept $C$ is a \emph{product} concept when there are effects $C_1, \dots, C_n$ on the factors $Z_1, \dots Z_n$ such that 
\[
\begin{tikzpicture}
	\begin{pgfonlayer}{nodelayer}
		\node [style=map] (0) at (-2.5, 0.75) {$C$};
		\node [style=none] (1) at (-2.5, 0.5) {};
		\node [style=none] (3) at (-2.5, -0.75) {};
		\node [style=label] (5) at (-2.5, -1.25) {$Z$};
		\node [style=none] (7) at (-0.5, 0) {$=$};
		\node [style=none] (9) at (1.75, 1.5) {};
		\node [style=none] (10) at (3.75, 1.5) {};
		\node [style=none] (11) at (1.75, 0.25) {};
		\node [style=none] (12) at (3.75, 0.25) {};
		\node [style=none] (13) at (2.75, 1) {$\dots$};
		\node [style=decompwide] (15) at (2.75, 0) {};
		\node [style=map] (16) at (1.75, 2) {$C_1$};
		\node [style=map] (17) at (3.75, 2) {$C_n$};
		\node [style=none] (18) at (2.75, -0.5) {};
		\node [style=none] (19) at (2.75, -1.5) {};
		\node [style=label] (20) at (2.75, -2) {$Z$};
		\node [style=label] (21) at (1, 0.75) {$Z_1$};
		\node [style=label] (22) at (4.5, 0.75) {$Z_n$};
	\end{pgfonlayer}
	\begin{pgfonlayer}{edgelayer}
		\draw (3.center) to (1.center);
		\draw (12.center) to (10.center);
		\draw (11.center) to (9.center);
		\draw (18.center) to (19.center);
	\end{pgfonlayer}
\end{tikzpicture}
\]
% \snew{A concept $C$ is \emph{separable} when it can be written as a (weighted) sum of product concepts $C = \sum $
% Todo: Could make this a convex sum 
A concept $C$ is \emph{separable} when its value on instances is equal to that given by a convex mixture of product concepts. That is, there are product concepts $C^{(1)},\dots,C^{(k)}$ such that 
\begin{equation} \label{eq:sep}
\begin{tikzpicture}[tikzfig]
	\begin{pgfonlayer}{nodelayer}
		\node [style=map] (1) at (-1, 1) {$C$};
		\node [style=none] (2) at (-1, 0.75) {};
		\node [style=none] (3) at (-1, -0.75) {};
		\node [style=point] (12) at (-1, -1) {$z$};
	\end{pgfonlayer}
	\begin{pgfonlayer}{edgelayer}
		\draw (3.center) to (2.center);
	\end{pgfonlayer}
\end{tikzpicture}
\quad 
=
\quad 
\sum^{k}_{j=1}  \ \ \ 
\begin{tikzpicture}[tikzfig]
	\begin{pgfonlayer}{nodelayer}
		\node [style=none] (5) at (4.75, 1.5) {};
		\node [style=none] (6) at (6.75, 1.5) {};
		\node [style=none] (7) at (4.75, 0.25) {};
		\node [style=none] (8) at (6.75, 0.25) {};
		\node [style=none] (9) at (5.75, 1) {$\dots$};
		\node [style=decompwide] (10) at (5.75, 0) {};
		\node [style=map] (11) at (4.75, 2) {$C^{(j)}_1$};
		\node [style=map] (12) at (6.75, 2) {$C^{(j)}_n$};
		\node [style=none] (13) at (5.75, -0.5) {};
		\node [style=point] (14) at (5.75, -1.5) {$z$};
		\node [style=label] (16) at (4, 0.75) {$Z_1$};
		\node [style=label] (17) at (7.5, 0.75) {$Z_n$};
	\end{pgfonlayer}
	\begin{pgfonlayer}{edgelayer}
		\draw (8.center) to (6.center);
		\draw (7.center) to (5.center);
		\draw (13.center) to (14);
	\end{pgfonlayer}
\end{tikzpicture}
\end{equation}
for all instances $z$, where the sum is taken in $\mathbb{R}^+$. If a concept $C$ is not separable we say that it is \emph{entangled}. 
\end{definition}

Intuitively, a product concept is one which treats the factors separately, being described by individual concepts on the factors acting independently. A separable concept generalises this by allowing for a `coarse-graining', given by the sum, over a number of product concepts. Entangled concepts describe correlations between the factors which cannot be explained in terms of any such classical combination of product concepts. 

Each of our classical categories $\Class, \Prob$ and $\ConSp$ contain product concepts as well as separable (but non-product) concepts. Here a concept will fail to form a product whenever it encodes correlations between the factors, being given by a function $C$ on $Z$ which is not of the form $C(z) = C_1(z_1) \dots C_n(z_n)$ for any $C_1,\dots,C_n$. In fact, in $\Class$ every concept is separable. Separability in $\Prob$ and $\ConSp$ is discussed further in Remark \ref{rem:cts-entanglement}. 

 However, from the definitions we can see that these categories do not contain any pure entangled concepts. Indeed, they share the property that every point of a  model $Z \subseteq Z_1 \times \dots \times Z_n$ forms an instance $z=(z_1,\dots, z_n)$. Thus every pure concept $z^\dagger$ on $Z$ is a product concept: 
\[
\begin{tikzpicture}
	\begin{pgfonlayer}{nodelayer}
		\node [style=decompflip] (0) at (-2.5, 0.75) {$z$};
		\node [style=none] (1) at (-2.5, 0.5) {};
		\node [style=none] (3) at (-2.5, -0.75) {};
		\node [style=label] (5) at (-2.5, -1.25) {$Z$};
		\node [style=none] (7) at (-0.5, 0) {$=$};
		\node [style=none] (9) at (1.75, 1.5) {};
		\node [style=none] (10) at (3.75, 1.5) {};
		\node [style=none] (11) at (1.75, 0.25) {};
		\node [style=none] (12) at (3.75, 0.25) {};
		\node [style=none] (13) at (2.75, 1) {$\dots$};
		\node [style=decompwide] (15) at (2.75, 0) {};
		\node [style=copoint] (16) at (1.75, 1.75) {$z_1$};
		\node [style=copoint] (17) at (3.75, 1.75) {$z_n$};
		\node [style=none] (18) at (2.75, -0.5) {};
		\node [style=none] (19) at (2.75, -1.5) {};
		\node [style=label] (20) at (2.75, -2) {$Z$};
		\node [style=label] (21) at (1, 0.75) {$Z_1$};
		\node [style=label] (22) at (4.5, 0.75) {$Z_n$};
	\end{pgfonlayer}
	\begin{pgfonlayer}{edgelayer}
		\draw (3.center) to (1.center);
		\draw (12.center) to (10.center);
		\draw (11.center) to (9.center);
		\draw (18.center) to (19.center);
	\end{pgfonlayer}
\end{tikzpicture}
\]
In contrast, quantum models contain both entangled and pure entangled concepts.  Consider a quantum model $\hilbH \subseteq \hilbH_1 \otimes \dots \otimes \hilbH_n$ and any unit vector $\psi \in \hilbH$ which is entangled in the usual sense, i.e. not equal to a tensor product $\psi_1 \otimes \dots  \otimes\textbf{}\psi_n$. Then the point $\ket{\psi}\bra{\psi}$ is not an instance of the model, and the corresponding pure concept of $\hilbH$ is entangled:
\[
\begin{tikzpicture}[tikzfig]
	\begin{pgfonlayer}{nodelayer}
		\node [style=wide copoint] (0) at (1.25, 0.5) {$\psi$};
		\node [style=none] (1) at (0.25, 0) {};
		\node [style=none] (2) at (2.25, 0) {};
		\node [style=none] (3) at (0.25, -1) {};
		\node [style=none] (4) at (2.25, -1) {};
		\node [style=label] (5) at (0.25, -1.5) {$\hilbH_1$};
		\node [style=label] (6) at (2.25, -1.5) {$\hilbH_n$};
		\node [style=none] (7) at (4, 0) {$\neq$};
		\node [style=copoint] (8) at (6, 0.5) {$\psi_1$};
		\node [style=copoint] (9) at (8.5, 0.5) {$\psi_n$};
		\node [style=none] (10) at (6, -0.75) {};
		\node [style=none] (11) at (8.5, -0.75) {};
		\node [style=label] (12) at (6, -1.25) {$\hilbH_1$};
		\node [style=label] (13) at (8.5, -1.25) {$\hilbH_n$};
		\node [style=none] (14) at (7.25, -0.25) {$\dots$};
		\node [style=none] (15) at (1.25, -0.5) {$\dots$};
	\end{pgfonlayer}
	\begin{pgfonlayer}{edgelayer}
		\draw (3.center) to (1.center);
		\draw (4.center) to (2.center);
		\draw (11.center) to (9);
		\draw (10.center) to (8);
	\end{pgfonlayer}
\end{tikzpicture}
\]

\begin{example}
Consider a Hilbert space $\hilbH$ with orthonormal basis $\{\ket{i}\}^n_{i=1}$.
An example of a (maximally) entangled pure concept on $\hilbH \otimes \hilbH$ is the so-called \emph{Bell effect}, depicted: 
\begin{equation} \label{eq:Bell-effect}
\begin{tikzpicture}[tikzfig]
	\begin{pgfonlayer}{nodelayer}
		\node [style=none] (6) at (6.75, 0) {};
		\node [style=none] (7) at (8.75, 0) {};
		\node [style=none] (8) at (6.75, -0.5) {};
		\node [style=none] (9) at (8.75, -0.5) {};
		\node [style=label] (10) at (6.75, -1) {$\hilbH$};
		\node [style=label] (11) at (8.75, -1) {$\hilbH$};
	\end{pgfonlayer}
	\begin{pgfonlayer}{edgelayer}
		\draw [bend left=90, looseness=1.75] (6.center) to (7.center);
		\draw (9.center) to (7.center);
		\draw (8.center) to (6.center);
	\end{pgfonlayer}
\end{tikzpicture}
\end{equation}
which is the effect induced by the (unnormalised) vector $\sum^{n-1}_{i=0} \ket{i i}$, where the sum denotes superposition, i.e. addition in $\hilbH \otimes \hilbH$. Explicitly, the Bell effect has operator 
\[
\sum^{n}_{i=0} \sum^n_{j=0} \ket{i, i}\bra{j, j}
\]
\end{example}

\subsection{Entangled vs classically correlated concepts} \label{sec:ent-vs-class-corr}

To explore the use of correlated and entangled concepts, let us consider the setting where we have a model $Z$ and a given collection of instances $z_1,\dots,z_n$, corresponding to the pure product concepts $z_1^\dagger,\dots,z_n^\dagger$. We will consider how we can combine these instances into a new concept in our main example categories. 

\paragraph{Classical combination}
In $\Class$ and $\Quant{}$ we can `coarse-grain' over the instances to yield a separable concept. Both categories in fact come with an addition operation $+$ on morphisms, given by element-wise addition of matrices and addition of CP maps, respectively. Given any pair of product concepts we can sum them to obtain a separable concept:
\begin{equation} \label{eq:class-correlation}
\begin{tikzpicture}
	\begin{pgfonlayer}{nodelayer}
		\node [style=map] (0) at (-2, 0) {$C$};
		\node [style=none] (1) at (-2, -0.5) {};
		\node [style=none] (3) at (-2, -1.25) {};
		\node [style=label] (5) at (-2, -1.75) {$Z$};
		\node [style=none] (7) at (0.25, 0) {$=$};
		\node [style=map] (8) at (2.25, 1.5) {$C_1$};
		\node [style=map] (9) at (3.75, 1.5) {$C_2$};
		\node [style=map] (10) at (6.75, 1.5) {$C_3$};
		\node [style=map] (11) at (8.25, 1.5) {$C_4$};
		\node [style=none] (12) at (2.25, 0.25) {};
		\node [style=none] (13) at (3.75, 0.25) {};
		\node [style=none] (14) at (6.75, 0.25) {};
		\node [style=none] (15) at (8.25, 0.25) {};
		\node [style=none] (20) at (5.25, 0) {$+$};
		\node [style=decompwide] (21) at (3, 0) {};
		\node [style=decompwide] (22) at (7.5, 0) {};
		\node [style=none] (23) at (3, -0.5) {};
		\node [style=none] (24) at (7.5, -0.5) {};
		\node [style=none] (25) at (3, -1.25) {};
		\node [style=none] (26) at (7.5, -1.25) {};
		\node [style=label] (27) at (3, -1.75) {$Z$};
		\node [style=label] (28) at (7.5, -1.75) {$Z$};
	\end{pgfonlayer}
	\begin{pgfonlayer}{edgelayer}
		\draw (3.center) to (1.center);
		\draw (15.center) to (11);
		\draw (14.center) to (10);
		\draw (13.center) to (9);
		\draw (12.center) to (8);
		\draw (25.center) to (23.center);
		\draw (26.center) to (24.center);
	\end{pgfonlayer}
\end{tikzpicture} 
\end{equation}
which we may call their \emph{classical combination}. More generally we can sum any number of concepts in this way. Summing the pure concepts $z_1^\dagger + \dots + z_n^\dagger$ yields a separable crisp concept which in $\Class$ forms their least upper bound $z_1^\dagger \vee \dots \vee z_n^\dagger$ in the order $\leq$ on concepts. This is interpreted as a coarse-graining, or logical `OR' of the instances.

% Todo: Should really say if this is entangled or not. 
\paragraph{Spatial combination} 
In $\ConSp$ addition of morphisms is not available, but one may still combine the instances by taking the least upper bound $z_1^\dagger \vee \dots \vee z_n^\dagger$.  Here this is given by forming the \emph{convex closure} of the instances, the smallest convex subset containing them all. We call this their \emph{spatial combination}. 

\paragraph{Quantum combination}
In $\Quant{}$ suppose the instances $z_1,\dots,z_n$ correspond to unit vectors $\psi_1,\dots,\psi_n$ respectively. As well as being able to form the classical combination \eqref{eq:class-correlation}, we can construct  a pure entangled concept $\ket{\psi}\bra{\psi}$ by forming a superposition: 
\begin{equation} \label{eq:pure-quant-comb}
\psi = \sum^n_{i=1} \psi_i
\end{equation}
where the sum denotes addition in $\hilbH$ (rather than the coarse-graining of CP maps). We call any such concept a \emph{quantum combination} of the  instances. Note that this superposition is only specified up to relative phase between the terms. 

For the purposes of conceptual modelling, it would be desirable to understand the differences between these combination methods, in particular the separable classical combination versus the entangled concept \eqref{eq:pure-quant-comb} in $\Quant{}$. For this we consider an example.

\begin{example} \label{ex:banana-running}
Consider the case where our model is given by two factors $C, T$ for \textsc{colour} and \textsc{taste} respectively, and consider a concept \emph{banana} which correlates both factors as in Example \ref{ex:fuzzy-concepts}. Two prototypical instances are given: a yellow $(Y)$ banana which is sweet $(S)$, and a green $(G)$ banana which is bitter $(B)$. Thus the banana concept should be an effect $B$ satisfying: 
\[
\begin{tikzpicture}[tikzfig]
	\begin{pgfonlayer}{nodelayer}
		\node [style=point] (1) at (1, -0.75) {$Y$};
		\node [style=point] (2) at (2.5, -0.75) {$S$};
		\node [style=none] (3) at (1, 0.75) {};
		\node [style=none] (4) at (2.5, 0.75) {};
		\node [style=point] (6) at (8.75, -0.75) {$G$};
		\node [style=point] (7) at (10.25, -0.75) {$B$};
		\node [style=none] (8) at (8.75, 0.75) {};
		\node [style=none] (9) at (10.25, 0.75) {};
		\node [style=medium map] (10) at (1.75, 1.25) {$B$};
		\node [style=medium map] (11) at (9.5, 1.25) {$B$};
		\node [style=none] (12) at (4.5, 0) {$=$};
		\node [style=none] (13) at (7, 0) {$=$};
		\node [style=none] (14) at (5.75, 0) {$1$};
		\node [style=label] (15) at (0.5, 0) {$C$};
		\node [style=label] (16) at (3, 0) {$T$};
		\node [style=label] (17) at (10.75, 0) {$T$};
		\node [style=label] (18) at (8.25, 0) {$C$};
	\end{pgfonlayer}
	\begin{pgfonlayer}{edgelayer}
		\draw (3.center) to (1);
		\draw (4.center) to (2);
		\draw (8.center) to (6);
		\draw (9.center) to (7);
	\end{pgfonlayer}
\end{tikzpicture}
\]
For simplicity, we suppose that yellowness and greenness are `orthogonal' in that $Y^\dagger \circ G = 0$.  In $\Class$ we can form the classical combination of the instances:
\begin{equation} \label{eq:classical-comb}
\begin{tikzpicture}[tikzfig]
	\begin{pgfonlayer}{nodelayer}
		\node [style=medium map] (0) at (0.75, 0.75) {$D$};
		\node [style=none] (1) at (0, 0.25) {};
		\node [style=none] (2) at (1.5, 0.25) {};
		\node [style=none] (3) at (0, -0.5) {};
		\node [style=none] (4) at (1.5, -0.5) {};
		\node [style=label] (5) at (0, -1) {$C$};
		\node [style=label] (6) at (1.5, -1) {$T$};
		\node [style=none] (7) at (2.75, 0) {$=$};
		\node [style=copoint] (8) at (4.25, 0.5) {$Y$};
		\node [style=copoint] (9) at (5.5, 0.5) {$S$};
		\node [style=copoint] (10) at (8, 0.5) {$G$};
		\node [style=copoint] (11) at (9.25, 0.5) {$B$};
		\node [style=none] (12) at (4.25, -0.5) {};
		\node [style=none] (13) at (5.5, -0.5) {};
		\node [style=none] (14) at (8, -0.5) {};
		\node [style=none] (15) at (9.25, -0.5) {};
		\node [style=label] (16) at (4.25, -1) {$C$};
		\node [style=label] (17) at (5.5, -1) {$T$};
		\node [style=label] (18) at (8, -1) {$C$};
		\node [style=label] (19) at (9.25, -1) {$T$};
		\node [style=none] (20) at (6.75, 0) {$+$};
	\end{pgfonlayer}
	\begin{pgfonlayer}{edgelayer}
		\draw (4.center) to (2.center);
		\draw (3.center) to (1.center);
		\draw (15.center) to (11);
		\draw (14.center) to (10);
		\draw (13.center) to (9);
		\draw (12.center) to (8);
	\end{pgfonlayer}
\end{tikzpicture} 
\end{equation}
which yields a crisp concept whose only points are the two instances themselves, illustrated in the left-hand side of Figure \ref{fig:combinations}. In $\ConSp$ the spatial combination of the instances instead corresponds to the line connecting the two points in the right-hand side of Figure \ref{fig:combinations}.
\begin{figure} 
\centering 
\[
\begin{tikzpicture}
	\begin{pgfonlayer}{nodelayer}
		\node [style=none] (0) at (-4, -2) {};
		\node [style=none] (1) at (-3.75, 2) {};
		\node [style=none] (2) at (1, -2) {};
		\node [style=none] (3) at (-3.75, -2.5) {};
		\node [style=none] (4) at (-3, -1.25) {x};
		\node [style=none] (5) at (0, 1.25) {x};
		\node [style=label] (6) at (0, 0.5) {y, s};
		\node [style=label] (7) at (-3, -0.5) {g, b};
		\node [style=label] (8) at (-3.75, 2.5) {\text{Colour}};
		\node [style=label] (9) at (0, -2.5) {\text{Taste}};
		\node [style=none] (10) at (-3, 1.25) {};
		\node [style=none] (11) at (0, -1.25) {};
	\end{pgfonlayer}
	\begin{pgfonlayer}{edgelayer}
		\draw (0.center) to (2.center);
		\draw (3.center) to (1.center);
	\end{pgfonlayer}
\end{tikzpicture}
\qquad 
\qquad 
\qquad 
\begin{tikzpicture}
	\begin{pgfonlayer}{nodelayer}
		\node [style=none] (0) at (-4, -2) {};
		\node [style=none] (1) at (-3.75, 2) {};
		\node [style=none] (2) at (1, -2) {};
		\node [style=none] (3) at (-3.75, -2.5) {};
		\node [style=none] (4) at (-3, -1.25) {x};
		\node [style=none] (5) at (0, 1.25) {x};
		\node [style=label] (6) at (0, 0.5) {y, s};
		\node [style=label] (7) at (-3, -0.5) {g, b};
		\node [style=label] (8) at (-3.75, 2.5) {\text{Colour}};
		\node [style=label] (9) at (0, -2.5) {\text{Taste}};
		\node [style=none] (10) at (-3, 1.25) {};
		\node [style=none] (11) at (0, -1.25) {};
	\end{pgfonlayer}
	\begin{pgfonlayer}{edgelayer}
		\draw (0.center) to (2.center);
		\draw (3.center) to (1.center);
		\draw (4.center) to (5.center);
	\end{pgfonlayer}
\end{tikzpicture}
\]
\caption{Classical combination (left) versus spatial combination (right).} 
\label{fig:combinations}
\end{figure}
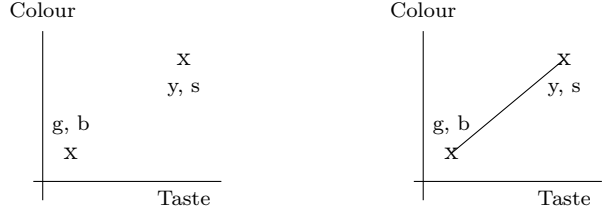
In $\Quant$ we can form the classical combination \eqref{eq:classical-comb} of the instances or a quantum combination $\ket{\psi}\bra{\psi}$ where: 
\[
\psi = |Y,S \rangle + | G,B \rangle \in C \otimes T
\]
Since $\ket{Y}$ and $\ket{G}$ are orthogonal, the equal amplitudes above ensures both instances are deemed to fit the concept equally well.

More generally any linear map $f \colon C \to T$ such that $f(\ket{Y}) = \ket{S}$, $f(\ket{G}) = \ket{B}$ defines a suitable entangled concept: 
% \scc{rephrase domain Hilbert spaces?}
\[
\begin{tikzpicture}
	\begin{pgfonlayer}{nodelayer}
		\node [style=none] (0) at (1.25, 0.75) {};
		\node [style=none] (1) at (3.25, 0.75) {};
		\node [style=medium map] (2) at (-3.75, 1) {$E$};
		\node [style=none] (3) at (-4.5, 0.5) {};
		\node [style=none] (4) at (-3, 0.5) {};
		\node [style=none] (5) at (-4.5, -0.5) {};
		\node [style=none] (6) at (-3, -0.5) {};
		\node [style=none] (7) at (1.25, -0.75) {};
		\node [style=none] (8) at (3.25, -0.75) {};
		\node [style=label] (9) at (-4.5, -1) {$C$};
		\node [style=label] (10) at (-3, -1) {$T$};
		\node [style=label] (11) at (1.25, -1.25) {$C$};
		\node [style=label] (12) at (3.25, -1.25) {$T$};
		\node [style=none] (13) at (-0.75, 0) {$=$};
		\node [style=map] (14) at (1.25, 0.25) {$f$};
	\end{pgfonlayer}
	\begin{pgfonlayer}{edgelayer}
		\draw [bend left=90, looseness=1.75] (0.center) to (1.center);
		\draw (6.center) to (4.center);
		\draw (3.center) to (5.center);
		\draw (8.center) to (1.center);
		\draw (7.center) to (0.center);
	\end{pgfonlayer}
\end{tikzpicture}
\]
The case $E = \psi^\dagger$ corresponds to setting $f(\phi) = 0$ on all vectors $\phi$ orthogonal to $\ket{Y}, \ket{G}$.

To compare the classical and quantum combinations within a quantum model, consider the case where $C = T = \mathbb{C}^2$, $\ket{Y} = \ket{S} = \ket{0}$ and $\ket{G} = \ket{B} = \ket{1}$. The classical combination $D$ as in \eqref{eq:classical-comb} acts on instances as follows:
\[
\begin{tikzpicture}[tikzfig]
	\begin{pgfonlayer}{nodelayer}
		\node [style=medium map] (0) at (1, 0.75) {$D$};
		\node [style=none] (1) at (0.25, 0.25) {};
		\node [style=none] (2) at (1.75, 0.25) {};
		\node [style=none] (3) at (0.25, -0.5) {};
		\node [style=none] (4) at (1.75, -0.5) {};
		\node [style=point] (15) at (0.25, -0.75) {$\psi$};
		\node [style=point] (16) at (1.75, -0.75) {$\phi$};
	\end{pgfonlayer}
	\begin{pgfonlayer}{edgelayer}
		\draw (4.center) to (2.center);
		\draw (3.center) to (1.center);
	\end{pgfonlayer}
\end{tikzpicture} 
 \ \ 
 =  \ \  |\braket{0|\psi}|^2 |\braket{0|\phi}|^2 + |\braket{1|\psi}|^2 |\braket{1|\phi}|^2 
\]
We can take the quantum combination $E$ to be the Bell effect \eqref{eq:Bell-effect}, which `generalises' from the points   $\ket{00}, \ket{11}$ into a structural relationship between the factors, with  
\begin{align*}
\begin{tikzpicture}[tikzfig]
	\begin{pgfonlayer}{nodelayer}
		\node [style=medium map] (4) at (1.25, 0.75) {$E$};
		\node [style=none] (5) at (0.5, 0.25) {};
		\node [style=none] (6) at (2, 0.25) {};
		\node [style=none] (7) at (0.5, -0.75) {};
		\node [style=none] (8) at (2, -0.75) {};
		\node [style=point] (9) at (0.5, -0.75) {$\psi$};
		\node [style=point] (10) at (2, -0.75) {$\phi$};
	\end{pgfonlayer}
	\begin{pgfonlayer}{edgelayer}
		\draw (8.center) to (6.center);
		\draw (5.center) to (7.center);
	\end{pgfonlayer}
\end{tikzpicture}
\ \ 
& = 
\ \ 
\begin{tikzpicture}[tikzfig]
	\begin{pgfonlayer}{nodelayer}
		\node [style=none] (0) at (5.25, 0.25) {};
		\node [style=none] (1) at (7.25, 0.25) {};
		\node [style=none] (2) at (5.25, -0.75) {};
		\node [style=none] (3) at (7.25, -0.75) {};
		\node [style=point] (12) at (5.25, -0.75) {$\psi$};
		\node [style=point] (13) at (7.25, -0.75) {$\phi$};
	\end{pgfonlayer}
	\begin{pgfonlayer}{edgelayer}
		\draw [bend left=90, looseness=1.75] (0.center) to (1.center);
		\draw (3.center) to (1.center);
		\draw (2.center) to (0.center);
	\end{pgfonlayer}
\end{tikzpicture}
  \ \ = \ \  |\braket{\psi^* \mid \phi}|^2 \\ 
&= 
\sum^{n}_{i=0} \sum^n_{j=0} \braket{\psi \mid i, i}\braket{j, j \mid \phi}
\end{align*} 
where $\psi^*$ denotes the conjugate of $\psi$ with respect to the basis $\ket{0}$, $\ket{1}$. 

Both concepts $D$ and $E$ have the exemplars $\ket{Y}\ket{S} = \ket{0, 0} $, $\ket{G}\ket{B} = \ket{1, 1}$ as prototypical instances, sending them to $1$. However, the classical combination $D$ simply compares any input to the two instances, with no further prototypical instances. As a result the structure of each space `between' $\ket{0}$ and $\ket{1}$ is lost, with the orthogonal states  $\ket{\pm} = \frac{1}{\sqrt{2}}(\ket{0} \pm \ket{1})$ treated identically and ${D(\ket{+} \otimes \ket{-})} = \frac{1}{2}$.

In contrast the quantum combination $E$ can be seen to encode a structural relationship between the domains. Any instance of the form $\ket{\phi^*} \otimes \ket{\phi}$, which we can think of as lying `in-between' the exemplars (up to conjugation on the left factor), will form a prototypical instance. Conversely, (conjugate) orthogonal instances will not fit the concept, for example $E(\ket{+} \otimes \ket{-}) = 0$. 
\end{example}

In the above example we see that entangled concepts can encode relationships between factors, rather than simply (weighted) collections of exemplars. Indeed we saw that any pure entangled concept on the factors $C, T$ corresponds to a pure linear map $f \colon C \to T$ which relates the behaviour of the instances on the factors.  

We can understand such a linear map as a generalisation from the two instances into a structural relationship between the factors, as the analogue of an affine (convexity-preserving) function between conceptual spaces. Hence the concept is akin to a concept of the form $\{(x,f(x)) \mid x \in C \}$ for some affine function $f \colon C \to T$ between conceptual spaces. In this sense such a quantum combination forms  an analogue of the spatial combination of concepts on a conceptual space. 

In summary, quantum models are able to store structured correlations between factors even within their pure concepts. In an implementation on a quantum computer, as explored in the next section, pure states and effects may be `cheaper' to prepare than mixed ones, allowing for efficient representation of correlations. Our conclusion is that entanglement should provide an effective way for concepts to encode relationships between factors in the quantum analogue of a conceptual space. 

% Todo: Discuss disjunction, conjunction etc? 

% Todo: Would be nicer to move this later if possible. 
\begin{remark} \label{rem:cts-entanglement}
To treat entanglement for non-pure concepts in $\Prob$ and $\ConSp$  the notion of separability should be generalised from the finite to the continuous setting, by replacing the finite sum in \eqref{eq:sep} with an appropriate integral. Under this definition each concept in $\Prob$ should remain separable. In future it would be interesting to explore whether any form of entanglement can be said to exist in conceptual space models. 
\end{remark}

\section{Experiments}
\label{sec:expts}

In this section we demonstrate how the quantum concept models described in Sections~\ref{sec:quantum-models} and \ref{sec:entangled_concepts} can be learned in practice.
As mentioned in the Introduction, our theoretical framework has been described at an abstract level independent of any particular implementation. However, the use-case we have in mind is one in which the models are (eventually) run on a quantum computer. Here the implementation is a classical simulation, in which we calculate the expected values of measurements analytically (rather than approximating them through sampling).
 The key idea is to use a probabilistic classifier to implement a concept as an effect, where the (binary) classifier learns to distinguish between positive and negative examples of the relevant concept.\footnote{The idea of word meanings as probabilistic classifiers has appeared in the computational linguistics literature \cite{kennington-schlangen-2015-simple}. In terms of classical machine learning, our classifier shares similarities with a ``nearest-centroid" classifier, which has been investigated recently in the quantum setting \cite{Johri2021}.}

We also describe the dataset used in the experiments; the  architecture used for the hybrid networks, including the classical neural network part and the parameterised quantum circuit part; followed by some analysis of our models, including a demonstration of how well the models perform as concept classifiers, as well as how quantum features such as entanglement and mixedness can be used to capture correlations across domains\footnote{Throughout this section, we refer to the factors \textsc{shape}, \textsc{size}, \textsc{colour} and \textsc{position} as `domains', as they would be described in the conceptual space framework.} and discard particular domains. 
Since we are using a hybrid network to perform classification, the experimental work in this section could also be considered part of the nascent field of quantum machine learning \shortcite{PhysRevA.101.032308}.

\subsection{The Shapes Dataset}
\label{sec:shapes}

We use the Spriteworld software \cite{spriteworld19} to generate simple images consisting of coloured shapes of particular sizes in particular positions in a 2D box. For the main dataset, there are three shapes: \{\emph{square, triangle, circle}\}; three colours: \{\emph{red, green, blue}\}; three sizes: \{\emph{small, medium, large}\}; and three positions: \{\emph{bottom, centre, top}\}. The \textsc{colour} attribute here refers to the hue, with the saturation and brightness varied randomly.
The position is relative to the vertical dimension, and the horizontal position is fixed to the centre. The background colour is always the same. 

\begin{figure}
\centering
    \includegraphics[]{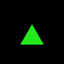}
    \includegraphics[]{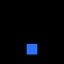}
    \includegraphics[]{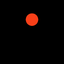}
    \includegraphics[]{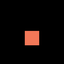}
    \includegraphics[]{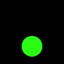}
    \caption{Example shapes, from left to right: (\emph{green, large, triangle, centre}); (\emph{blue, small, square, bottom}); 
    (\emph{red, medium, circle, top});
    (\emph{red, medium, square, centre});
    (\emph{green, large, circle, bottom}).}
    \label{fig:shapes}
\end{figure}

Figure~\ref{fig:shapes} shows some example shapes.
The examples nicely demonstrate the vagueness and variety inherent in the underlying concepts generating the data. For example, the red square is towards the orange end of redness (in contrast to the darker red circle), and the medium-sized circle is close in size to the small blue square. 
Appendix~\ref{sec:app_shapes} contains the parameters used in the Spriteworld software to generate the main dataset (which is the same as that used in \shortciteA{shaikh2022}). The parameters give a range of values for
each of the atomic labels (other than those for \textsc{shape} which are discrete). The appropriate ranges are sampled from uniformly, given a tuple of 4 uniformly sampled input labels, to give the particular values used to generate an instance. We ran the sampler to generate a training set of 3,000 instances, and development and test sets with 300 instances each.

\subsection{The Hybrid Network}
\begin{figure}
\begin{center}
\begin{tikzpicture}[tikzfig]
	\begin{pgfonlayer}{nodelayer}
		\node [style=map] (0) at (4, -0.75) {CNN};
		\node [style=map] (1) at (4, 1.75) {Encoder PQC};
		\node [style=large map] (2) at (5.5, 4.75) { \ \ \ \ Concept PQC \ \  \ \ };
		\node [style=none] (3) at (4, -2.5) {};
		\node [style=none] (4) at (4, -1) {};
		\node [style=label] (5) at (4, -3) {Image};
		\node [style=none] (6) at (4, -0.25) {};
		\node [style=none] (7) at (4, 1.5) {};
		\node [style=label] (8) at (3.25, 0.5) {$\theta$};
		\node [style=none] (9) at (2.5, 2.25) {};
		\node [style=none] (10) at (2.5, 4.25) {};
		\node [style=none] (11) at (8, 4.25) {};
		\node [style=none] (12) at (8, 2.5) {};
		\node [style=label] (13) at (8, 2) {Concept };
		\node [style=label] (14) at (8, 1.25) {params};
		\node [style=label] (15) at (3.75, 3.25) {$\dots$};
		\node [style=none] (16) at (4, 7.75) {};
		\node [style=none] (17) at (5.5, 2.25) {};
		\node [style=none] (18) at (5.5, 4.25) {};
		\node [style=none] (19) at (4, 8.5) {};
		\node [style=none] (20) at (4, 7.75) {};
		\node [style=none] (21) at (4, 8.5) {};
		\node [style=label] (22) at (4, 9) {Yes/no};
		\node [style=label] (23) at (2, 3.25) {$\hilbH_1$};
		\node [style=label] (24) at (5, 3.25) {$\hilbH_n$};
		\node [style=none] (25) at (2.5, 5.25) {};
		\node [style=none] (26) at (2.5, 7) {};
		\node [style=label] (27) at (4, 6.25) {$\dots$};
		\node [style=none] (28) at (5.5, 5.25) {};
		\node [style=none] (29) at (5.5, 7) {};
		\node [style=map] (32) at (4, 7.25) {Measurement};
	\end{pgfonlayer}
	\begin{pgfonlayer}{edgelayer}
		\draw (4.center) to (3.center);
		\draw (7.center) to (6.center);
		\draw [style=thick] (10.center) to (9.center);
		\draw [in=-90, out=90] (12.center) to (11.center);
		\draw [style=thick] (18.center) to (17.center);
		\draw (21.center) to (20.center);
		\draw [style=thick] (26.center) to (25.center);
		\draw [style=thick] (29.center) to (28.center);
	\end{pgfonlayer}
\end{tikzpicture}
\caption{A CNN produces classical parameters for an encoder PQC which prepares a state which is fed into a Concept PQC.}
\label{fig:PQCs}
\end{center}
\end{figure}
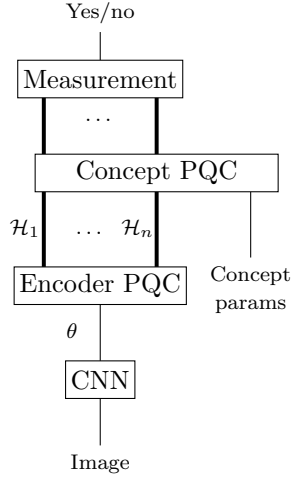

% Todo: 'z' for quantum state or psi? 
An input image is first processed by a convolutional neural network (CNN) \shortcite[Ch.9]{deep_learning} which outputs classical parameters which are fed into a parameterised quantum circuit (PQC) \cite{Benedetti2019}. This PQC we call the \emph{encoder PQC}; it implements a quantum state $z$ which is the representation of the image in our model. Given a concept $C$, a separate \emph{concept PQC} implements a quantum effect corresponding to $C$ which can be applied to the instance $z$, as described in Sections~\ref{sec:cat-con-spaces} and \ref{sec:quantum-models}. 
We assume that the factorisation of the model into the domains $\hilbH_1,\dots,\hilbH_n$ is known by the model; in our experiments these will be the four domains \textsc{shape}, \textsc{colour}, \textsc{size}, \textsc{position}.\footnote{The question of whether, and how, the domains could be learned automatically in the classical setting is an ongoing debate \shortcite{beta-vae,Locatello2019,Brehmer2022}.} The overall setup is shown in Figure~\ref{fig:PQCs}, with thin wires denoting classical data and each thick wire denoting a Hilbert space given by some number of qubits. 

% Sean: By concept label, perhaps it should just say concept paramters. I mean the actual choice of which concept to run (effectively this means inputting paramters in the concept PQC), so this is present whenever we run any particular concept (e.g. twike). 

Given an input image and the parameters encoding a concept, a single run of the circuit produces a `yes' or `no' to determine whether the concept has been deemed to fit the image. The probability of each outcome is obtained either by sampling the circuit many times (on a physical device) or direct calculation (in simulation). With the probabilities for each concept one can then classify which concept best fits the input image.  

In more detail, each instance $z$ is a pure quantum state given by passing an image $X$ into the CNN and then using the resulting parameters in the encoder PQC network:
\[
\begin{tikzpicture}[tikzfig]
	\begin{pgfonlayer}{nodelayer}
		\node [style=map] (0) at (3.75, -0.75) {CNN};
		\node [style=map] (1) at (3.75, 0.75) {Encoder PQC};
		\node [style=none] (3) at (3.75, -2) {};
		\node [style=none] (4) at (3.75, -1) {};
		\node [style=map] (5) at (3.75, -2.25) {$X$};
		\node [style=none] (6) at (3.75, -0.25) {};
		\node [style=none] (7) at (3.75, 0.5) {};
		\node [style=none] (9) at (2.25, 1.25) {};
		\node [style=none] (10) at (2.25, 2.25) {};
		\node [style=label] (15) at (3.75, 2) {$\dots$};
		\node [style=none] (17) at (5.25, 1.25) {};
		\node [style=none] (18) at (5.25, 2.25) {};
		\node [style=label] (23) at (2.25, 2.75) {$\hilbH_1$};
		\node [style=label] (24) at (5.25, 2.75) {$\hilbH_n$};
		\node [style=none] (25) at (-2.75, -0.5) {};
		\node [style=wide point] (26) at (-2.75, -0.75) {$z$};
		\node [style=none] (27) at (-3.5, -0.5) {};
		\node [style=none] (28) at (-3.5, 1) {};
		\node [style=label] (29) at (-2.75, 0.25) {$\dots$};
		\node [style=none] (30) at (-2, -0.5) {};
		\node [style=none] (31) at (-2, 1) {};
		\node [style=label] (32) at (-3.5, 1.5) {$\hilbH_1$};
		\node [style=label] (33) at (-2, 1.5) {$\hilbH_n$};
		\node [style=none] (34) at (-0.25, 0) {$=$};
	\end{pgfonlayer}
	\begin{pgfonlayer}{edgelayer}
		\draw (4.center) to (3.center);
		\draw (7.center) to (6.center);
		\draw [style=thick] (10.center) to (9.center);
		\draw [style=thick] (18.center) to (17.center);
		\draw [style=thick] (28.center) to (27.center);
		\draw [style=thick] (31.center) to (30.center);
	\end{pgfonlayer}
\end{tikzpicture}
\]

Each specific concept $C$ can be understood as a measurement with two outcomes `yes' and `no', such that outcome `yes' means the instance has been deemed to fit the concept. The measurement is given by a Pauli-Z measurement on each qubit, with the overall outcome `yes' identified with obtaining outcome $0$ on every qubit individually, and all other outcomes labelled as `no'. Diagrammatically this is expressed as follows:
\begin{equation} \label{eq:concept-impl}
\begin{tikzpicture}
	\begin{pgfonlayer}{nodelayer}
		\node [style=map] (0) at (9.5, 0) {\ \ \ \ \ Concept PQC \ \ \ \ \ };
		\node [style=none] (1) at (7, -1.75) {};
		\node [style=none] (2) at (7, -0.5) {};
		\node [style=none] (3) at (12.25, -0.5) {};
		\node [style=none] (4) at (12.25, -1.75) {};
		\node [style=label] (5) at (14.25, -1.25) {};
		\node [style=none] (6) at (8.75, -1.25) {$\dots$};
		\node [style=none] (7) at (7, 0.5) {};
		\node [style=none] (8) at (10.5, -1.75) {};
		\node [style=none] (9) at (10.5, -0.5) {};
		\node [style=none] (10) at (7, 1.5) {};
		\node [style=none] (11) at (7, 0.5) {};
		\node [style=none] (12) at (7, 1.5) {};
		\node [style=label] (14) at (7, -2.25) {$\hilbH_1$};
		\node [style=label] (15) at (10.5, -2.25) {$\hilbH_n$};
		\node [style=large map] (16) at (1.75, 0) {$C$};
		\node [style=none] (17) at (0.5, -1.75) {};
		\node [style=none] (18) at (0.5, -0.5) {};
		\node [style=none] (19) at (1.75, -1.25) {$\dots$};
		\node [style=none] (20) at (3, -1.75) {};
		\node [style=none] (21) at (3, -0.5) {};
		\node [style=label] (22) at (0.5, -2.25) {$\hilbH_1$};
		\node [style=label] (23) at (3, -2.25) {$\hilbH_n$};
		\node [style=none] (24) at (4.75, 0) {$:=$};
		\node [style=copoint] (25) at (7, 1.75) {$0$};
		\node [style=point] (26) at (12.25, -2) {$\thetac$};
		\node [style=none] (27) at (10.5, 0.75) {};
		\node [style=none] (28) at (10.5, 1.5) {};
		\node [style=none] (29) at (10.5, 0.5) {};
		\node [style=none] (30) at (10.5, 1.5) {};
		\node [style=copoint] (31) at (10.5, 1.75) {$0$};
	\end{pgfonlayer}
	\begin{pgfonlayer}{edgelayer}
		\draw [style=thick] (2.center) to (1.center);
		\draw [in=-90, out=90] (4.center) to (3.center);
		\draw [style=thick] (9.center) to (8.center);
		\draw [style=thick] (12.center) to (11.center);
		\draw [style=thick] (18.center) to (17.center);
		\draw [style=thick] (21.center) to (20.center);
		\draw [style=thick] (30.center) to (29.center);
	\end{pgfonlayer}
\end{tikzpicture}
\end{equation}
where $\thetac$ are the parameters encoding the concept $C$. Each concept $C$ can be either pure or mixed, depending on whether a pure or mixed circuit is chosen for the concept PQC, which we discuss in the next subsection.

The probability of obtaining `yes' for the concept on an instance $z$ is the value $C \circ z \in [0,1]$. Since all experiments are being carried out in simulation, this probability can be calculated exactly without the need to sample from the model.

% Finally, given a particular concept (and hence set of concept PQCs), a PauliZ measurement is made on each qubit and the expected value of that measurement returned as the output. Since all experiments are being carried out in simulation, this expectation can be calculated exactly without the need to sample from the model. \scc{need more here defining the operator exactly}

% Sean: this could go into the next subsection perhaps. 
% Hence, in the basic model, both the encoder and the concept PQCs are effectively 4 separate PQCs, one for each domain, and so for the basic dataset there are effectively 12 concept PQCs, 1 for each of the 3 values in the 4 domains. \scc{we need some pictures here}

\subsubsection{The CNN and PQCs}
\label{cnn_pqcs}

We use the same CNN from \shortciteA{shaikh2022} for the image processing, which consists of 4 convolutional layers followed by a fully-connected layer. In \shortciteA{shaikh2022} the CNN predicted the means and variances of a multivariate Gaussian, whereas here the CNN predicts the parameters of the encoder PQC. The ReLU activation function is used throughout (including the final layer which predicts the parameters). Appendix~\ref{sec:app_neural_nets} contains more details of the CNN architecture used in our experiments, including the various hyperparameter choices.

The PQCs make use of the parameterised circuit ansatz shown below, defined over any finite collection of qubits. The ansatz $U(\theta)$ is given by performing parameterised $X, Y, Z$ rotations on each qubit, followed by entangling pairs of adjacent qubits using controlled $Z$ gates (with an additional gate operating on the two outermost qubits to complete the chain). Multiple layers of this ansatz can be composed to give a more complex circuit.
We define another ansatz $V(\theta)$ in the same way but with initial rotations in the reverse order $Z, Y, X$. An important special case is that, when given on a single qubit, $U(\theta)$ is simply equal to sequential parameterised $X, Y$ and $Z$ rotations. Similarly $V(\theta)$ on a single qubit amounts to rotating in the order $Z, Y, X$. 

\begin{equation}
%\begin{center}
\begin{tikzpicture}
	\begin{pgfonlayer}{nodelayer}
		\node [style=none] (6) at (5.75, -6.25) {};
		\node [style=none] (7) at (5.75, -0.25) {};
		\node [style=map] (10) at (5.75, -5) {$\RX{1}$};
		\node [style=map] (11) at (5.75, -3.25) {$\RY{1}$};
		\node [style=map] (12) at (5.75, -1.5) {$\RZ{1}$};
		\node [style=none] (22) at (-3, -2.25) {};
		\node [style=none] (23) at (-3, 0.75) {};
		\node [style=large map] (24) at (-1, -0.75) {$ \ \ \  \ \ \ \ U(\theta) \ \ \  \ \ \ \ $};
		\node [style=none] (25) at (1, -2.25) {};
		\node [style=none] (26) at (1, 0.75) {};
		\node [style=none] (27) at (3, -0.75) {$:=$};
		\node [style=none] (28) at (5.75, -0.25) {};
		\node [style=none] (29) at (5.75, 5.25) {};
		\node [style=none] (30) at (17.25, -0.25) {};
		\node [style=none] (31) at (17.25, 5.25) {};
		\node [style=none] (32) at (8.5, -0.25) {};
		\node [style=none] (33) at (8.5, 5.25) {};
		\node [style=blackdot] (34) at (8.5, 0.75) {};
		\node [style=blackdot] (35) at (5.75, 0.75) {};
		\node [style=none] (36) at (-1, 0.25) {$\dots$};
		\node [style=none] (37) at (-1, -1.75) {$\dots$};
		\node [style=blackdot] (38) at (11.25, 1.5) {};
		\node [style=blackdot] (39) at (8.5, 1.5) {};
		\node [style=blackdot] (40) at (17.25, 3.25) {};
		\node [style=blackdot] (41) at (14.5, 3.25) {};
		\node [style=none] (42) at (13, 2.25) {$\dots$};
		\node [style=none] (43) at (11.25, 5.25) {};
		\node [style=none] (44) at (11.25, -0.25) {};
		\node [style=none] (45) at (14.5, 5.25) {};
		\node [style=none] (46) at (14.5, -0.25) {};
		\node [style=blackdot] (47) at (17.25, 4.25) {};
		\node [style=blackdot] (48) at (5.75, 4.25) {};
		\node [style=none] (49) at (17.25, -6.25) {};
		\node [style=none] (50) at (17.25, -0.25) {};
		\node [style=none] (54) at (14.5, -6.25) {};
		\node [style=none] (55) at (14.5, -0.25) {};
		\node [style=none] (59) at (8.5, -6.25) {};
		\node [style=none] (60) at (8.5, -0.25) {};
		\node [style=none] (64) at (11.25, -6.25) {};
		\node [style=none] (65) at (11.25, -0.25) {};
		\node [style=map] (69) at (8.5, -5) {$\RX{2}$};
		\node [style=map] (70) at (8.5, -3.25) {$\RY{2}$};
		\node [style=map] (71) at (8.5, -1.5) {$\RZ{2}$};
		\node [style=map] (72) at (11.25, -5) {$\RX{3}$};
		\node [style=map] (73) at (11.25, -3.25) {$\RY{3}$};
		\node [style=map] (74) at (11.25, -1.5) {$\RZ{3}$};
		\node [style=map] (75) at (14.5, -5) {$\RX{n-1}$};
		\node [style=map] (76) at (14.5, -3.25) {$\RY{n-1}$};
		\node [style=map] (77) at (14.5, -1.5) {$\RZ{n-1}$};
		\node [style=map] (78) at (17.25, -5) {$\RX{n}$};
		\node [style=map] (79) at (17.25, -3.25) {$\RY{n}$};
		\node [style=map] (80) at (17.25, -1.5) {$\RZ{n}$};
		\node [style=none] (81) at (13, -3.25) {$\dots$};
	\end{pgfonlayer}
	\begin{pgfonlayer}{edgelayer}
		\draw [style=thick] (7.center) to (6.center);
		\draw [style=thick] (23.center) to (22.center);
		\draw [style=thick] (26.center) to (25.center);
		\draw [style=thick] (29.center) to (28.center);
		\draw [style=thick] (31.center) to (30.center);
		\draw [style=thick] (33.center) to (32.center);
		\draw (35) to (34);
		\draw (39) to (38);
		\draw (41) to (40);
		\draw [style=thick] (43.center) to (44.center);
		\draw [style=thick] (46.center) to (45.center);
		\draw (48) to (47);
		\draw [style=thick] (50.center) to (49.center);
		\draw [style=thick] (55.center) to (54.center);
		\draw [style=thick] (60.center) to (59.center);
		\draw [style=thick] (65.center) to (64.center);
	\end{pgfonlayer}
\end{tikzpicture}
%\end{center}
%\caption{General circuit anzatz for the PQCs.}
\label{fig:ansatz}
\end{equation}
In the above, $\theta_{j,X}$, $\theta_{j,Y}$, $\theta_{j,Z}$ are single parameters passed to the $X, Y, Z$ rotations on qubit $j = 1,\dots,n$, all contained in the parameters vector $\theta$. In fact this ansatz is universal in that with sufficient layers of the form $U(\theta)$ one may implement any unitary circuit.\footnote{The entangling layer is self-inverse, so that two layers allow us to implement a rotation on any qubit. A swap operation on any pair of qubits can be implemented using three layers, and from this any CX gate. Hence we may implement the universal gate set given by single-qubit phase and Clifford gates; see, for example, \citeA{van2021constructing}.}

Now let us describe the encoder and concept PQCs in more detail. Both consist of some number of qubits per domain $\hilbH_i$. The form of the encoder PQC is the following: 
\[
\begin{tikzpicture}
	\begin{pgfonlayer}{nodelayer}
		\node [style=none] (0) at (-1, 0) {};
		\node [style=none] (1) at (-1, 1.5) {};
		\node [style=large map] (2) at (0.5, 0) {Encoder PQC};
		\node [style=label] (3) at (-1, 2) {$\hilbH_1$};
		\node [style=none] (5) at (2, 0) {};
		\node [style=none] (6) at (2, 1.5) {};
		\node [style=label] (7) at (2, 2) {$\hilbH_n$};
		\node [style=none] (10) at (3.75, 0) {$=$};
		\node [style=none] (11) at (6.5, -1.25) {};
		\node [style=none] (12) at (6.5, 1.5) {};
		\node [style=map] (13) at (6.5, 0) {$U(\theta_1)$};
		\node [style=label] (14) at (6.5, 2) {$\hilbH_1$};
		\node [style=none] (16) at (11.5, -1.25) {};
		\node [style=none] (17) at (11.5, 1.5) {};
		\node [style=label] (18) at (11.5, 2) {$\hilbH_n$};
		\node [style=map] (20) at (11.5, 0) {$U(\theta_n)$};
		\node [style=none] (21) at (9, 0) {$\dots$};
		\node [style=none] (22) at (0.5, 1.25) {$\dots$};
		\node [style=none] (23) at (0.5, -0.5) {};
		\node [style=none] (24) at (0.5, -1) {};
		\node [style=point] (25) at (0.5, -1.25) {$\theta$};
		\node [style=point] (26) at (6.5, -1.5) {$0$};
		\node [style=point] (27) at (11.5, -1.5) {$0$};
	\end{pgfonlayer}
	\begin{pgfonlayer}{edgelayer}
		\draw [style=thick] (1.center) to (0.center);
		\draw [style=thick] (6.center) to (5.center);
		\draw [style=thick] (12.center) to (11.center);
		\draw [style=thick] (17.center) to (16.center);
		\draw (24.center) to (23.center);
	\end{pgfonlayer}
\end{tikzpicture} 
\]
% \[
% \tikzfig{encoder-general-4}
% \]
More generally we can compose multiple layers of such $U$ circuits on each domain. Here the $\ket{0}$ states denote product states $\ket{0 \dots 0}$ on each $\hilbH_i$. Thus by construction the encoder never involves entanglement across domains, and can be viewed as a single encoder per domain. Since the ansatz $U$ is universal, the encoder is able to prepare an arbitrary quantum instance.

In the initial basic setup used, beginning in Section~\ref{sec:states_effects}, we only have one qubit per domain $\hilbH_i$, and only use one layer in the encoder. In this case the encoder simply carries an $X, Y$ and $Z$ rotation per qubit, involving no entanglement. In this  basic setup, the concept PQC also involves no entanglement, taking the following form. 
\begin{equation} \label{eq:concept-initial}
\begin{tikzpicture}
	\begin{pgfonlayer}{nodelayer}
		\node [style=none] (0) at (-1.25, -1.5) {};
		\node [style=none] (1) at (-1.25, 1.5) {};
		\node [style=large map] (2) at (0.5, 0) {\ \ Concept PQC \ \  };
		\node [style=label] (3) at (-1.25, 2) {$\hilbH_1$};
		\node [style=none] (5) at (1.75, -1.5) {};
		\node [style=none] (6) at (1.75, 1.5) {};
		\node [style=label] (7) at (1.75, 2) {$\hilbH_n$};
		\node [style=none] (10) at (5.25, 0) {$:=$};
		\node [style=none] (11) at (8.25, -1.5) {};
		\node [style=none] (12) at (8.25, 1.5) {};
		\node [style=map] (13) at (8.25, 0) {$V(\thetac_1)$};
		\node [style=label] (14) at (8.25, 2) {$\hilbH_1$};
		\node [style=none] (16) at (13.75, -1.5) {};
		\node [style=none] (17) at (13.75, 1.5) {};
		\node [style=label] (18) at (13.75, 2) {$\hilbH_n$};
		\node [style=map] (20) at (13.75, 0) {$V(\thetac_n)$};
		\node [style=none] (21) at (11, 0) {$\dots$};
		\node [style=none] (22) at (0.25, 1) {$\dots$};
		\node [style=none] (28) at (0.25, -1) {$\dots$};
		\node [style=label] (29) at (-1.25, -2) {$\hilbH_1$};
		\node [style=label] (30) at (1.75, -2) {$\hilbH_n$};
		\node [style=label] (31) at (8.25, -2) {$\hilbH_1$};
		\node [style=label] (32) at (13.75, -2) {$\hilbH_n$};
		\node [style=none] (43) at (3, 0) {};
		\node [style=none] (44) at (3, -1.25) {};
		\node [style=point] (45) at (3, -1.5) {$\thetac$};
	\end{pgfonlayer}
	\begin{pgfonlayer}{edgelayer}
		\draw [style=thick] (1.center) to (0.center);
		\draw [style=thick] (6.center) to (5.center);
		\draw [style=thick] (12.center) to (11.center);
		\draw [style=thick] (17.center) to (16.center);
		\draw [in=-90, out=90, looseness=1.25] (44.center) to (43.center);
	\end{pgfonlayer}
\end{tikzpicture}
\end{equation}
% \[
% \tikzfig{concepts-initial-4}
% \]
Concretely, with four domains and one qubit per domain, in this setup the application of a concept $C$ to an instance $z$ amounts to the (probability of the) circuit shown below with post-selection, where $\theta$ is the encoding of the image from the CNN, $\thetac$ are the learned concept parameters and each wire is a single qubit.

\begin{equation}
%\begin{center}
\begin{tikzpicture}
	\begin{pgfonlayer}{nodelayer}
		\node [style=none] (0) at (6.5, 6) {};
		\node [style=none] (1) at (6.5, 0) {};
		\node [style=map] (2) at (6.5, 4.75) {$\RXC{1}$};
		\node [style=map] (3) at (6.5, 3) {$\RYC{1}$};
		\node [style=map] (4) at (6.5, 1.25) {$\RZC{1}$};
		\node [style=none] (5) at (6.5, 0) {};
		\node [style=none] (6) at (9.25, 0) {};
		\node [style=none] (7) at (12, 0) {};
		\node [style=none] (8) at (14.75, 0) {};
		\node [style=none] (9) at (14.75, 6) {};
		\node [style=none] (10) at (14.75, 0) {};
		\node [style=none] (11) at (9.25, 6) {};
		\node [style=none] (12) at (9.25, 0) {};
		\node [style=none] (13) at (12, 6) {};
		\node [style=none] (14) at (12, 0) {};
		\node [style=map] (15) at (9.25, 4.75) {$\RXC{2}$};
		\node [style=map] (16) at (9.25, 3) {$\RYC{2}$};
		\node [style=map] (17) at (9.25, 1.25) {$\RZC{2}$};
		\node [style=map] (18) at (12, 4.75) {$\RXC{3}$};
		\node [style=map] (19) at (12, 3) {$\RYC{3}$};
		\node [style=map] (20) at (12, 1.25) {$\RZC{3}$};
		\node [style=map] (21) at (14.75, 4.75) {$\RXC{4}$};
		\node [style=map] (22) at (14.75, 3) {$\RYC{4}$};
		\node [style=map] (23) at (14.75, 1.25) {$\RZC{4}$};
		\node [style=copoint] (24) at (6.5, 6.25) {$0$};
		\node [style=copoint] (25) at (9.25, 6.25) {$0$};
		\node [style=copoint] (26) at (12, 6.25) {$0$};
		\node [style=copoint] (27) at (14.75, 6.25) {$0$};
		\node [style=none] (28) at (4, 0) {$=$};
		\node [style=wide copoint] (29) at (1.25, 1.75) {$C$};
		\node [style=none] (30) at (1, 1.5) {};
		\node [style=none] (31) at (2, 1.5) {};
		\node [style=none] (32) at (1.5, 1.5) {};
		\node [style=none] (33) at (0.5, 1.5) {};
		\node [style=none] (34) at (1, 0) {};
		\node [style=none] (35) at (2, 0) {};
		\node [style=none] (36) at (1.5, 0) {};
		\node [style=none] (37) at (0.5, 0) {};
		\node [style=none] (38) at (6.5, -6) {};
		\node [style=none] (39) at (6.5, 0) {};
		\node [style=map] (40) at (6.5, -4.75) {$\RX{1}$};
		\node [style=map] (41) at (6.5, -3) {$\RY{1}$};
		\node [style=map] (42) at (6.5, -1.25) {$\RZ{1}$};
		\node [style=none] (43) at (6.5, 0) {};
		\node [style=none] (44) at (9.25, 0) {};
		\node [style=none] (45) at (12, 0) {};
		\node [style=none] (46) at (14.75, 0) {};
		\node [style=none] (47) at (14.75, -6) {};
		\node [style=none] (48) at (14.75, 0) {};
		\node [style=none] (49) at (9.25, -6) {};
		\node [style=none] (50) at (9.25, 0) {};
		\node [style=none] (51) at (12, -6) {};
		\node [style=none] (52) at (12, 0) {};
		\node [style=map] (53) at (9.25, -4.75) {$\RX{2}$};
		\node [style=map] (54) at (9.25, -3) {$\RY{2}$};
		\node [style=map] (55) at (9.25, -1.25) {$\RZ{2}$};
		\node [style=map] (56) at (12, -4.75) {$\RX{3}$};
		\node [style=map] (57) at (12, -3) {$\RY{3}$};
		\node [style=map] (58) at (12, -1.25) {$\RZ{3}$};
		\node [style=map] (59) at (14.75, -4.75) {$\RX{4}$};
		\node [style=map] (60) at (14.75, -3) {$\RY{4}$};
		\node [style=map] (61) at (14.75, -1.25) {$\RZ{4}$};
		\node [style=point] (62) at (6.5, -6.25) {$0$};
		\node [style=point] (63) at (9.25, -6.25) {$0$};
		\node [style=point] (64) at (12, -6.25) {$0$};
		\node [style=point] (65) at (14.75, -6.25) {$0$};
		\node [style=wide point] (67) at (1.25, -1.75) {$z$};
		\node [style=none] (68) at (1, -1.5) {};
		\node [style=none] (69) at (2, -1.5) {};
		\node [style=none] (70) at (1.5, -1.5) {};
		\node [style=none] (71) at (0.5, -1.5) {};
		\node [style=none] (72) at (1, 0) {};
		\node [style=none] (73) at (2, 0) {};
		\node [style=none] (74) at (1.5, 0) {};
		\node [style=none] (75) at (0.5, 0) {};
		\node [style=none] (76) at (-0.5, 0) {};
		\node [style=none] (77) at (3, 0) {};
		\node [style=none] (78) at (5.5, 0) {};
		\node [style=none] (79) at (16, 0) {};
	\end{pgfonlayer}
	\begin{pgfonlayer}{edgelayer}
		\draw [style=thick] (1.center) to (0.center);
		\draw [style=thick] (10.center) to (9.center);
		\draw [style=thick] (12.center) to (11.center);
		\draw [style=thick] (14.center) to (13.center);
		\draw [style=thick] (33.center) to (37.center);
		\draw [style=thick] (30.center) to (34.center);
		\draw [style=thick] (32.center) to (36.center);
		\draw [style=thick] (31.center) to (35.center);
		\draw [style=thick] (39.center) to (38.center);
		\draw [style=thick] (48.center) to (47.center);
		\draw [style=thick] (50.center) to (49.center);
		\draw [style=thick] (52.center) to (51.center);
		\draw [style=thick] (71.center) to (75.center);
		\draw [style=thick] (68.center) to (72.center);
		\draw [style=thick] (70.center) to (74.center);
		\draw [style=thick] (69.center) to (73.center);
		\draw [style=dashed] (76.center) to (77.center);
		\draw [style=dashed] (79.center) to (78.center);
	\end{pgfonlayer}
\end{tikzpicture}
%\end{center}
%\caption{Circuit applying a concept PQC to an instance $z$, with post-selection. The probability of this circuit is the probability of the concept applying to $z$.}
\label{fig:basic-setup-concrete}
\end{equation}

In order to capture mixed and entangled concepts, in  Section \ref{sec:mixedconcepts} we use a richer form for the concept PQC. Entanglement is provided by using the full ansatz $V(\theta)$ over all domains. To introduce mixing, we use an ancilliary copy of each domain $\hilbH_1,\dots\hilbH_n$, prepared in initial state $\ket{0}$, and then discard the original domains as in the following circuit:
\begin{equation} \label{eq:concepts-pqc-general}
\begin{tikzpicture}
	\begin{pgfonlayer}{nodelayer}
		\node [style=none] (0) at (-1.25, -1.5) {};
		\node [style=none] (1) at (-1.25, 1.5) {};
		\node [style=large map] (2) at (0.5, 0) {\ \ \ Concept PQC \ \ \ };
		\node [style=label] (3) at (-1.25, 2) {$\hilbH_1$};
		\node [style=none] (5) at (1.75, -1.5) {};
		\node [style=none] (6) at (1.75, 1.5) {};
		\node [style=label] (7) at (1.75, 2) {$\hilbH_n$};
		\node [style=none] (10) at (5.25, 0) {$:=$};
		\node [style=none] (11) at (8, -1.5) {};
		\node [style=upground] (12) at (8, 1.75) {};
		\node [style=label] (14) at (8, 2.25) {$\hilbH_1$};
		\node [style=none] (16) at (10.5, -1.5) {};
		\node [style=upground] (17) at (10.5, 1.75) {};
		\node [style=label] (18) at (10.5, 2.25) {$\hilbH_n$};
		\node [style=none] (21) at (9.25, -1) {$\dots$};
		\node [style=none] (22) at (0.25, 1) {$\dots$};
		\node [style=none] (28) at (0.25, -1) {$\dots$};
		\node [style=label] (29) at (-1.25, -2) {$\hilbH_1$};
		\node [style=label] (30) at (1.75, -2) {$\hilbH_n$};
		\node [style=label] (31) at (8, -2) {$\hilbH_1$};
		\node [style=label] (32) at (10.5, -2) {$\hilbH_n$};
		\node [style=none] (43) at (2.75, 0) {};
		\node [style=none] (44) at (2.75, -1.25) {};
		\node [style=point] (45) at (2.75, -1.5) {$\thetac$};
		\node [style=medium map] (46) at (11, 0) {$ \ \ \ \ \ \ \ \ \ \ \ \ \ V(\thetac) \ \ \ \ \ \ \ \ \ \ \ \  \ $};
		\node [style=point] (47) at (11.5, -2.75) {$0$};
		\node [style=none] (48) at (11.5, 1.75) {};
		\node [style=label] (49) at (11.5, 2.25) {$\hilbH_1$};
		\node [style=label] (50) at (12.25, -2) {$\hilbH_1$};
		\node [style=none] (51) at (13.75, -2.75) {};
		\node [style=none] (52) at (13.75, 1.75) {};
		\node [style=label] (53) at (13.75, 2.25) {$\hilbH_n$};
		\node [style=label] (54) at (14.25, -2) {$\hilbH_n$};
		\node [style=none] (55) at (9.25, 1) {$\dots$};
		\node [style=none] (56) at (12.75, 1) {$\dots$};
		\node [style=none] (57) at (12.75, -1) {$\dots$};
		\node [style=point] (58) at (13.75, -2.75) {$0$};
	\end{pgfonlayer}
	\begin{pgfonlayer}{edgelayer}
		\draw [style=thick] (1.center) to (0.center);
		\draw [style=thick] (6.center) to (5.center);
		\draw [style=thick] (12) to (11.center);
		\draw [style=thick] (17) to (16.center);
		\draw [in=-90, out=90, looseness=1.25] (44.center) to (43.center);
		\draw [style=thick] (48.center) to (47);
		\draw [style=thick] (52.center) to (51.center);
	\end{pgfonlayer}
\end{tikzpicture}
\end{equation}
% \[
% \tikzfig{concepts-general-4}
% \]
More generally one can include multiple layers of the form $V(\theta)$ prior to discarding. Note that since this ansatz is universal we can implement any unitary with sufficient layers of the form $V(\theta)$, and thus any (sub-normal) quantum concept as outlined in \eqref{eq:concept-from-unitary}.

\subsubsection{Training and Loss}

In \shortciteA{shaikh2022} the classical concepts model is a variant of a VAE \cite{kingma14}, which is a generative model consisting of an encoder and a decoder. Here we choose to train the quantum model to perform binary classification; hence the basic model is a discriminative model with an encoder only.\footnote{In Section~\ref{sec:decoder} below we investigate how the addition of a decoder can affect the instance and concept representations.}

The loss function is the standard binary cross entropy (BCE) loss for binary classification:

\begin{equation}
    %\text{BCE}(D, \psi, \phi) = - \sum_{i} L_i \log p(C_i(X_i) = \text{T}) + (1 - L_i) \log(1 - p(C_i(X_i) = \text{F}))
    \text{BCE}(D, \psi, \phi) = - \frac{1}{N} \sum_{i} L_i \log C_i(X_i) + (1 - L_i) \log(1 - C_i(X_i))
    \label{eqn:loss}
\end{equation}
where $N$ is the size of the data.

The full set of parameters to be learned is $\psi \cup \phi$, where $\psi$ is the set of parameters in the classical encoder CNN and $\phi$ is the set of PQC parameters associated with the set of 12 basic concepts. $D$ is the training data, which consists of 3,000 positive training examples (described in Section~\ref{sec:shapes}) and 3,000 negative examples.\footnote{The negative examples are sampled afresh for each epoch, so strictly speaking (the negative half of) $D$ varies for each epoch.} Each example $D_i$ consists of an image $X_i$, a concept $C_i$, and a label $L_i$, where $L_i$ has the value 1 or 0 depending on whether $C_i$ applies to $X_i$ or not. $C(X_i)$ is the probability that concept $C$ applies to image $X_i$.
Each negative example is created from a positive one by randomly sampling an incorrect concept for each domain; for example, if the positive example is (\emph{green, large, triangle, centre}) then a negative example could be (\emph{blue, medium, square, bottom}). Since we are effectively learning each domain independently in the basic model, a negative example disagrees on every domain. Later models  will use variations on this data (Sections~\ref{sec:correlations}, \ref{sec:mixedconcepts}, \ref{sec:logical_operators} describe these variations). 

% Todo: Come back to this and compre with the description of the loss when over multiple domains. 
We equate $C(X)$, for a concept $C$ on a single domain, with the probability of obtaining the zero state when performing a Pauli-Z measurement on the corresponding qubit, as shown earlier in (\ref{fig:basic-setup-concrete}). For the basic model with no entanglement, the $\log C_i(X_i)$ term in (\ref{eqn:loss}), where $C_i$ is over all 4 domains, is then just the sum of the log probabilities over the domains.

The implementation is in Tensorflow Quantum \cite{tfq}, and the whole hybrid network---both the quantum and the classical parts---are trained end-to-end in simulation on a GPU. The training was run for 100 epochs (unless stated otherwise), with a batch size of 64, and the Adam optimizer was used.\footnote{Each batch has 32 images, but with a positive and negative example for each image.}

% Todo: Call this Instance States and Pure Concepts?
% Todo: Could add some clear explanation that we are using the basic form and reference the concrete equation abobve in the PQCs section. 

\subsection{Instance States and Concept Effects}
\label{sec:states_effects}

We trained a quantum model, using the circuit shown in (\ref{fig:basic-setup-concrete}) above, on the basic dataset of 3,000 labelled images described in Section~\ref{sec:shapes} (together with 3,000 negative examples randomly sampled for each epoch), and tested it on the 300 examples in the development set. The model was trained to perform binary classification, but at test time we choose the concept for each domain which has the highest probability of applying to the input image.

The classification model performed with almost perfect accuracy, obtaining 100\% on the \textsc{colour} and \textsc{shape} domains, and 99\% and 97\% on the \textsc{position} and \textsc{size} domains, respectively.\footnote{The training is relatively stable, but sometimes requires more than one run (because of the random parameter initialisation) to obtain a model which performs this well on the development data. Note also that some examples on the \textsc{position} and \textsc{size} domains would be difficult for humans to classify, given how we have parameterised the data generation process, with some medium-sized shapes being extremely close to some large-sized shapes, for example.} This high accuracy carried over to the 300 examples in the test set, obtaining 100\% on the \textsc{colour} and \textsc{shape} domains, and 96\% and 97\% on the \textsc{position} and \textsc{size} domains, respectively.

\begin{figure}[t!]
    \hspace*{-2.0cm}
    \includegraphics[scale=0.4]{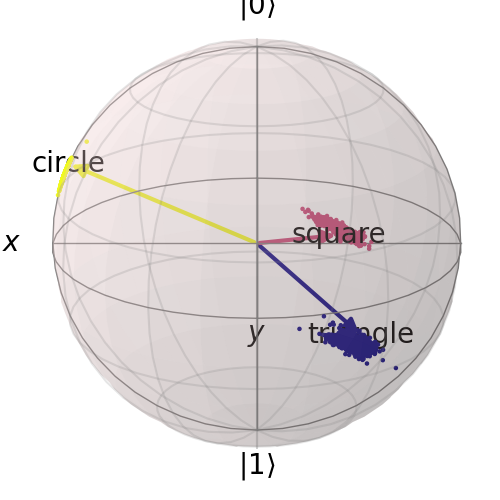}
    \includegraphics[scale=0.4]{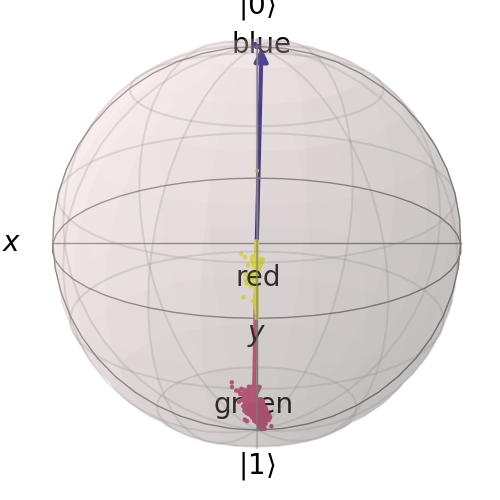}
    \includegraphics[scale=0.4]{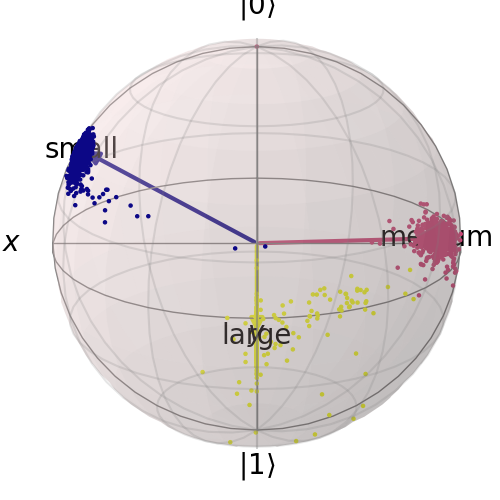}
    \hspace*{2cm}\includegraphics[scale=0.4]{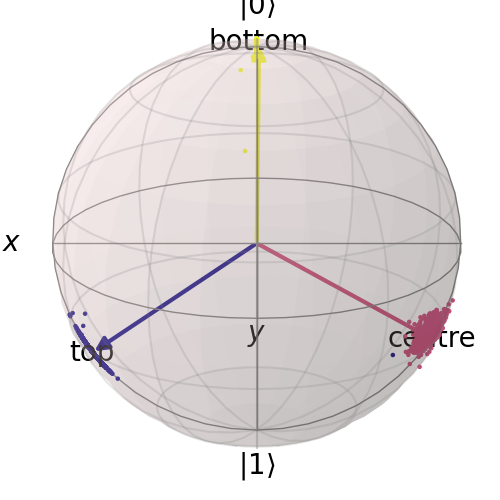}
    \caption{Visualisation of the pure concept effects and instance states on the Bloch sphere, for \textsc{shape}, \textsc{colour}, \textsc{position} and \textsc{size} (clockwise from top-left).}
    \label{fig:basic_bloch}
\end{figure}

Figure~\ref{fig:basic_bloch} visualises the pure effects for each of the 3 concepts on the 4 domains, by plotting the corresponding pure states on a Bloch sphere (we are able to perform the visualisation for this basic model since only one qubit is being used per domain, with no entanglement). The clusters of dots around each concept are the corresponding instances (pure states) in the training data. This visualisation is for the model which performs as described above on the classification task; a model trained from a different random initialisation would have the concepts and instances distributed differently around the sphere, but this visualisation is representative in terms of how the concepts are typically separated and the instances clustered. Note how the 3 concepts on each domain are being pushed apart (strikingly so in the case of the \textsc{position} domain) and how the concepts sit neatly in the centre of each cluster of instances. This is precisely the behaviour we would expect given the loss function in (\ref{eqn:loss}).

\subsubsection{The Rainbow Dataset}
\label{sec:rainbow}

\begin{figure}[t!]
\begin{center}
    \hspace*{-0.0cm}
    \includegraphics[scale=0.3]{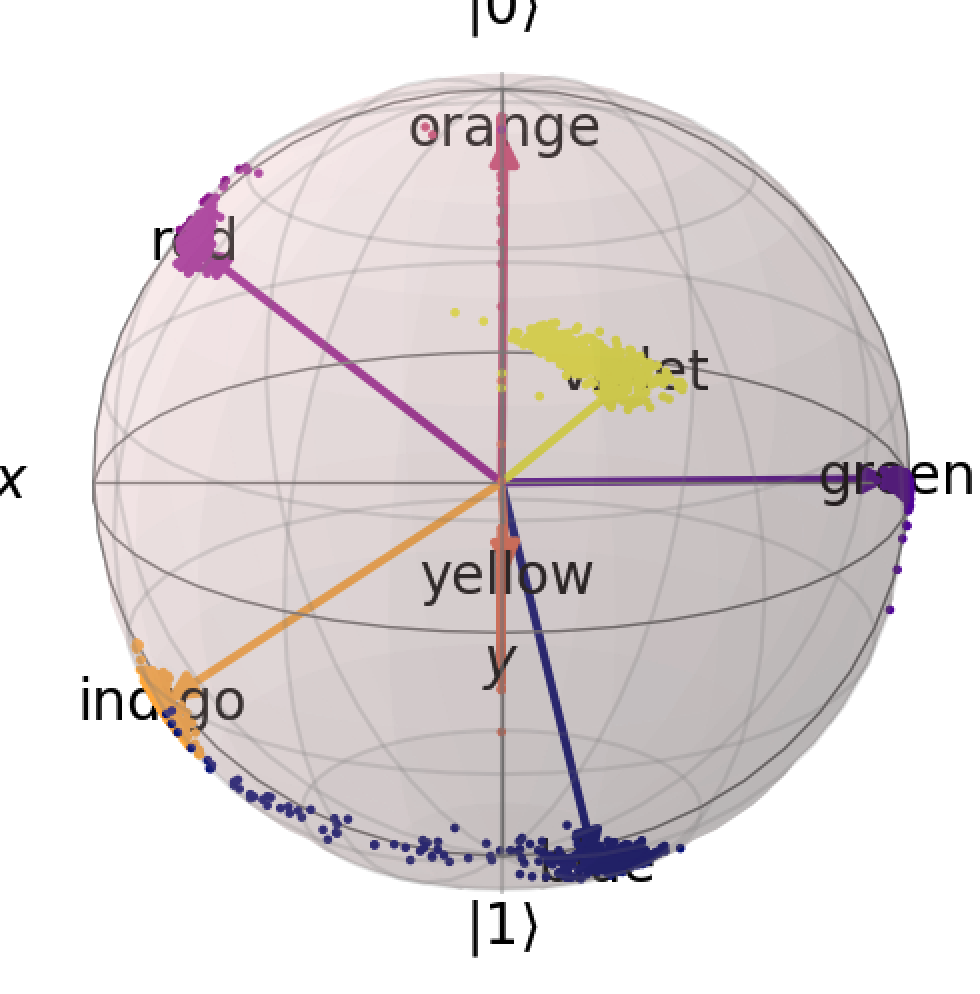}
    \hspace*{1.3cm}
    \includegraphics[scale=0.3]{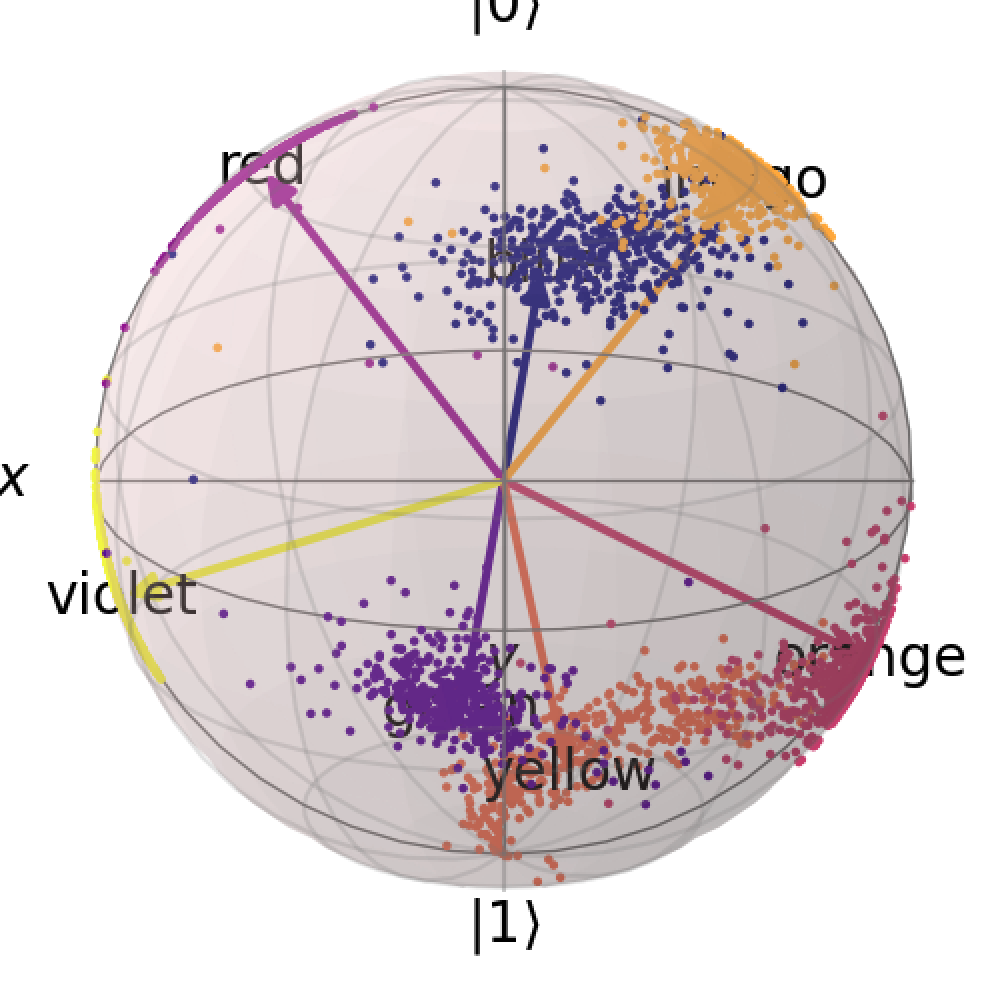}
    \includegraphics[scale=0.35]{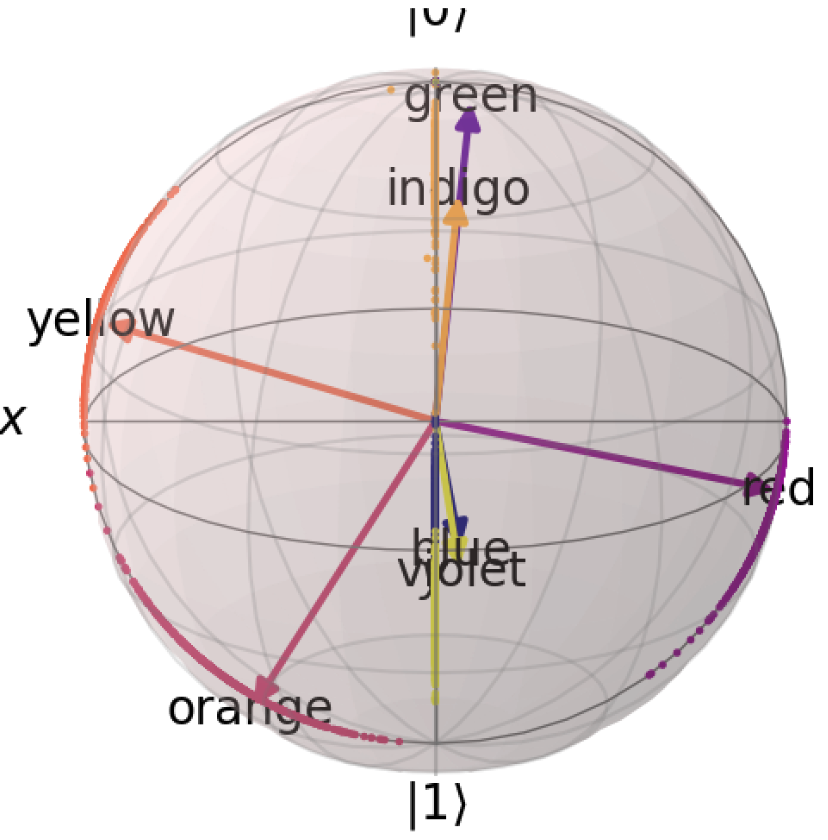}
    
    \caption{Visualisation of the concept effects and instance states on the Bloch sphere, for 3 trained models, for the \textsc{colour} domain on the rainbow dataset.}
    \label{fig:rainbow_bloch}
    \end{center}
\end{figure}

In order to test our model further, we ran the classification training on the rainbow dataset from \shortciteA{shaikh2022}. This data has the same shapes, sizes and positions as the basic dataset, but with an additional 4 colours to give the 7 colours of the rainbow. Appendix~\ref{sec:ext_shapes} contains the parameters used in the Spriteworld software to generate the additional colours. The dataset consists of 3,000 positive images for training, with a further 3,000 negative examples (for each epoch) randomly generated as before, and 300 positive examples each for development and test.

Perhaps unsurprisingly, it was more difficult with this data to obtain a clean separation of the colours on a single qubit.\footnote{Of course there is nothing to prevent us from using more than one qubit per domain, in order to provide a larger Hilbert space in which to represent the additional colours, but the visualisation is harder with more qubits.} However, with a weighting of 0.5 applied to the negative examples in the loss (equation~\ref{eqn:loss}) and running the training for 200 epochs, we were able to obtain the distribution of colours around the Bloch sphere shown in Figure~\ref{fig:rainbow_bloch} (with instances again taken from the training data). The three visualisations are for three separately trained models (i.e. with three different random initialisations of the model parameters).

In terms of accuracy on the development data, the classification model for the Bloch sphere at the top left achieved similar scores on the non-colour domains as before, and an overall accuracy of 95\% on \textsc{colour}, with F1-scores ranging from 91\% to 100\% for the individual colours. The Bloch sphere at the top right is for a model with similar performance, and is shown to demonstrate the variation in models. The example at the bottom is cherry-picked as an example of how the training is able to neatly represent the various colours on the Bloch sphere: note how the \emph{yellow}, \emph{orange} and \emph{red} instances are beautifully placed on the circumference of a circle, with the \emph{red} instances leading into \emph{orange} and then \emph{yellow}.

\subsubsection{Adding a Decoder Loss}
\label{sec:decoder}

One notable feature of the visualisations in Figure~\ref{fig:basic_bloch} is how ``tight" the instance clusters are, despite the variation in the images for a single concept (for example the variation in red shapes in Figure~\ref{fig:shapes}). However, given the loss function in equation~\ref{eqn:loss}, it is not surprising that the model chooses to cluster the instances in this way: the job of the CNN is to produce angle parameters which place an instance (as a pure state) close to the corresponding concept (as a pure effect), but at the same time place the instance far away from the other concepts which do not apply to it.

\begin{figure}[t!]
    \hspace*{-2.0cm}
    \includegraphics[scale=0.3]{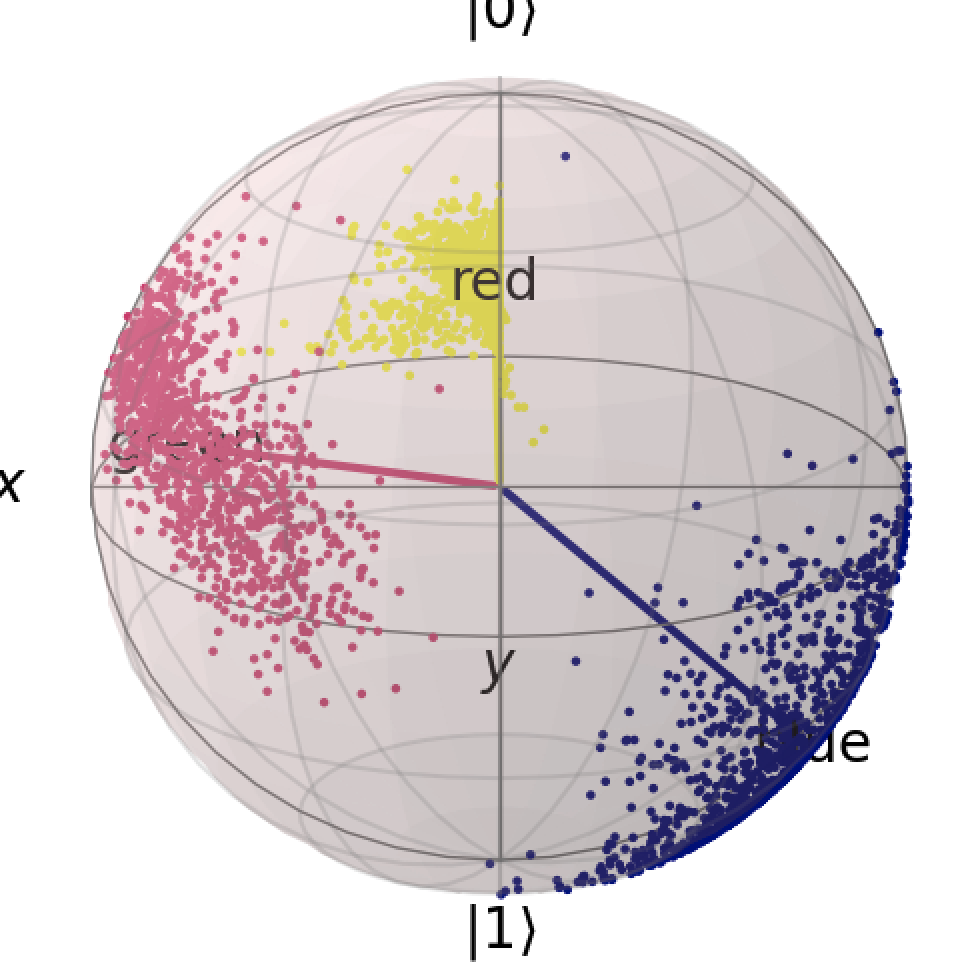}
    \includegraphics[scale=0.3]{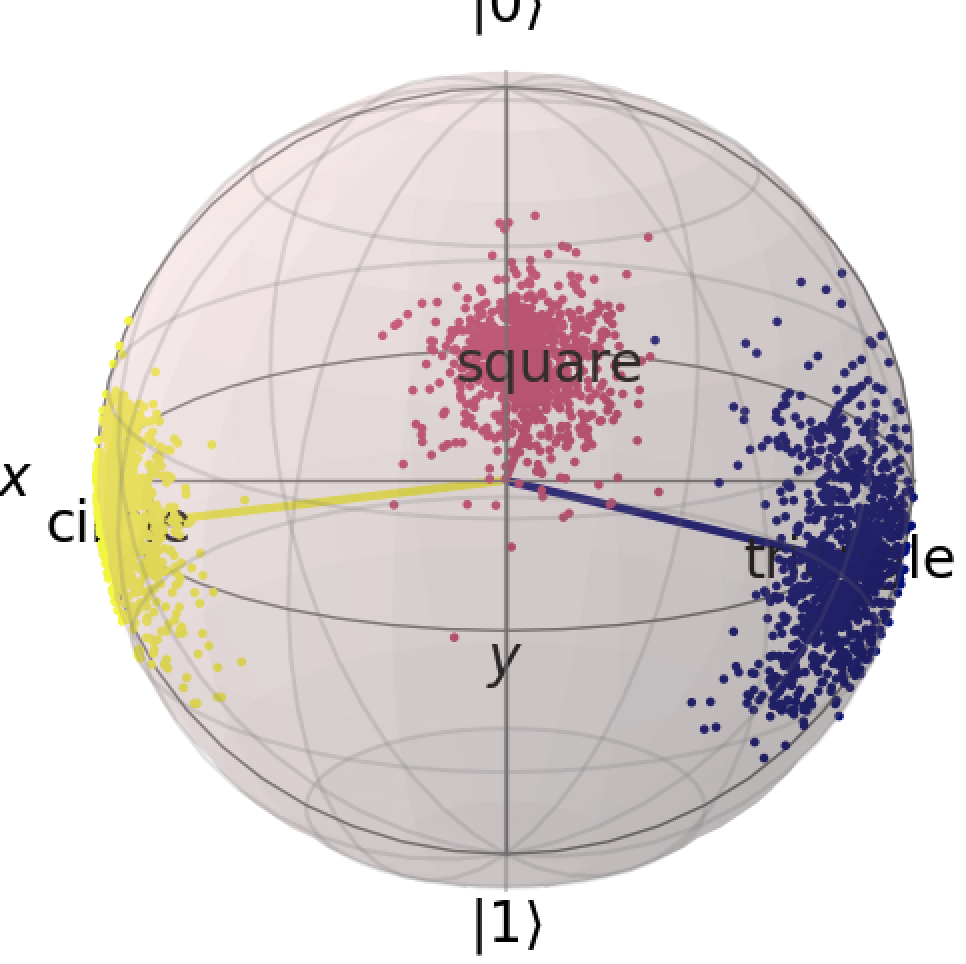}
    \includegraphics[scale=0.41]{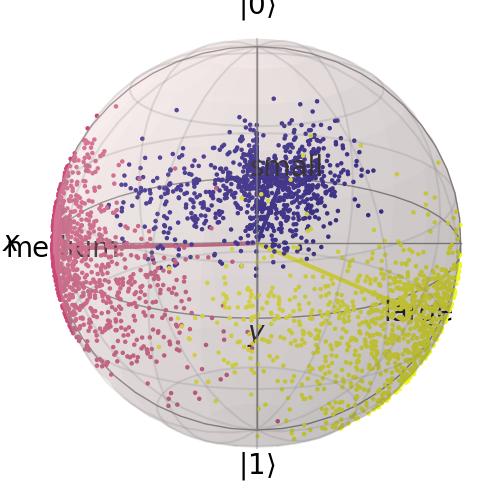}
    \hspace*{2cm}
    \includegraphics[scale=0.41]{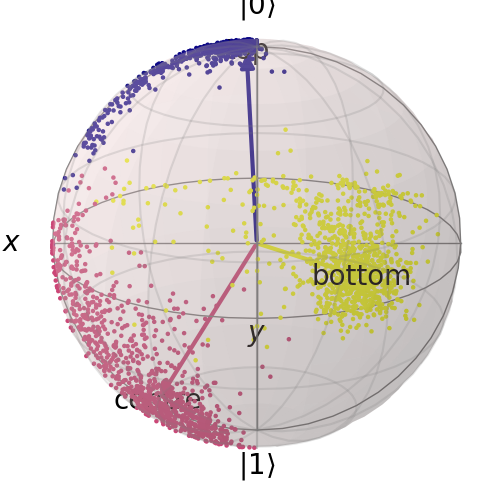}
    \caption{Visualisation of the concept effects and instance states for all 4 domains, for the basic dataset with an additional decoder loss.}
    \label{fig:decoder_bloch}
\end{figure}

There may be use-cases where we would like the representation of instances to better reflect the variation in the underlying images, for example in order to better capture correlations across domains (see Section~\ref{sec:correlations} below). In order to provide more of a ``spread" of the instances, we experimented with an additional decoder loss in the loss function:

\begin{equation}
    \text{Loss}(D, \psi, \phi, \chi) = \text{BCE}(D, \psi, \phi) + \frac{\lambda}{N} \sum_i \text{SE}(\text{DeCNN}(\chi,\text{CNN}(\psi,X_i)),X_i)
    \label{eqn:loss_dec}
\end{equation}

The decoder is a deconvolutional neural network (DeCNN), with parameters $\chi$, which essentially is the CNN ``in reverse": it takes as input the angles output by the CNN, given an image $X_i$, and outputs RGB values for each pixel in the image. SE is the sum of squared errors across all RGB values in the image, and $\lambda$ is a weighting term in the overall loss. The intuition is that, in order to obtain a low SE loss, the encoder CNN has to output angles which are sufficiently informative in order for the DeCNN to accurately reconstruct the original image. This means that, even for two shapes which have similar but different shades of red, for example, the PQC parameters output by the CNN for the \textsc{colour} domain have to be sufficiently different for the two examples so that they can both be accurately reconstructed by the DeCNN. The overall effect is for the instances to be spread more around the Bloch sphere, with the amount of spread controlled by the $\lambda$ hyperparameter. Now the model is similar to the Conceptual VAE model that we presented in \citeA{shaikh2022} (albeit without the generative model interpretation), in that it has both ``encoder" and ``decoder" parts to the loss.\footnote{One possibility for future work is to develop and implement a ``quantum VAE" \cite{khoshaman2018quantum} for concept modelling, and have a generative model in which all parts of the model are quantum.}

Figure~\ref{fig:decoder_bloch} shows how the instances can be distributed more broadly around the Bloch sphere, using the additional decoder loss (with $\lambda = 0.1$). This model still performs well as a classification model on the development data, achieving 
98\% accuracy on \textsc{size}, 99\% on \textsc{colour}, 100\% on \textsc{shape}, and 98\% on \textsc{position}. As a qualitative demonstration of this approach, note how the instances for \emph{centre} and \emph{top} start to merge into each other (blue and red instance dots bottom right), and also for \emph{medium} and \emph{small} (blue and red instance dots bottom left), which is what we would expect for a less discrete representation. 

\subsection{Capturing Correlations}
\label{sec:correlations}

\begin{figure}[t!]
\begin{center}
    \hspace*{-0.0cm}
    \includegraphics[scale=0.3]{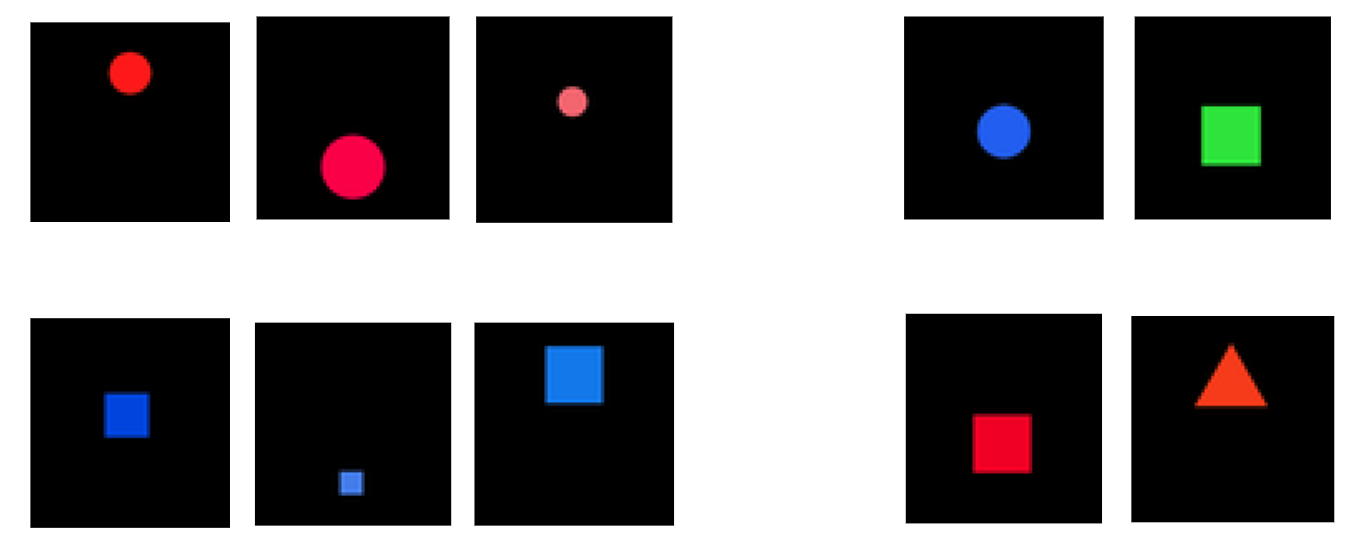}
    \caption{Examples of twikes (on the left) and non-twikes (on the right).}
    \label{fig:twike}
    \end{center}
\end{figure}

Here we show how one of the characteristic features of quantum theory, namely entanglement, can be used to capture correlations across domains. In order to test whether our model can handle concepts which contain correlations, we define a new concept which we call \emph{twike}, which is defined as \emph{(red and circle) or (blue and square)} (i.e. it applies to images containing red circles or blue squares). Figure~\ref{fig:twike} shows some examples of twikes and non-twikes.

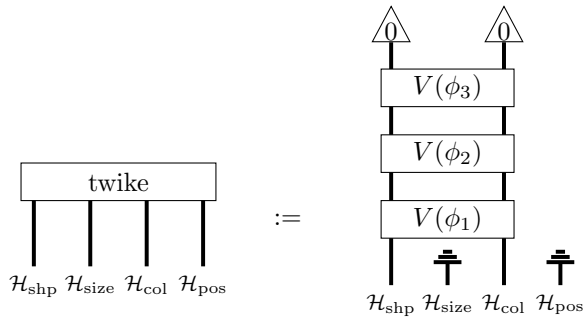
\begin{figure}
\begin{center}
\begin{tikzpicture}
	\begin{pgfonlayer}{nodelayer}
		\node [style=none] (0) at (-2.25, -2.25) {};
		\node [style=none] (1) at (-2.25, -0.25) {};
		\node [style=large map] (2) at (0, 0) {$ \ \ \ \ \ \ $ twike $ \ \ \ \ \ \ $};
		\node [style=none] (3) at (2.25, -2.25) {};
		\node [style=none] (4) at (2.25, -0.25) {};
		\node [style=none] (5) at (4.5, -1) {$:=$};
		\node [style=none] (6) at (7.25, -2.75) {};
		\node [style=copoint] (7) at (7.25, 4) {$0$};
		\node [style=none] (10) at (8.75, -2.75) {};
		\node [style=upground] (11) at (8.75, -2) {};
		\node [style=copoint] (21) at (10.25, 4) {$0$};
		\node [style=none] (22) at (10.25, -2.75) {};
		\node [style=medium map] (27) at (8.75, -1) {$ \ \ \ V(\phi_1) \ \ \ $};
		\node [style=none] (35) at (11.75, -2.75) {};
		\node [style=upground] (36) at (11.75, -2) {};
		\node [style=none] (37) at (-0.75, -2.25) {};
		\node [style=none] (38) at (-0.75, -0.25) {};
		\node [style=none] (39) at (0.75, -2.25) {};
		\node [style=none] (40) at (0.75, -0.25) {};
		\node [style=medium map] (41) at (8.75, 0.75) {$ \ \ \ V(\phi_2) \ \ \ $};
		\node [style=medium map] (42) at (8.75, 2.5) {$ \ \ \ V(\phi_3) \ \ \ $};
		\node [style=label] (43) at (-2.25, -2.75) {$\Hdom{shp}$};
		\node [style=label] (44) at (-0.75, -2.75) {$\Hdom{size}$};
		\node [style=label] (45) at (0.75, -2.75) {$\Hdom{col}$};
		\node [style=label] (46) at (2.25, -2.75) {$\Hdom{pos}$};
		\node [style=label] (47) at (7.25, -3.25) {$\Hdom{shp}$};
		\node [style=label] (48) at (8.75, -3.25) {$\Hdom{size}$};
		\node [style=label] (49) at (10.25, -3.25) {$\Hdom{col}$};
		\node [style=label] (50) at (11.75, -3.25) {$\Hdom{pos}$};
	\end{pgfonlayer}
	\begin{pgfonlayer}{edgelayer}
		\draw [style=thick] (1.center) to (0.center);
		\draw [style=thick] (4.center) to (3.center);
		\draw [style=thick] (7) to (6.center);
		\draw [style=thick] (11) to (10.center);
		\draw [style=thick] (21) to (22.center);
		\draw [style=thick] (36) to (35.center);
		\draw [style=thick] (38.center) to (37.center);
		\draw [style=thick] (40.center) to (39.center);
	\end{pgfonlayer}
\end{tikzpicture}
\end{center}
\caption{Encoder PQC for learning \emph{twike}, here shown with 3 layers of the rotation and entangling $V$ ansatz.}
\label{fig:twike-pqc}
\end{figure}

The concept PQCs we have considered so far, of the form in (\ref{eq:concept-initial}), are unable to learn the concept \emph{twike}, since the domains have been treated independently, with each of the 4 domains effectively containing its own independent concept. 
In order to create connections between the domains in the concept PQC, we can apply our full ansatz $V$ from Section~\ref{cnn_pqcs}, involving controlled-Z gates between wires, across multiple domains. In this first experiment we assume knowledge of the fact that, for the twike concept, the correlations are across the \textsc{shape} and \textsc{colour} domains, with entangling gates only between the qubits for \textsc{shape} and \textsc{colour}.\footnote{This assumption will be relaxed for some of the experiments below.} We also assume that the remaining domains are not relevant and so are not measured, thus effectively being discarded in the concept. We apply potentially multiple layers of ansatz $V$ to the relevant domains, and so the resulting form of the twike concept over the four domains is as shown in Figure~\ref{fig:twike-pqc}, where $\phi$ are the learned parameters for the twike concept.
% Todo: Consistency in wire labels in diagrams.

The training of this model only updates the rotation parameters of the concept PQC; the parameters of the encoder (i.e. the CNN) are kept fixed from the earlier training of the basic model. The loss function is binary cross entropy, as before, with the 3,000 examples from Section~\ref{sec:shapes} used as training data. Roughly 20\% of these instances are positive examples of \emph{twike}, with the remaining being negative examples. We trained this model for 50 epochs, using 2 layers of the rotation and entangling $V$ ansatz for the concept PQC, and obtained 100\% accuracy on the unseen test examples. It was only through the introduction of the entangling gates that we were able to learn the \emph{twike} concept at all.

In terms of the discussion of entanglement and classical correlation in Section \ref{sec:entangled_concepts}, we can say that the twike concept can be naturally described without entanglement, as a classical combination of the pure concepts \emph{red circle} and \emph{blue square} (at least in the case where these pure effects are orthogonal). However, such correlations are not always immediately implementable in many conventional classical models. 
In terms of a concepts model based on a variational autoencoder \cite{beta-vae,shaikh2022}, for example, it would be possible to capture correlations using the covariance matrix of the multivariate Gaussian which is used as the concept representation. However, a standard assumption in that work is to assume a multivariate Gaussian with a diagonal covariance matrix (and so no correlations across domains). Whether a concept like \emph{twike} could be easily modelled using the Conceptual VAE from \shortciteA{shaikh2022}, as it was here, is left as a question for future work.

In contrast to these classical models, the quantum model, via entanglement, provides a relatively straightforward way to learn correlated concepts, even when restricting to pure effects. 

Another question for future work is whether our model could capture continuous correlations across domains. For \emph{twike}, the correlation is ``discrete" in the sense that a discrete value for the \textsc{shape} domain is strongly correlated with a discrete value for the \textsc{colour} domain. An example of a continuously correlated concept would be one in which the size of the shape varies continuously with position, so that large objects are at the top, medium objects in the middle, and small objects at the bottom, with size and position varying monotonically across the whole range. Our discussion of `quantum combinations' of concepts in Section \ref{sec:ent-vs-class-corr}, such as a Bell effect as a combination of $\ket{00}$ and $\ket{11}$, suggests that such continuous correlations should be capturable in the quantum model. We leave the investigation of such concepts for future work.

\subsection{Learning General Mixed and Entangled Concepts} \label{sec:mixedconcepts}

One assumption made above in the \emph{twike} experiments was that the relevant domains---in this case \textsc{shape} and \textsc{colour}---are known in advance, so that the concept PQC can effectively ignore the wires corresponding to the other domains. One interesting question is whether the concept PQC could also learn which domains are relevant, as well as which of those domains should be correlated, if provided with all 4 wires as input.

To allow for such correlations between arbitrary domains, the concept PQC should allow for entanglement between any of its domains. Furthermore, to treat concepts such as \emph{red} in which certain domains are irrelevant, note also that pure effects will not be adequate. As was effectively taking place in the twike concept (Figure~\ref{fig:twike-pqc}), for a domain to be treated as irrelevant the concept should effectively discard it, which is a mixed quantum effect. 
Both of these features can be included by using our most general form of the concept PQC \eqref{eq:concepts-pqc-general} which explicitly includes discarding and is able to describe concepts which are mixed and include entanglement between arbitrary domains. 

In order to test the learning of these general concepts, we set up a similar experiment to \emph{twike}, but this time with just \emph{red} as the concept to be learned. Of course the encoder had already learned \emph{red} when trained to perform classification as part of the basic setup, but in this new experiment we remove the knowledge of which wire the \textsc{colour} domain lives on, and see whether a new concept PQC can learn \emph{red}, given red and non-red instances as input.

Again the training of this model only updates the rotation parameters of the concept PQC; the parameters of the CNN are kept fixed. The loss function is again binary cross entropy, with the 3,000 examples from Section~\ref{sec:shapes} used as training data. Roughly 33\% of these instances are positive examples of \emph{red}, with the remaining being negative examples. We trained this model for 50 epochs, using 2 layers of rotation and entangling gates for the concept PQC, and obtained 100\% accuracy on the unseen test examples. It was only through the introduction of the discarding (plus entangling gates) that we were able to obtain these high accuracies.

The extent to which the concept PQC uses mixing, and whether the learned concept agrees with the intuition given above (for example discarding all domains besides colour for the \emph{red} concept), is a question left for future work. 

\subsection{Concepts containing Logical Operators}
\label{sec:logical_operators}

For one final set of experiments, we investigated whether the entangling and discarding PQC \eqref{eq:concepts-pqc-general} could learn concepts built from logical operators, with concepts such as \emph{red or blue}. As well as the obvious connection with concept representation and learning in general, logical operators are connected to quantum theory through quantum logic \cite{birkhoff_1936}. 

\subsubsection{Conjunction across Domains}

The first concept with a logical operator that we consider is \emph{red and circle}, firstly with the knowledge of which domains are relevant for the concept (in this case \textsc{colour} and \textsc{shape}). The encoder PQC is the simple one back in (\ref{fig:basic-setup-concrete}), but with only the \textsc{colour} and \textsc{shape} wires (so the other two are effectively discarded). We used the same 3,000 training examples as previously, of which roughly 17\% are positive examples and 83\% negative examples. In this case the learning is particularly easy, and the model obtains 100\% accuracy with only a single layer of rotations for the PQC, without any entangling gates or discarding of any ancilliary qubits. The reason is that the factorisation of the domains through the tensor product has effectively provided all the structure required to use conjunction; in some sense, the tensor product is conjunction.

When the knowledge of which domains are relevant is removed, and the more general encoder PQC in (\ref{eq:concepts-pqc-general}) is used, learning becomes harder but an encoder PQC with 4 layers of rotation and entangling gates is able to learn the concept with 100\% accuracy.

\subsubsection{Disjunction within Domains}

Next we consider disjunction, but within rather than across domains, with the concept to be learned being \emph{red or blue}. Of the 3,000 training examples, 61\% are positive examples and 39\% negative. Again, when knowledge of which domains are relevant is provided to the concept PQC, the learning is easy, with 100\% accuracy obtained with a single layer of rotations. 

If each point on the Bloch sphere were to correspond to an instance of the  \textsc{colour} domain, i.e. a single colour, as in our model, then the PQC learning such a pure effect for \emph{red or blue} will in fact be simply learning a single colour, intuitively somewhere ``in between" \emph{red} and \emph{blue}. When the domain only comes with a few concepts, such as the 3 concepts used here, this single instance may do well in approximating \emph{red or blue}, as with the 100\% accuracy. However, in the presence of more concepts, we expect that a concept for \emph{red or blue} should involve mixing. 

When knowledge of which domains are relevant is not provided to the PQC, \emph{red or blue} can also be successfully learned with the more general PQC in (\ref{eq:concepts-pqc-general}) with 3 layers of rotation and entangling gates, including discarding.

Conjunction within domains---e.g. \emph{red and blue}---is not applicable with the base concepts we have in this report, since they are mutually exclusive. Disjunction across domains---e.g. \emph{red or circle}---does make sense (although perhaps a little unnatural), and indeed we can successfully learn \emph{red or circle} using the general PQC in (\ref{eq:concepts-pqc-general}) with 3 layers of rotation and entangling gates, including discarding.
Intuitively, {mixing (and thus discarding) will be necessary for the learning of \emph{red or circle} since it is more general and so greater than both the concepts for \emph{red} and \emph{circle} in the hierarchy on concepts, and each of these concepts in turn require the use of (mixed) discarding effects to ignore irrelevant domains (e.g. \emph{red} ignores all domains aside from \textsc{colour}).

% One intuition for how mixing enables the learning of such a concept is that, when the image contains a circle, the \textsc{colour} domain requires a mixed effect in order to allow any colour; and when the image contains a red shape, the \textsc{shape} domain requires a mixed effect to allow any shape; and these constraints can be satisfied with a suitable superposition in the space spanned by the relevant domains. However, whether the model in fact learns such an effect is left as investigation for future work. \scc{sean: does this intuition make sense in theory?}

\section{Is a Quantum Model a Conceptual Space?} \label{sec:Is-Quantum-Conceptual}

In comparing conceptual space models with our proposed quantum models, it is natural to ask to what extent we may view the latter as an instance of the former. That is, we ask: \emph{is a quantum model a conceptual space}? In fact this question is non-trivial and so we discuss it in some depth here. 

We will begin by leaving aside the factorisation of a model, and simply consider the case of a model with a single factor described by a Hilbert space. 

\paragraph{Hilbert space as a convex space.}

Naively we can first observe that any Hilbert space $\hilbH$ does count as a convex space. Indeed $\hilbH$ forms a complex vector space and so in particular allows us to take convex combinations 
\begin{equation} \label{eq:conv-comb-hilb}
\sum^n_{i=1} p_i \psi_i 
\end{equation} 
of elements $\psi_i \in \hilbH$, for any weights $p_i \in [0,1]$ with $\sum p_i = 1$. 

However, arbitrary vectors in $\hilbH$ do not have a direct physical interpretation, but only the unit vectors $\psi \in \hilbH$ with $\| \psi \| = 1$, which describe the pure states of the space (and these in turn are identified when equal up to global phase $e^{i \theta}$).  Thus the correct question is whether this resulting set of pure states forms a convex space. 
In fact this is not the case. Taking convex combinations \eqref{eq:conv-comb-hilb} of unit vectors does not yield a unit vector in general (and could even yield the vector $0 \in \hilbH$). Thus we cannot straightforwardly view the pure states as a convex space in the sense of Definition \ref{def:convex-space}. 

\paragraph{Pure states as a betweenness space. }

We can nonetheless view the pure states of a Hilbert space as a geometric space, akin to a conceptual space, in a different way. This is most evident for a qubit $\hilbH=\mathbb{C}^2$, whose pure states as we have seen may be visualised via the Bloch sphere. Each point on the surface of the Bloch sphere describes a unique pure quantum state. 
Though the Bloch sphere does not come with the straightforward style of convex mixing from Definition \ref{def:convex-space}, we can view it as an instance of a broader notion of convex space, known as a \emph{Betweenness space}. This is a set $Z$ along with a ternary operation $B(x,y,z)$ which intuitively states that the point $y$ is `in-between' the points $x$ and $z$. 

In \citeA{gardenfors2004conceptual} and \citeA{gardenfors2014geometry}, and in formalisations such as \citeA{aisbett2001general}, it is suggested that one may formalise all of conceptual space theory entirely in terms of such spaces. Indeed they allow one to define the central notion of \emph{convexity} of a subset by saying that $S$ is convex if, whenever $x, z \in S$ and $B(x,y,z)$ holds, then $y \in S$ also. 

The Bloch sphere may be seen as a Betweenness space by stating that $B(x,y,z)$ holds whenever a geodesic from $x$ to $z$ passes through $y$. Figure \ref{fig:bloch-convex} illustrates the betweenness relations and a convex subset in this space. Our question is now the following: to what extent is the quantum model of concepts on $\mathbb{C}^2$ the same as this view of the Bloch sphere as a Betweenness space $Z$? Our answer here is that in fact they are fundamentally different. 

\begin{figure} \label{fig:bloch-convex}
\centering 
\includegraphics[scale=0.18]{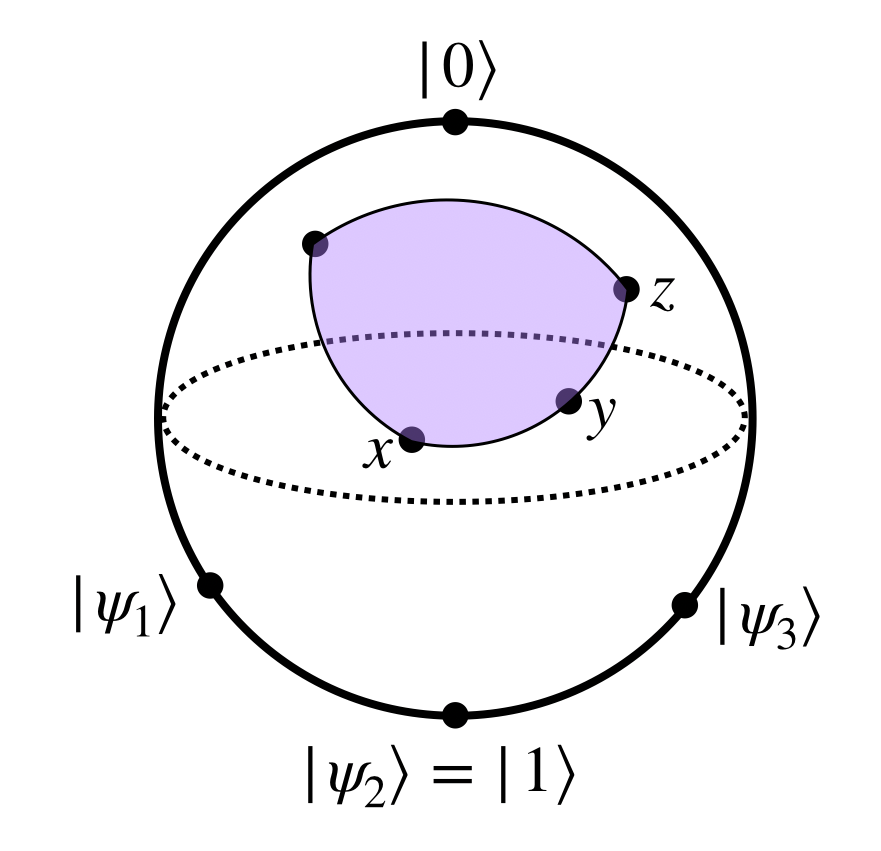}
\caption{Modelling of Bloch sphere as a Betweenness space, including an example of betweenness $B(x,y,z)$, and an example convex region shown in purple. The states $\ket{\psi_i}$ are used to show that the effect $\bra{0}$ is not quasi-concave. }
\end{figure} 

\paragraph{Comparing concepts.}

The most important reason for this difference is that the descriptions of concepts in the two models are different. 

Firstly, as we have seen, crisp concepts in the quantum model correspond to subspaces of $\hilbH$, which in the case of the Bloch sphere amount to either single points (dimension 1 subspaces) or the entire surface of the sphere (dimension 2 subspace). As such, most convex regions on the sphere, which form the crisp concepts in the Betweenness space $Z$, do not form valid quantum concepts.

% TOdo: Ideally cite the original quasi-concavity statement
Conversely, most quantum concepts do not qualify as valid fuzzy concepts in the Betweenness space $Z$. As argued in \citeA{tull2021categorical}, a fuzzy concept $C \colon Z \to [0,1]$ should at least satisfy the notion of `quasi-concavity', which states that each set 
\[
C_t := \{x \in X \mid C(x) \geq t \}
\]
is convex. However, Example \ref{ex:quasi-conc-fails} below illustrates that even pure quantum concepts may fail to satisfy this condition. 

Hence the quantum model on $\mathbb{C}^2$, and the Bloch sphere viewed as a Betweenness space $Z$, have distinct sets of concepts. We conclude that a quantum model is not a special case of a conceptual space.

\begin{example} \label{ex:quasi-conc-fails}
Consider the pure concept $C = \ket{0}\bra{0}$. Let $\ket{\psi_i} = \cos(\frac{\theta_i}{2}) \ket{0} + \sin(\frac{\theta_i}{2}) \ket{1}$ for $i=1,2,3$, as shown in Figure \ref{fig:bloch-convex}. Setting $\theta_1 = \frac{2 \pi}{3}$, $\theta_2 = \pi$, $\theta_3 = \frac{4 \pi}{3}$ we have that $\ket{\psi_2}\bra{\psi_2} = \ket{1}\bra{1}$ is in between $\ket{\psi_1} \bra{\psi_1}$ and $\ket{\psi_3} \bra{\psi_3}$, but this means that $C$ is not quasi-concave, since:
% since 
% \[
% C(\ket{\psi_i}\bra{\psi_i}) = \cos^2(\frac{\theta_i}{2})
% \]
% we see that $C$ is not quasi-concave:
\[
C(\ket{\psi_1}\bra{\psi_1}) = 
C(\ket{\psi_3}\bra{\psi_3}) =
\frac{1}{4} > 0 =
C(\ket{\psi_2}\bra{\psi_2})
\]
\end{example}

\paragraph{Spaces of mixed states.}

In fact one may view quantum concepts as valid fuzzy concepts on a different convex space, the space $\St(\hilbH)$ of all (pure and mixed) density matrices of $\hilbH$. Quasi-concavity of each quantum concept $C$ follows from the fact that, as it is a linear functional, it is affine, satisfying:
\[
C (\sum_i p_i \rho_i) = \sum_i p_i C(\rho_i)
\]
for density matrices $\rho_1, \dots, \rho_n$ and $\sum_i p_i = 1$. 

From this perspective, one may be tempted to view a quantum conceptual model as a convex space $Z=\St(
\hilbH)$ with arbitrary density matrices as its instances $z \in Z$.       However, since density matrices are interpreted as states of uncertainty over pure quantum states, it is more natural to view them as the analogues of \emph{distributions over} a conceptual space, rather than the instances themselves. 

\paragraph{Comparing tensors.}

Finally, even if one attempts to view a quantum model as a convex space $\St(\hilbH)$, the manner in which we compose such models via the tensor is fundamentally different in both cases, since we have:  
\[
\St(\hilbH \otimes \hilbK) = \St(\hilbH) \otimes \St(\hilbK) \neq \St(\hilbH) \times \St(\hilbK) 
\]   
The composition rule for factors can be considered an important ingredient of any class of models, and so this makes both classes distinct.

In summary, for all of these reasons, it is most natural to view quantum models as a distinct class of models from conceptual spaces. Nonetheless they possess the same benefits for learnability, with convex structure replaced by linear structure, and thanks to entanglement may be even more natural for describing correlated concepts.

\section{Future Work}
\label{sec:conclusion}

In this report we have presented a new model of concepts based on the quantum formalism, in which a concept is defined as an effect, showing how it can be implemented as a hybrid network trained to perform concept classification on images of simple shapes. We have also provided a thorough category-theoretic formalisation of the model using string diagrams. Through the use of entanglement and discarding, the model is able to successfully learn concepts containing correlations, and determine which domains are relevant for a concept. One avenue for future analysis of the learned concepts is to use some standard measures from quantum information theory to measure the level of entanglement and purity in the quantum effects \cite{entanglemeasures}. 

Another interesting avenue for future work is to apply the quantum concepts model to data generated from a conceptual hierarchy---for example having shades of
colour such as dark-red---making use of the natural ordering on effects. This would include exploring to what extent our learned quantum concepts already respect this ordering, for example whether a \emph{red or blue} concept is above each of \emph{red} and \emph{blue} as an effect. 
And continuing with the representation of colour, here we have only modelled
the hue---along a single dimension---whereas colour is more appropriately represented using something like a colour spindle. Whether the spindle can be appropriately represented on one or more qubits is an interesting theoretical and practical question. A quantum model of colour space close to our own can be found in \shortciteA{yan2021qhsl}. 

Further representational aspects of quantum models could also be explored. Along with the hierarchy, (sub-normal) concepts in $\Quant{}$ (as well as $\Class$) come with a \emph{negation} operation $C^\bot := \discard{} - C$, which has been studied in natural language \shortcite{rodatz2021conversational,shaikh2021composing,lewis2020}. In contrast, negation is harder to define for concepts in conceptual spaces; for example the complement of a convex region is generally non-convex.

In Tull (2021) the definition of fuzzy concepts is extended to define a symmetric monoidal category of fuzzy
conceptual processes between conceptual spaces. It would be interesting to explore the learning of such processes, including ``metaphorical" mappings between
domains, in our quantum framework. Beyond learning effects for concepts, this would involve learning general completely positive maps between quantum models.  

% Fuzzy conceptual processes also form a symmetric monoidal category. \scc{maybe one line more?}

%Much of the work in the hybrid network has been carried out by a classical CNN. Given the early stage of the field of quantum machine learning, using a hybrid network was entirely appropriate. However, it is an interesting question whether the whole network could be made quantum, perhaps using a quantum CNN \shortcite{quantum_cnn}, or even a form of quantum VAE \shortcite{khoshaman2018quantum}.

Finally, even though all the practical work here has been carried out in simulation on a classical computer, the number of qubits is relatively small, and the circuits are relatively shallow, and so the running of these models on real quantum hardware is a distinct possibility, and left for future work. Also left for future work is the search for tasks which could demonstrate advantages for our quantum representations, for example establishing whether non-separable effects in the theory do provide an advantage over classical correlation in modelling conceptual structure.

\section*{Acknowledgements}
Thanks to Lia Yeh, Robin Lorenz and Douglas Brown for extremely detailed and helpful comments on an earlier draft, and also to the rest of the Oxford Quantum Compositional Intelligence team.

\bibliographystyle{apacite}
\bibliography{cog}

\appendix 
\newpage

\section{The Shapes Dataset}
\label{sec:app_shapes}

The parameters used in the Spriteworld software to generate the Shapes dataset in Section~\ref{sec:shapes}:\\

\noindent
\includegraphics[width=12cm]{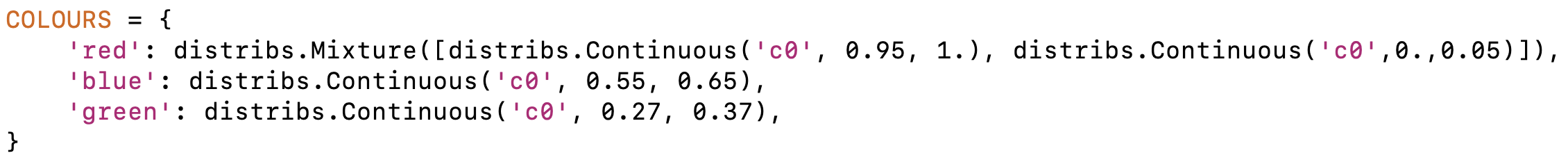}

\noindent
\includegraphics[width=7cm]{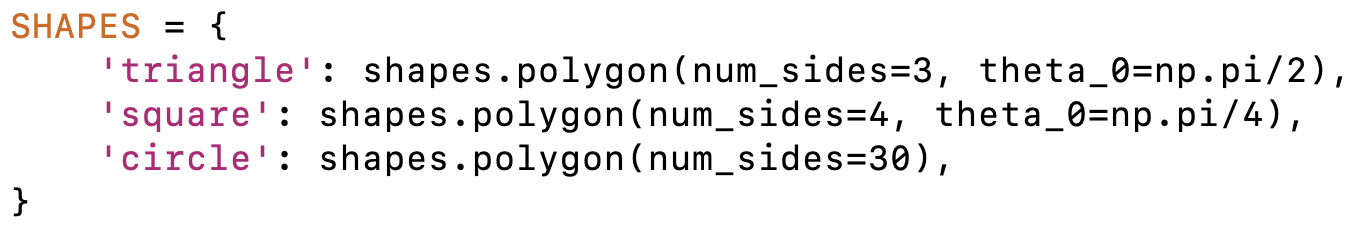}

\noindent
\includegraphics[width=6.5cm]{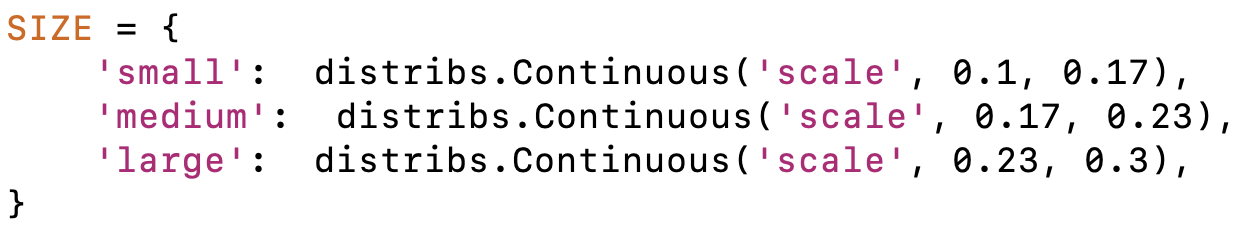}

\noindent
\includegraphics[width=12cm]{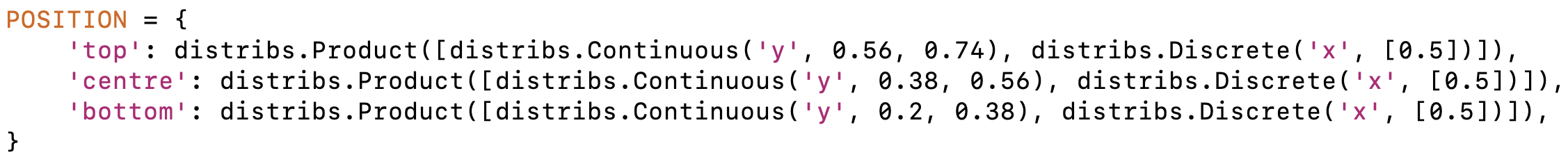}

\noindent
Additional parameters for the \textsc{colour} domain:\\

\noindent
\includegraphics[width=6cm]{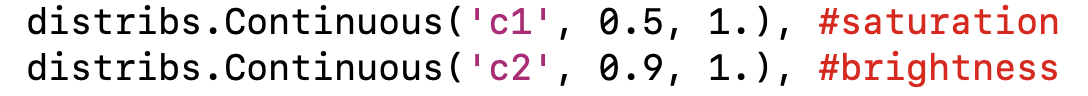}

\subsection{The Rainbow Colour Dataset}
\label{sec:ext_shapes}

The parameters used in the Spriteworld software to generate the Shapes dataset with more colours in Section~\ref{sec:rainbow}:\\

\noindent
\includegraphics[width=12cm]{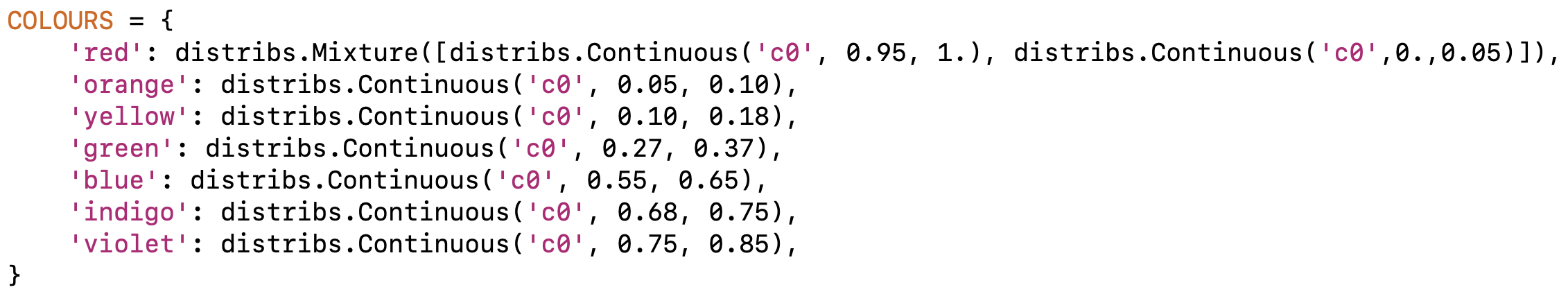}

\section{Network Architectures and Hyperparameters}
\label{sec:app_neural_nets}

\begin{center}
\begin{tabular}{@{}lr@{}}
\toprule
image width & 64\\
image height & 64\\
image channels & 3\\
\midrule
CNN kernel size & $4 \times 4$\\
CNN stride & $2 \times 2$\\
CNN layers & 4\\
CNN filters & 64\\
CNN dense layers & 2\\
CNN dense layer size & 256\\
\midrule
initialization interval for Concept PQC params & $[0.0, 1.0]$\\
\midrule
batch size & 64\\
Adam learning rate & $10^{-3}$\\
Adam $\beta_1$ & 0.9 \\
Adam $\beta_2$ & 0.999\\
Adam $\epsilon$ & $10^{-7}$\\
\bottomrule
\end{tabular}
\end{center}

\end{document}